\documentclass[]{egpubl}
\usepackage{eg2019}

\Tutorial
\electronicVersion      

\ifpdf \usepackage[pdftex]{graphicx} \pdfcompresslevel=9
\else \usepackage[dvips]{graphicx} \fi

\PrintedOrElectronic

\usepackage{t1enc,dfadobe}

\usepackage{egweblnk}
\usepackage{cite}

\title[SPH Techniques for the Physics Based Simulation of Fluids and Solids]%
{Smoothed Particle Hydrodynamics\\Techniques for the Physics Based Simulation of Fluids and Solids}

\author[D. Koschier, J. Bender, B. Solenthaler \& M. Teschner]
{
	\parbox{\textwidth}{\centering 
		Dan Koschier$^{1}$\orcid{0000-0002-2376-9475},
		Jan Bender$^{2}$, 
		Barbara Solenthaler$^{3}$, 
		and
		Matthias Teschner$^{4}$
	}
	\\
	{\parbox{\textwidth}{\centering 
			$^1$University College London, UK\\
			$^2$RWTH Aachen University, Germany\\
			$^3$ETH Zurich, Switzerland\\
			$^4$University of Freiburg, Germany
		} 
	}
	\\	
	\smallskip
	\\
	{\parbox{0.8\textwidth}{\centering 
	\textbf{Disclaimer}: This document is work-in-progress and will be maintained and updated over a longer period. Please find the current version of the document under \url{InteractiveComputerGraphics.github.io/SPH-Tutorial}.
 	}}
}

\usepackage{amsmath}
\usepackage{amssymb}
\usepackage{booktabs}
\usepackage{tabularx}
\usepackage{siunitx}
\usepackage[super]{nth}
\usepackage{mathtools}
\usepackage{suffix}
\usepackage{subfig}
\usepackage{algpseudocode}
\usepackage[plain]{algorithm}
\usepackage{overpic}
\usepackage{pgfplots}
\usepackage{upgreek}
\usepackage[utf8]{inputenc}
\usepackage{hyperref}

\usepackage{pgf,pgfplots}
\usepgfplotslibrary{colormaps,external,units} 
\usetikzlibrary{external,3d,fit,calc,patterns,arrows,decorations.pathmorphing,backgrounds,positioning,shapes,quotes,arrows.meta}

\pgfplotsset{compat=newest}

\definecolor{plotColor1}{RGB}{215,25,28}
\definecolor{plotColor2}{RGB}{253,174,97}
\definecolor{plotColor3}{RGB}{254,224,144}
\definecolor{plotColor4}{RGB}{171,221,164}
\definecolor{plotColor5}{RGB}{55,94,144}

\definecolor{fluid}{RGB}{55,94,144}
\definecolor{boundary}{RGB}{128,128,128}

\newcommand\mathcomma{\ensuremath{\,,}}
\newcommand\mathpoint{\ensuremath{\,.}}
\newcommand\ie{\textit{i.e.},\ }
\newcommand\Ie{\textit{I.e.},\ }
\newcommand\cf{\textit{cf.},\ }
\newcommand\eg{\textit{e.g.},\ }
\newcommand\Eg{\textit{E.g.},\ }
\newcommand\etc{\textit{etc}.\ }
\WithSuffix\newcommand\etc*{\textit{etc}.}

\usepackage{xparse}

\DeclareMathAlphabet{\mymathbb}{U}{BOONDOX-ds}{m}{n}

\renewcommand{\vec}[1]{\ensuremath{\boldsymbol{\mathbf{#1}}}}
\newcommand{\mat}[1]{\ensuremath{\vec{#1}}}

\newcommand{\Landau}[1]{\ensuremath{\mathcal O}({#1})}

\newcommand{\Time}{\ensuremath{t}}
\newcommand{\Dt}{\ensuremath{\Delta} \ensuremath{t}}
\newcommand{\D}{\ensuremath{\text{D}} }

\newcommand{\DynamicViscosity}{\ensuremath{\mu}}
\newcommand{\KinematicViscosity}{\ensuremath{\nu}}
\newcommand{\Gravity}{\ensuremath{\vec{g}}}

\newcommand{\NormalizeQuantity}[1]{\ensuremath{\tilde{#1}}}
\newcommand{\KernelNormalizationFactor}{\ensuremath{\sigma}}
\newcommand{\AverageQuantity}[1]{\ensuremath{\overline{#1}}}
\newcommand{\SPHDiscrete}[1]{\ensuremath{\langle {#1} \rangle}}
\newcommand{\FluidParticleSet}{\ensuremath{\mathcal F}}
\newcommand{\Neighborhood}{\ensuremath{\mathcal N}}
\newcommand{\RigidParticles}{\ensuremath{\mathcal R}}

\newcommand{\Laplace}{\ensuremath{{\nabla}^2}}
\newcommand{\Gradient}{\ensuremath{{\nabla}}}
\newcommand{\Divergence}{\ensuremath{{\nabla}\cdot}}
\newcommand{\Curl}{\ensuremath{{\nabla}\times}}
\newcommand{\Hessian}{\ensuremath{{\nabla\nabla}}}

\newcommand{\SpatialDimension}{\ensuremath{d}}
\newcommand{\SmoothingLength}{\ensuremath{h}}
\newcommand{\KernelSupportRadius}{\ensuremath{\hbar}}
\newcommand{\ParticleSize}{\ensuremath{\tilde h}}

\newcommand{\Predicted}{\ensuremath{\ast}}

\ExplSyntaxOn
\DeclareExpandableDocumentCommand{\IfNoValueOrEmptyTF}{ m m m }{\IfNoValueTF{#1}{#2}{\tl_if_empty:nTF {#1}{#2}{#3}}}
\DeclareExpandableDocumentCommand{\IfValueAndNotEmptyTF}{ m m m }{\IfNoValueOrEmptyTF{#1}{#3}{#2}}
\ExplSyntaxOff

\DeclareDocumentCommand{\DefineQuantity}{ o o o m }{%
  \ensuremath{
    \IfValueAndNotEmptyTF{#1}
    {
      \IfValueAndNotEmptyTF{#2}
      {
    	\IfValueAndNotEmptyTF{#3}
    	{{#4}_{#1}^{#3}{\left(#2\right)}}
    	{{#4}_{#1}{\left(#2\right)}}
	  }
      {
        \IfValueAndNotEmptyTF{#3}
        {{#4}_{#1}^{#3}}
        {{#4}_{#1}}
      }
    }
    {
      \IfValueAndNotEmptyTF{#2}
      {
    	\IfValueAndNotEmptyTF{#3}
    	{{#4}^{#3}{\left(#2\right)}}
    	{{#4}{\left(#2\right)}}
      }
      {
    	\IfValueAndNotEmptyTF{#3}
    	{{#4}^{#3}}
    	{{#4}}
      }    	
    }
  }
}

\DeclareDocumentCommand{\SPHVectorQuantity}{ o }{\DefineQuantity[#1]{\vec{A}}}
\DeclareDocumentCommand{\Position}{ o o o }{\DefineQuantity[#1][#2][#3]{\vec{x}}}
\DeclareDocumentCommand{\DistanceVector}{ o o o }{\DefineQuantity[#1][#2][#3]{\vec{r}}}
\DeclareDocumentCommand{\Velocity}{ o o o }{\DefineQuantity[#1][#2][#3]{\vec{v}}}
\DeclareDocumentCommand{\AngularVelocity}{ o o o }{\DefineQuantity[#1][#2][#3]{\vec{\omega}}}
\DeclareDocumentCommand{\MicroInertia}{ o o o }{\DefineQuantity[#1][#2][#3]{\Theta}}
\DeclareDocumentCommand{\BodyTorque}{ o o o }{\DefineQuantity[#1][#2][#3]{\vec{\tau}}}
\DeclareDocumentCommand{\Torque}{ o o o }{\DefineQuantity[#1][#2][#3]{\vec{\uptau}}}
\DeclareDocumentCommand{\Acceleration}{ o o o }{\DefineQuantity[#1][#2][#3]{\vec{a}}}
\DeclareDocumentCommand{\Force}{ o o o }{\DefineQuantity[#1][#2][#3]{\vec{F}}}
\DeclareDocumentCommand{\BodyForce}{ o o o }{\DefineQuantity[#1][#2][#3]{\vec{f}}}
\DeclareDocumentCommand{\PressureForce}{ o o }{\Force[#1][#2][p]}
\DeclareDocumentCommand{\PressureAcceleration}{ o o }{\Acceleration[#1][#2][\text{p}]}
\DeclareDocumentCommand{\NonPressureAcceleration}{ o o }{\Acceleration[#1][#2][\text{nonp}]}
\DeclareDocumentCommand{\InertiaTensor}{ o o o }{\DefineQuantity[#1][#2][#3]{\mat{I}}}

\DeclareDocumentCommand{\SPHScalarQuantity}{ o }{\DefineQuantity[#1]{Q}}
\DeclareDocumentCommand{\Quantity}{ o o o }{\DefineQuantity[#1][#2][#3]{\Phi}}
\DeclareDocumentCommand{\Mass}{ o }{\DefineQuantity[#1]{m}}
\DeclareDocumentCommand{\PseudoMass}{ o }{\DefineQuantity[#1]{\Psi}}
\DeclareDocumentCommand{\Volume}{ o o o }{\DefineQuantity[#1][#2][#3]{V}}
\DeclareDocumentCommand{\RestVolume}{ o o }{\Volume[#1][#2][0]}
\DeclareDocumentCommand{\NormalizedVolume}{ o o o }{\DefineQuantity[#1][#2][#3]{\NormalizeQuantity{V}}}
\DeclareDocumentCommand{\NormalizedRestVolume}{ o o }{\NormalizedVolume[#1][#2][0]}
\DeclareDocumentCommand{\Pressure}{ o o o }{\DefineQuantity[#1][#2][#3]{p}}
\DeclareDocumentCommand{\PressureVector}{ o o o }{\DefineQuantity[#1][#2][#3]{\vec{p}}}
\DeclareDocumentCommand{\Density}{ o o o }{\DefineQuantity[#1][#2][#3]{\rho}}
\DeclareDocumentCommand{\RestDensity}{ o o }{\Density[#1][#2][0]}
\DeclareDocumentCommand{\NumberDensity}{ o o o }{\DefineQuantity[#1][#2][#3]{\delta}}
\DeclareDocumentCommand{\NormalizedNumberDensity}{ o o o }{\DefineQuantity[#1][#2][#3]{\NormalizeQuantity{\delta}}}
\DeclareDocumentCommand{\ScalarAuxiliaryFunction}{ o o o }{\DefineQuantity[#1][#2][#3]{A}}
\DeclareDocumentCommand{\VectorAuxiliaryFunction}{ o o o }{\DefineQuantity[#1][#2][#3]{\vec{A}}}
\DeclareDocumentCommand{\DiracDelta}{ o o o }{\DefineQuantity[#1][#2][#3]{\delta}}

\DeclareDocumentCommand{\Kernel}{ o o o o }{\DefineQuantity[{#1}{#2}][#3][#4]{W}}
\DeclareDocumentCommand{\KernelGradient}{ o o o o }{\DefineQuantity[{#1}{#2}][#3][#4]{\Gradient W}}
\DeclareDocumentCommand{\KernelLaplace}{ o o o o }{\DefineQuantity[{#1}{#2}][#3][#4]{\Laplace W}}
\DeclareDocumentCommand{\NormalizedKernel}{ o o o o }{\DefineQuantity[{#1}{#2}][#3][#4]{\NormalizeQuantity{W}}}
\DeclareDocumentCommand{\NormalizedKernelGradient}{ o o o o }{\DefineQuantity[{#1}{#2}][#3][#4]{\Gradient\NormalizeQuantity{W}}}
\DeclareDocumentCommand{\LinearExactGradient}{ o o o }{\DefineQuantity[#1][#2][#3]{\mat{L}}}
\DeclareDocumentCommand{\LinearExactKernelGradient}{ o o o o }{\DefineQuantity[{#1}{#2}][#3][#4]{\tilde{\Gradient} W}}
\DeclareDocumentCommand{\KernelCorrection}{ o o o }{\DefineQuantity[#1][#2][#3]{\mat{L}}}
\DeclareDocumentCommand{\CorrectedKernelGradient}{ o o o o }{\DefineQuantity[{#1}{#2}][#3][#4]{\tilde{\Gradient} W}}
\DeclareDocumentCommand{\CorotatedKernelGradient}{ o o o o }{\DefineQuantity[{#1}{#2}][#3][#4]{\overset{\ast}{\Gradient} W}}

\DeclareDocumentCommand{\MatrixResidual}{ o o o }{\DefineQuantity[#1][#2][#3]{r}}
\DeclareDocumentCommand{\MatrixPreconditioner}{ o o o }{\DefineQuantity[#1][#2][#3]{P}}
\DeclareDocumentCommand{\Matrix}{ o o o }{\DefineQuantity[#1][#2][#3]{\mat{A}}}
\DeclareDocumentCommand{\MatrixElement}{ o o }{\DefineQuantity[{#1}{#2}]{a}}
\DeclareDocumentCommand{\MatrixSourceTerm}{ o o o }{\DefineQuantity[#1][#2][#3]{s}}
\DeclareDocumentCommand{\MatrixSourceTermVector}{ o o o }{\DefineQuantity[#1][#2][#3]{\vec{s}}}
\DeclareDocumentCommand{\MatrixPressureRow}{ o o o o  }{\ensuremath{(\Matrix\PressureVector[#2][#3][#4])_{#1}}}
\DeclareDocumentCommand{\CauchyStress}{ o o o }{\DefineQuantity[#1][#2][#3]{\mat T}}
\DeclareDocumentCommand{\FirstPiolaKirchhoffStress}{ o o o }{\DefineQuantity[#1][#2][#3]{\mat P}}
\DeclareDocumentCommand{\StrainRate}{ o o o }{\DefineQuantity[#1][#2][#3]{\mat E}}
\DeclareDocumentCommand{\Identity}{ o o o }{\DefineQuantity[#1][#2][#3]{\mymathbb{1}}}
\DeclareDocumentCommand{\ZeroVector}{ o o o }{\DefineQuantity[#1][#2][#3]{\vec 0}}
\DeclareDocumentCommand{\ShearRate}{ o o o }{\DefineQuantity[#1][#2][#3]{\mat S}}
\DeclareDocumentCommand{\ExpansionRate}{ o o o }{\DefineQuantity[#1][#2][#3]{\mat V}}
\DeclareDocumentCommand{\SpinRate}{ o o o }{\DefineQuantity[#1][#2][#3]{\mat R}}
\DeclareDocumentCommand{\Displacement}{ o o o }{\DefineQuantity[#1][#2][#3]{\vec u}}
\DeclareDocumentCommand{\RefPosition}{ o o o }{\DefineQuantity[#1][#2][#3]{\vec X}}
\DeclareDocumentCommand{\DeformationGradient}{ o o o }{\DefineQuantity[#1][#2][#3]{\vec J}}
\DeclareDocumentCommand{\Strain}{ o o o }{\DefineQuantity[#1][#2][#3]{\mat \epsilon}}
\DeclareDocumentCommand{\Rotation}{ o o o }{\DefineQuantity[#1][#2][#3]{\mat R}}

\begin{document}
	
	\maketitle

\begin{abstract}
Graphics research on Smoothed Particle Hydrodynamics (SPH) has produced fantastic visual results that are unique across the board of research communities concerned with SPH simulations.
Generally, the SPH formalism serves as a spatial discretization technique, commonly used for the numerical simulation of continuum mechanical problems such as the simulation of fluids, highly viscous materials, and deformable solids.
Recent advances in the field have made it possible to efficiently simulate massive scenes with highly complex boundary geometries on a single PC~\cite{Video:500million,Video:DFSPH}.
Moreover, novel techniques allow to robustly handle interactions among various materials~\cite{Video:Elastic,Video:Viscous}.
As of today, graphics-inspired pressure solvers, neighborhood search algorithms, boundary formulations, and other contributions often serve as core components in commercial software for animation purposes~\cite{Video:RealFlow} as well as in computer-aided engineering software~\cite{Video:PreonLab}.

This tutorial covers various aspects of SPH simulations.
Governing equations for mechanical phenomena and their SPH discretizations are discussed.
Concepts and implementations of core components such as neighborhood search algorithms, pressure solvers, and boundary handling techniques are presented.
Implementation hints for the realization of SPH solvers for fluids, elastic solids, and rigid bodies are given.
The tutorial combines the introduction of theoretical concepts with the presentation of actual implementations.

\keywords{Physically-based animation, Smoothed Particle Hydrodynamics, fluids, elastic solids, rigid bodies }

\begin{classification} 
  \CCScat{I.3.7}{Computer Graphics}{Three-Dimensional Graphics and Realism}{Animation}
\end{classification}

\end{abstract}


\section{Introduction}
\label{sec:introduction}

The SPH concept is increasingly popular in a large variety of application areas that range from entertainment technologies to engineering.
On the one hand, this popularity is based on the fact that Lagrangian approaches in general -- and SPH in particular -- can naturally handle scenarios that would be rather involved for Eulerian approaches.
A favorable example is a free-surface fluid with geometrically complex and dynamic solid boundaries.
Such settings are especially relevant for special effects productions in industry.
The scenario, however, has the same relevance in engineering, \eg for the analysis of vehicles in water passages, for the prediction of rain water evacuation on a vehicle with moving wipers or for the design of gear boxes with optimized lubrication. 

A second important aspect for the impressive advances in SPH based techniques is the fact that various research communities contribute to different aspects of SPH simulations. \Eg the simulation community has a strong focus on the accuracy of SPH discretizations or on specific properties of the discretizations.
Kernel functions and the effect of the size of kernel support domain are investigated. 
Effects of the sampling quality onto SPH approximations are analyzed, leading to concepts such as kernel gradient correction, particle shift, ambient pressure or density diffusion, just to name a few.
The computer science community -- the graphics community in particular -- focuses on efficient algorithms for neighborhood searches, efficient pressure solvers, and flexible boundary handling. Also, pre- and post-processing is a typical graphics topic, \eg boundary sampling and visualization.
The graphics community also experiments with combinations of different discretization concepts.
\Eg some projects have started to investigate combinations of SPH and MLS discretizations which is less typical in the simulation community, where we currently see a strong focus on SPH within Lagrangian approaches with exclusive SPH conferences and SPH initiatives.

Still, simulation and computer science are different communities, but there is a growing acceptance of advances across communities.
Graphics papers use state-of-the-art kernel functions, ranging from cubic spline to Wendland kernel types.
The kernel gradient correction is employed in a growing number of approaches.
Vice versa, the simulation community adopts efficient data structures for neighborhood searches, concepts for non-uniformly boundary samplings, and efficient pressure solvers. 

This tutorial aims at a practical introduction of the SPH concept and its application in the simulation of fluids, elastic solids, and rigid solids.
It starts with a description of the SPH concept and its usage for the interpolation of field quantities and for the computation of spatial derivatives.
Then, the governing equations for fluids and solids are stated and the SPH concepts for  the simulation of fluids and solids are outlined.
In the following, various aspects of SPH simulations are explained in more detail.
One of these aspects is the neighborhood search that is required for all SPH computations, as the interpolation of a quantity or a spatial derivative is always computed as a sum over adjacent particles.
Another important aspect is incompressibility which is not only relevant for fluids, but also, \eg in the case of elastic solids.
Next, boundary handling concepts are explained, \eg the interaction for fluid particles at solid walls, at free surfaces, \ie at the interface between fluid and air, or the interaction of particles from different fluids, \ie multiphase fluids.
Other topics are viscosity, surface tension, and vorticity. Further, the SPH simulation of elastic solids and SPH-based contact handling between rigid bodies is described.
Moreover, the techniques for the usage of SPH discretizations in data driven fluid simulations are presented. 
Finally, SPlisHSPlasH, an open-source library for the physically-based SPH simulation of fluids and solids, is introduced.
The most important quantities that will be used throughout this tutorial are summarized in Tab.~\ref{tab:intro:notation}.

\begin{table}[t]
	\centering
	{
		\small
		\begin{tabularx}{\linewidth}{clc}
			\toprule
			\textbf{Variable} & \textbf{Description} & \textbf{Unit} \\
			\midrule
			$\SpatialDimension$ & Spatial dimension & -- \\
			$\ScalarAuxiliaryFunction$ & Auxiliary function & -- \\
			$\Time$ & Time & \si{\second} \\
			$\Density$   & Volumetric mass density              & \si{\kilo\gram\per\cubic\meter}    \\
			$\Pressure$ & Pressure & \si{\pascal} \\
			$\Mass$   & Mass              & \si{\kilo\gram}    \\
			$\PseudoMass$   & Pseudo-mass              & \si{\kilo\gram}    \\
			$\DynamicViscosity$ & Dynamic viscosity & \si{\pascal\second} \\
			$\KinematicViscosity$ & Kinematic viscosity & \si{\square\meter\per\second} \\
			$\SmoothingLength$ & Smoothing length & \si{\meter}\\
			$\KernelSupportRadius$ & Kernel support radius & \si{\meter}\\
			$\ParticleSize$ & Particle size & \si{\meter}\\
			$\KernelNormalizationFactor$ & Kernel normalization factor & \si{\per\cubic\meter} \\
			$\Position$ & Position vector of material particle & \si{\meter}                     \\
			$\DistanceVector$ & Distance vector between two material particles & \si{\meter}                     \\
			$\Displacement$ & Displacement of a material particle & \si{\meter}                     \\
			$\Velocity$ & Velocity vector of material particle & \si{\meter\per\second} \\
			$\Acceleration$ & Acceleration vector of material particle & \si{\meter\per\square\second} \\
			$\AngularVelocity$ & Angular velocity vector of material particle & \si{\radian\per\second} \\
			$\BodyForce$ & Body force & \si{\newton\per\cubic\meter} \\
			$\Force$ & Force & \si{\newton} \\
			$\BodyTorque$ & Body torque & \si{\newton\meter\per\cubic\meter} \\
			$\Torque$ & Torque & \si{\newton\meter} \\
			$\MicroInertia$ & Microinertia & \si{\square\meter\per\second} \\			
			$\CauchyStress$ & Cauchy stress tensor					& \si{\newton\per\square\meter} \\
			$\FirstPiolaKirchhoffStress$ & \nth{1} Piola-Kirchhoff stress tensor & \si{\newton\per\square\meter} \\
			$\DeformationGradient$ & Deformation gradient & -- \\
			$\Strain$ & Strain tensor & -- \\
			$\StrainRate$ & Strain rate tensor & \si{\per\second} \\
			\bottomrule
		\end{tabularx}
	}
	\caption{Table of notation.}
	\label{tab:intro:notation}
\end{table}

	\section{Foundations}
\label{sec:foundations}

In this section, we introduce the fundamental concept of SPH for the phenomenological simulation of fluids and solids.
The section is primarily based on the excellent work of Price~\cite{Pri12a} and Monaghan~\cite{Mon05} but, moreover, includes important theoretical and practical insights that we have gained over the years working on SPH based techniques.

We first show how the SPH formalism discretizes spatial quantities using a set of particles equipped with a \emph{kernel} function.
Secondly, we discuss the approximation quality that can be expected and provide practical examples to illustrate the consequences for physics-based simulations targeting computer graphics applications.
Thirdly, we show how \nth{1}- and \nth{2}-order differential operators are discretized and present specialized variants of the discrete operators tailored to specific circumstances.
Finally, we give a brief introduction of the conservation law of linear momentum and the concept of stress in order to derive the governing equations for fluids and elastic solids and present a simple approach to simulate weakly compressible fluids using the knowledge that we have gained up to this point.

\subsection{SPH Discretization}

The concept of SPH can be generally understood as a method for the discretization of \emph{spatial field quantities} and \emph{spatial differential operators}, \eg gradient, divergence, curl, \etc*
In order to understand the basic idea, we first have to introduce the Dirac-$\DiracDelta$ distribution and the corresponding Dirac-$\DiracDelta$ identity.
$\DiracDelta$ is a generalized function defined as
\begin{equation}
	\DiracDelta(\DistanceVector) = \begin{cases}
		\infty & \text{if} \quad \DistanceVector = \vec 0 \\
		0 & \text{otherwise}
	\end{cases}
\end{equation}
and satisfies $\int \DiracDelta(\DistanceVector) \, dv = 1$.

\begin{figure}[t]
	\centering
	\includegraphics[width=\linewidth]{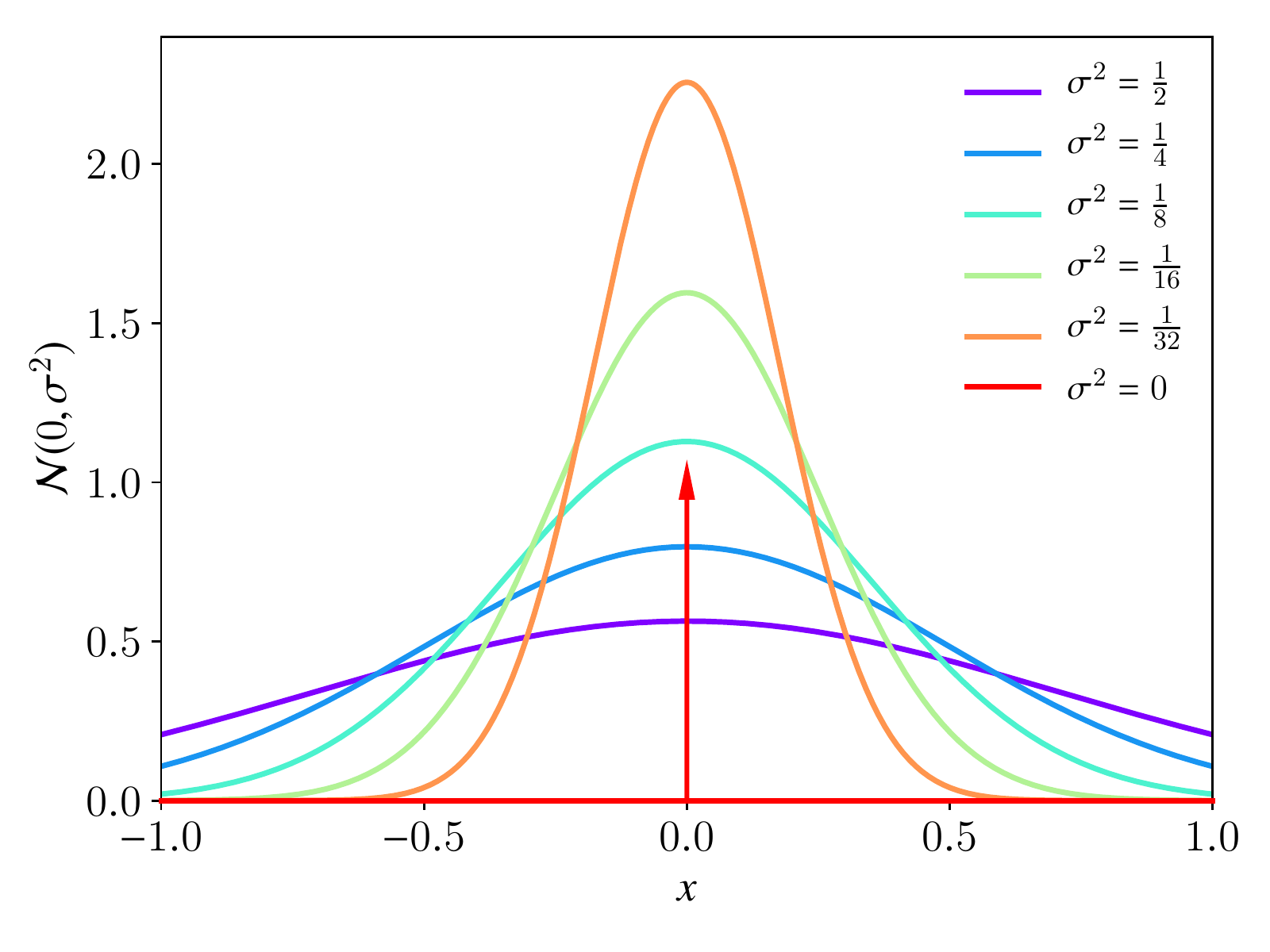}
	\caption{Gaussian bell function (normal distribution) $\mathcal N(0, \sigma^2)$ with varying variance $\sigma^2$. For $\sigma^2 \rightarrow 0$ the function approaches the Dirac-\DiracDelta distribution. The arrow indicates a function value of $\infty$. The function family has non-compact support.}
	\label{fig:gaussian_dirac}
\end{figure}

To provide a physical intuition of what this distribution describes, consider the following example.
In physics the mass of a body is usually defined as the spatial integral in the volumetric mass density, \ie $\Mass \eqqcolon \int \Density(\Position) \, dv$.
However, if an idealized point mass is considered, the concept of a density function loses its meaning as the point mass has no spatial extents.
In this case the density can not be described as a function, anymore, but collapses to the Dirac-$\DiracDelta$ distribution scaled using the point mass.
Another intuition of interpreting the Dirac-$\DiracDelta$ distribution is to understand it as the limit of the Gaussian normal distribution as the variance approaches zero (see Fig.~\ref{fig:gaussian_dirac}).

Now that we have understood the Dirac-$\delta$ distribution, we can apply the Dirac-$\delta$ identity as the basis for the discretization.
The identity states that the convolution of a continuous compactly supported function $\ScalarAuxiliaryFunction(\Position)$ with the Dirac-$\DiracDelta$ distribution is identical to $\ScalarAuxiliaryFunction$ itself, \ie
\begin{equation}
	\ScalarAuxiliaryFunction[][\Position] = \left( \ScalarAuxiliaryFunction * \DiracDelta \right)(\Position) = \int \ScalarAuxiliaryFunction[][\Position[][][\prime]] \DiracDelta[][\Position - \Position[][][\prime]] dv^\prime \mathcomma \label{eq:dirac_delta_identity}
\end{equation}
where $dv^\prime$ denotes the (volume) integration variable corresponding to $\Position[][][\prime]$.

\subsection{Continuous Approximation}

We will later approximate the integral of Eq.~\eqref{eq:dirac_delta_identity} using a sum for numerical quadrature.
Since $\DiracDelta(\DistanceVector)$ is, however, neither a function nor can be discretized, we first make a continuous approximation to the Dirac-$\DiracDelta$ distribution as a preparation to the discrete approximation of the integral.
A natural choice to approximate $\DiracDelta$ is to use a normalized Gaussian since $\DiracDelta$ is equal to the normal distribution with zero variance.
Consequently, convolving a field quantity $\ScalarAuxiliaryFunction$ with a Gaussian effectively smoothes $\ScalarAuxiliaryFunction$.
We will later see that the Gaussian is, however, not an optimal choice due to its non-compact support domain and will therefore consider more general smoothing functions $\Kernel[][]: \mathbb{R}^d \times \mathbb{R}^+ \to \mathbb{R}$ which we will refer to as \emph{kernel functions} or \emph{smoothing kernels}.
Formally the continuous approximation to $\ScalarAuxiliaryFunction(\Position)$ with $\Kernel[][][\DistanceVector, h]$ is
\begin{equation}
\begin{split}
	\ScalarAuxiliaryFunction[][\Position] &\approx \left( \ScalarAuxiliaryFunction * \Kernel[][] \right)(\Position) \\
	&= \int \ScalarAuxiliaryFunction[][\Position[][][\prime]] \Kernel[][][\Position - \Position[][][\prime], \SmoothingLength][] dv^\prime \mathcomma \label{eq:continuous_approximation}
\end{split}
\end{equation}
where $\SmoothingLength$ denotes the kernel's smoothing length.
The smoothing length controls the amount of smoothing and consequently how strongly the value of $\ScalarAuxiliaryFunction$ at position $\Position$ is influenced by the values in its close proximity.  
This means the smoothing effect increases with growing smoothing lengths.
The following properties are furthermore desired:
\begin{gather}
	\int_{\mathbb{R}^\SpatialDimension} \Kernel[][][\DistanceVector[][][\prime], \SmoothingLength] dv^\prime = 1 \; \quad \; \tag{\small normalization condition} \label{eq:normalization_condition} \\
	\lim_{\SmoothingLength^\prime \to 0} \Kernel[][][\DistanceVector[][][], \SmoothingLength^\prime] = \DiracDelta[][\DistanceVector] \vphantom{\int} \qquad \quad \; \tag{\small Dirac-$\DiracDelta$ condition} \\
	\Kernel[][][\DistanceVector[][][], \SmoothingLength] \geq 0 \vphantom{\int} \quad \quad \; \tag{\small positivity condition}\\
	\Kernel[][][\DistanceVector[][][], \SmoothingLength] = \Kernel[][][-\DistanceVector[][][], \SmoothingLength] \; \quad \; \vphantom{\int} \tag{\small symmetry condition} \\
	\Kernel[][][\DistanceVector[][][], \SmoothingLength] = 0 \ \text{for} \ \Vert \DistanceVector \Vert \geq \KernelSupportRadius, \vphantom{\int} \tag{\small compact support condition}
\end{gather} 
$\forall \; \DistanceVector \in \mathbb{R}^
\SpatialDimension, \SmoothingLength \in \mathbb{R}^+$, where $\KernelSupportRadius$ denotes the support radius of the kernel function.
Moreover, the kernel should be at least twice continuously differentiable to enable a consistent discretization of \nth{2}-order partial differential equations (PDEs).
It is essential to use a kernel that satisfies the first two conditions (normalization and Dirac-$\DiracDelta$), in order to ensure that the approximation in Eq.~\eqref{eq:continuous_approximation} remains valid.
The positivity condition is not strongly required (there are also kernels that do not have this property).
However, in the context of physical simulations kernels that take negative values may lead to physically inconsistent estimates of field quantities, \eg negative mass density estimates, and should therefore be avoided.
We will later see that the symmetry condition ensures \nth{1}-order consistency of the continuous approximation.
Finally, ensuring that the kernel is compactly supported is a purely practical consideration that will come into play after discretizing the continuous integral and will be discussed later.
To keep this tutorial practical, we refrain from discussing how to construct SPH kernels and would like to refer the reader to the review of Liu and Liu~\cite{LL10} for a discussion on kernel construction and an overview over a range of smoothing kernels suitable for SPH.

\begin{figure}[t]
	\centering
	\includegraphics[width=\linewidth]{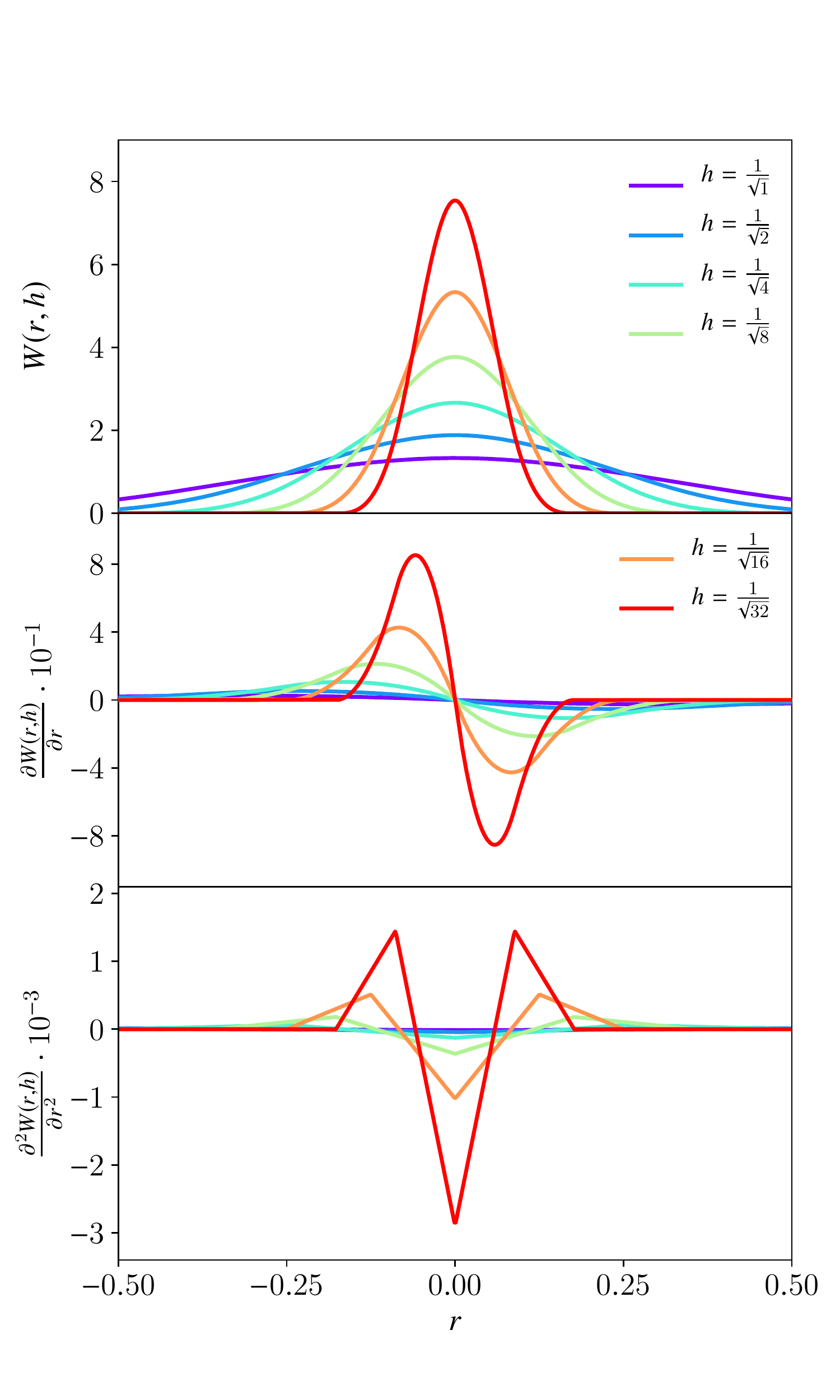}
	\caption{Graph of the cubic spline kernel (see Eq.~\eqref{eq:cubic_spline_kernel}) and its derivatives.}
	\label{fig:kernel}
\end{figure}

A typical choice for the smoothing kernel is the cubic spline kernel
\begin{equation}
	\Kernel[][][\DistanceVector, h] = \KernelNormalizationFactor_\SpatialDimension \begin{cases}
		6 (q^3 - q^2) + 1 & \text{for} \;  0 \leq q \leq \frac 1 2 \\
		2 (1 - q)^3 & \text{for} \; \frac 1 2 < q \leq 1 \\
		0 & \text{otherwise} \mathcomma
	\end{cases}
	\label{eq:cubic_spline_kernel}
\end{equation}
with $q = \frac 1 \SmoothingLength \Vert \DistanceVector \Vert$.
The kernel normalization factors for the respective dimensions $\SpatialDimension = 1,2,3$ are $\KernelNormalizationFactor_1 = \frac 4 {3 \SmoothingLength}$, $\KernelNormalizationFactor_2 = \frac {40} {7 \pi \SmoothingLength^2}$, and $\KernelNormalizationFactor_3 = \frac 8 {\pi \SmoothingLength^3}$.
Please note that there exist different formulations of the cubic spline kernel throughout SPH literature that are differently parametrized with respect to $\SmoothingLength$.
This kernel fulfills all of the discussed kernel properties and has the particular advantage that its smoothing length is identical to the kernel support radius, \ie $\SmoothingLength = \KernelSupportRadius$, which helps to avoid confusions in the implementation.
For a graphical illustration please see Fig.~\ref{fig:kernel}.
The plots demonstrate that the kernel is $C^2$-continuous.
Therefore, derivatives of order $> 1$ are not really useful in practice due to the lack of smoothness.
That is, however, not a major issue as there are more sophisticated approximations for \nth{2}-order derivatives solely based on the kernel gradient.
Otherwise, if desired, kernels of higher regularity can be found in the literature, \eg in the work of \cite{LL10}.

Let us consider the field $\ScalarAuxiliaryFunction: \mathbb{R}^\SpatialDimension \to \mathbb{R}$.
In order to investigate the accuracy of the continuous approximation, a
Taylor series expansion of $\ScalarAuxiliaryFunction$ in $\Position[][][\prime]$ about $\Position$ can be applied, \ie
\begin{align}
	\begin{split}
		(\ScalarAuxiliaryFunction * \Kernel[][])(\Position) &= \int \left[ \vphantom{\frac12} \right. \ScalarAuxiliaryFunction[][\Position] + \left. \nabla \ScalarAuxiliaryFunction \right|_{\Position} \cdot (\Position[][][\prime] - \Position) \; + \\
		&\qquad\quad \frac 1 2 (\Position[][][\prime] - \Position) \cdot \left. \Hessian \ScalarAuxiliaryFunction \right|_{\Position} (\Position[][][\prime] - \Position) \; + \\ 
		&\qquad\quad \left. \vphantom{\frac12} \Landau{(\Position[][][\prime] - \Position)^3} \right] \Kernel[][][\Position - \Position[][][\prime], \SmoothingLength][] dv^\prime
	\end{split} \\
	\begin{split}
		&= \ScalarAuxiliaryFunction[][\Position] \int \Kernel[][][\Position - \Position[][][\prime]] dv^\prime + \\
		& \quad\left. \Gradient\ScalarAuxiliaryFunction \right|_{\Position} \cdot \int (\Position - \Position[][][\prime]) \Kernel[][][\Position - \Position[][][\prime]] dv^\prime + \\
		& \quad \; \Landau{(\Position - \Position[][][\prime])^2} \mathpoint
	\end{split} \label{eq:continuous_approximation_taylor}
\end{align}
It is trivial to see that the approximation of $(\ScalarAuxiliaryFunction * \Kernel[][][])$ to $\ScalarAuxiliaryFunction$ is \nth{1}-order accurate if the integral in the first term of Eq.~\eqref{eq:continuous_approximation_taylor} becomes $1$, and if the integral in the second term vanishes.
The first condition is automatically fulfilled if the kernel is normalized (\cf normalization condition).
The second condition is met if the kernel is symmetric (\cf symmetry condition).
Consequently, given a normalized, symmetric kernel we can expect that the approximation is (at least) able to exactly reproduce functions up to \nth{1}-order.

\subsection{Discretization}
\label{sub:basics_discretization}

The remaining step to realize the SPH discretization is to replace the analytic integral in Eq.~\eqref{eq:continuous_approximation} by a sum over discrete sampling points as follows:
\begin{align}
	(\ScalarAuxiliaryFunction * \Kernel[][])(\Position[i]) 	&= \int \frac{\ScalarAuxiliaryFunction[][\Position[][][\prime]]}{\Density[][\Position[][][\prime]]} \Kernel[][][\Position - \Position[][][\prime], \SmoothingLength][] \underbrace{\Density[][\Position[][][\prime]]\, dv^\prime}_{dm^{\prime}} \\
	&\approx \sum_{j \in \FluidParticleSet} \ScalarAuxiliaryFunction_j \; \frac{\Mass[j]}{\Density[j]} \;\Kernel[][][\Position[i] - \Position[j], \SmoothingLength] \eqqcolon \SPHDiscrete{\ScalarAuxiliaryFunction[][\Position[i]]} \mathcomma \label{eq:discrete_sph}
\end{align}
where $\FluidParticleSet$ is the set containing all point samples and where all field quantities indexed using a subscript denote the field evaluated at the respective position, \ie $\ScalarAuxiliaryFunction[j] = \ScalarAuxiliaryFunction[][\Position[j]]$.
For improved readability, we will drop the second argument of the kernel function and use the abbreviation $\Kernel[i][j] = \Kernel[][][\Position[i] - \Position[j], \SmoothingLength]$ in the remainder of this tutorial.
The physical interpretation of this is that we keep track of a number of points that "carry" field quantities.
In this particular case, each point $j$ has a certain location $\Position[j]$ and carries a mass sample $\Mass[j]$ and a field sample $\ScalarAuxiliaryFunction[j]$.
It is not mandatory that the particle keeps track of its density $\Density[j]$ as this field can be reconstructed from its location and mass as explained later.
Due to the analogy to physical particles the term \emph{smoothed particle} has been coined in the pioneering work of Gingold and Monaghan~\cite{GM77}.
Nevertheless, we would like to stress the fact that a set of SPH particles \underline{must not} be misunderstood as discrete physical particles but simply as a spatial function discretization.

Analogously to the brief error analysis for the continuous approximation, a Taylor series expansion of $\SPHDiscrete{\ScalarAuxiliaryFunction}$ in $\Position[j]$ about $\Position_i$ reveals the accuracy of the discretization
\begin{equation}
\begin{split}
	\SPHDiscrete{\ScalarAuxiliaryFunction[][\Position[i]]} &= \ScalarAuxiliaryFunction[i] \sum_j \frac{\Mass[j]}{\Density[j]} \Kernel[i][j] \ +  \\
	& \quad \left. \Gradient\ScalarAuxiliaryFunction \right|_{\Position[i]} \cdot \sum_j \frac{\Mass[j]}{\Density[j]} (\Position[j] - \Position[i]) \Kernel[i][j] + \Landau{(\Position[j] - \Position[i])^2}. 
\end{split} \label{eq:sph_discrete_taylor}
\end{equation}
Due to the discretization the resulting approximation is only \nth{1}-order accurate if
\begin{equation}
	\sum_j \frac{\Mass[j]}{\Density[j]} \Kernel[i][j] = 1 \quad \text{and} \quad \sum_j \frac{\Mass[j]}{\Density[j]} (\Position[j] - \Position[i]) \Kernel[i][j] = \vec 0 \mathpoint \label{eq:discrete_consistency_conditions}
\end{equation}
Even presuming that a normalized symmetric kernel is used, the conditions are highly dependent on the sampling pattern leading to the fact that not even a \nth{0}-order consistent discretization can be guaranteed.
In practice, however, the approximation is sufficiently accurate to approximate physical field functions to obtain realistic simulations.
If desired, \nth{0}-order consistency can be easily restored by normalizing the SPH approximation with $\sum_j \frac{\Mass[j]}{\Density[j]} \Kernel[i][j]$ or even \nth{1}-order consistency can be restored by the cost of a small matrix inversion (see \cite{Pri12a}).

\begin{figure}[ht]
	\centering
	\begin{overpic}[width=\linewidth]{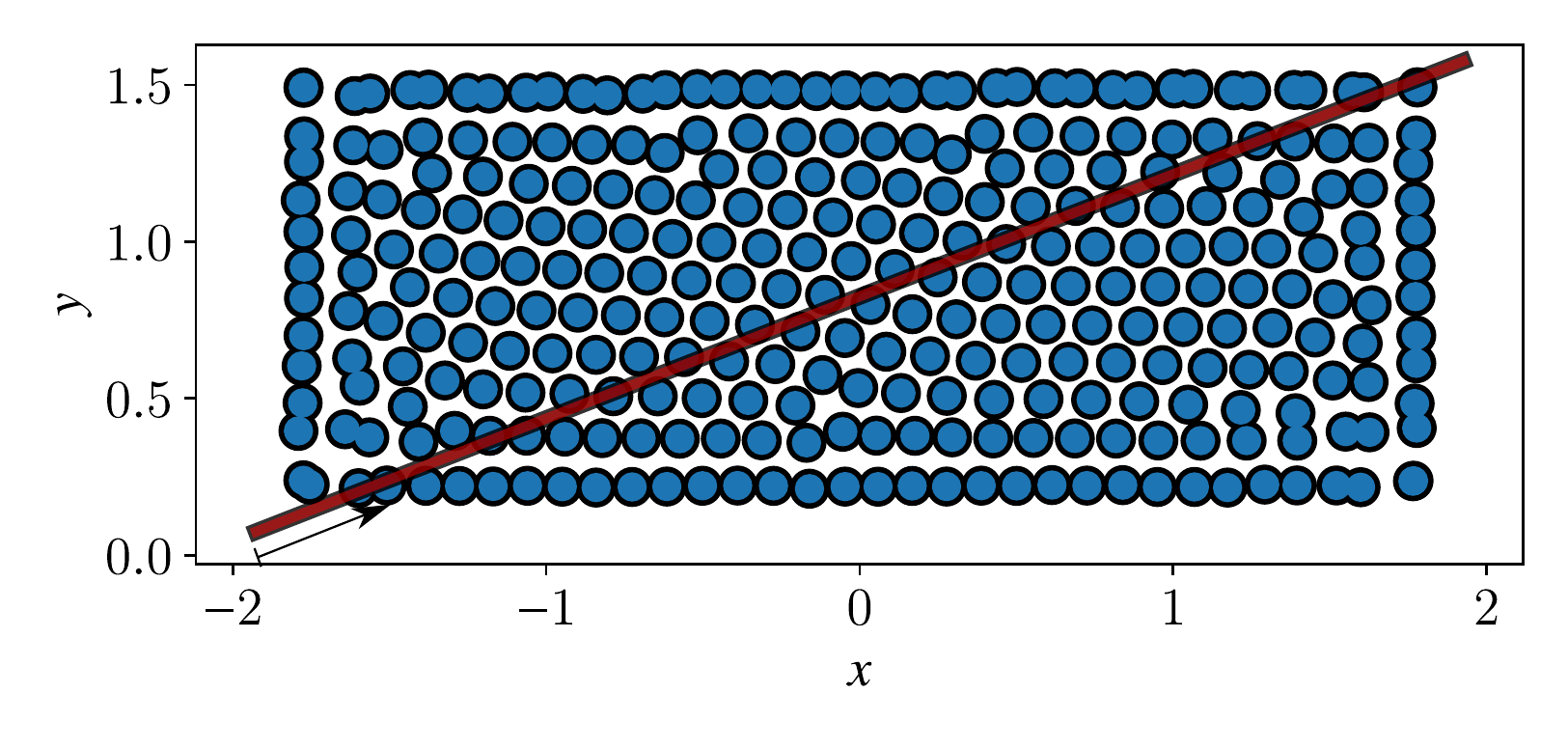}
		\put(20,11.8){$s$}
	\end{overpic}
	\caption{Point sampling of rectangular domain. Test functions are discretized using SPH. Function values are sampled along the red path parametrized by $s$.}
	\label{fig:test_pattern}
\end{figure}

\begin{figure}[ht]
	\centering
	\includegraphics[width=\linewidth]{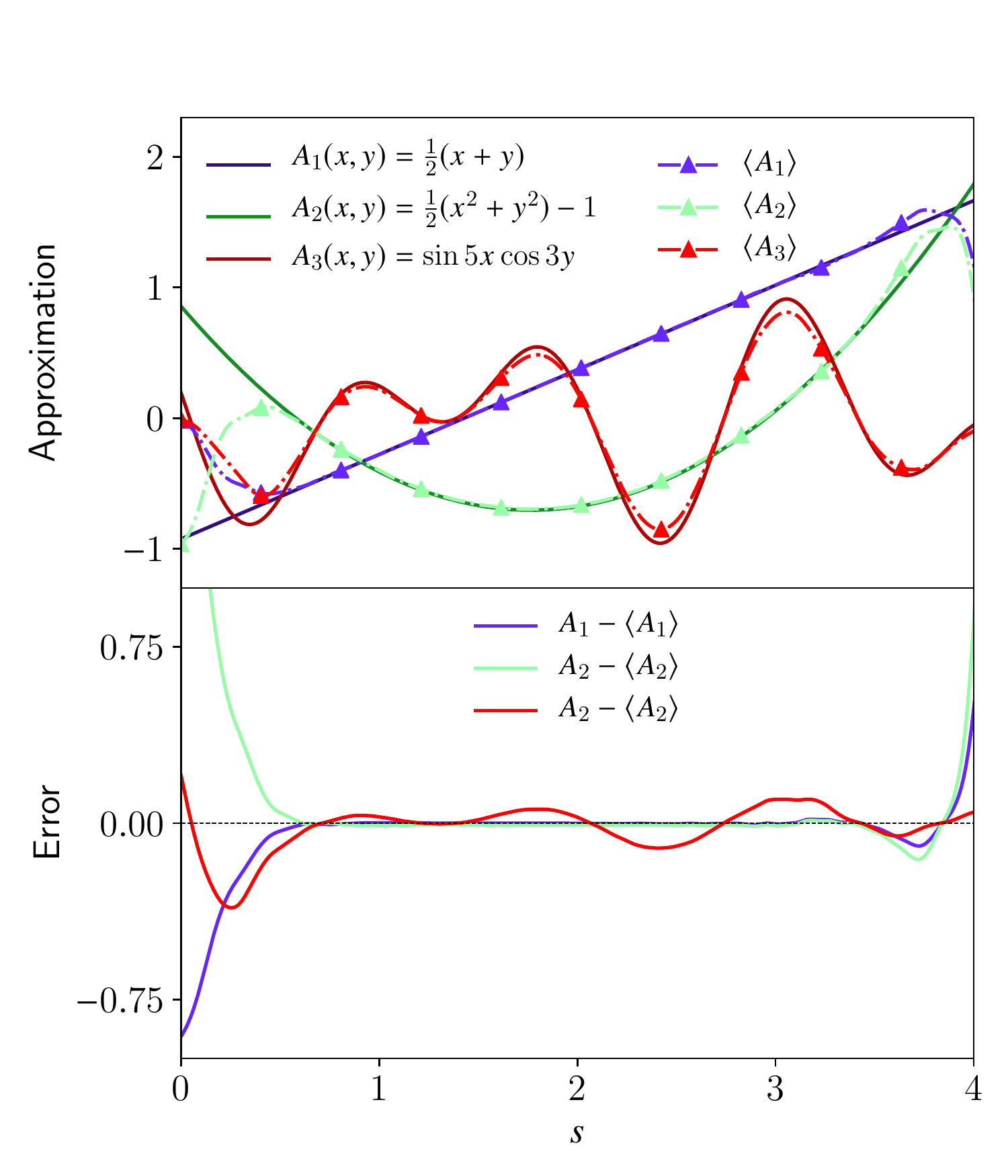}
	\caption{Comparison between analytic test functions and according SPH discretization using the sampling pattern illustrated in Fig.~\ref{fig:test_pattern}.}
	\label{fig:SPH_dis_samples}
\end{figure}

To give the reader a notion of the quality of the discrete approximation of functions, we have discretized a linear and a quadratic polynomial as well as a trigonometric function using a fairly coarse SPH discretization equipped with the cubic spline kernel. The sampling pattern is illustrated in Fig.~\ref{fig:test_pattern} while the the function and approximation graphs are depicted in Fig.~\ref{fig:SPH_dis_samples}.
In this example we have used the cubic spline kernel with a smoothing length of $\SmoothingLength=0.3 \si{\meter}$ and particle masses $\Mass[i]=18\si{\kilo\gram}$.
In order to find a suitable smoothing length given a dense (but not overlapping) sampling, we heuristically set the smoothing length to four times the particle radius, \ie $\SmoothingLength = 2 \ParticleSize$.
We, also recommend this heuristic to estimate a good smoothing length in practice.
In three-dimensional discretizations this leads to a number of approx.\ $30-40$ particles in a fully populated neighborhood.

Although no consistency can be strongly guaranteed in the absence of certain particle configurations that strongly fulfill the conditions in Eq.~\eqref{eq:discrete_consistency_conditions}, the graphs demonstrate that even a coarse sampling results in a discretization with good accuracy away from the boundary of the particle set.
The phenomenon of decreasing approximation quality in the close proximity of the domain boundary can be simply explained by the lack of sampling points outside the domain and is usually referred to as \emph{boundary deficiency}.
In the course of this tutorial practical solutions to this particular problem will be discussed.
We would also like to assure the reader that even without further considerations to recover the consistency order, SPH based approaches are able to produce robust and highly-realistic results as demonstrated in countless publications that have been published within recent decades.

\subsection{Mass Density Estimation}
\label{sec:mass_density_estimation}

As previously mentioned, it is not required that the particles "carry" the mass density field as it can be reconstructed.
Evaluating the density field at position $\Position[i]$ using the SPH discretization in Eq.~\eqref{eq:discrete_sph} results in
\begin{equation}
	\Density[i] = \sum_j \Mass[j] \Kernel[i][j] \label{eq:density_reconstruction}
\end{equation}
and is therefore solely dependent on the sample position and the mass field.
Alternatively, the density can be tracked by discretizing the mass density field using the SPH sampling and by numerical integration of the continuity equation which describes the density evolution, \ie $\dot{\Density} = -\Density (\Divergence\Velocity)$.
However, as also discussed by Randles and Libersky~\cite{RL96}, this approach is less robust and leads to accumulating errors in the density field due to the errors of the underlying numerical integration of the continuity equation. 

\begin{figure}[t]
	\centering
	\includegraphics[width=0.8\linewidth]{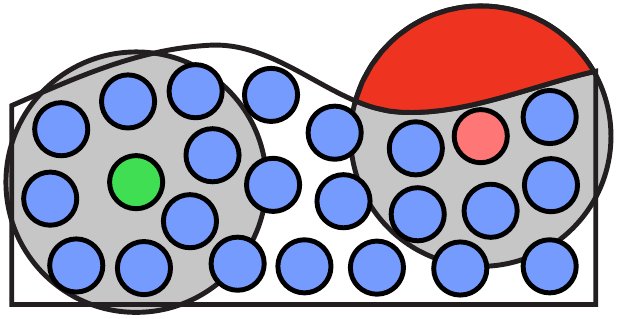}
	\caption{Particle deficiency problem. The green particle has a full neighborhood while the red particle at the free surface has no neighbors in the air. Therefore, its density is underestimated by Eq.~\eqref{eq:density_reconstruction}.
	}
	\label{fig:particle_deficiency}
\end{figure}

Note that the density can be reconstructed at any position by Eq.~\eqref{eq:density_reconstruction} but the reconstructed density is typically underestimated at the free surface due to particle deficiency (\cf Fig.~\ref{fig:particle_deficiency}).
This must be considered when implementing a pressure solver as discussed in-depth in Section~\ref{sec:pressure_solver}.

\subsection{Discretization of Differential Operators}
\label{sub:discretizatin_differential_operators}

Besides the discretization of field quantities, it is usually necessary to discretize spatial differential operators in order to numerically solve physical conservation laws.
In the remainder of this tutorial, we will assume that the smoothing length $\SmoothingLength$ is constant in space (and time).
Based on the discrete SPH approximation in Eq.~\eqref{eq:discrete_sph} the gradient of the underlying field can be approximated straightforwardly using
\begin{align}
	\Gradient\ScalarAuxiliaryFunction[i] &\approx \sum_j \ScalarAuxiliaryFunction[j] \frac{\Mass[j]}{\Density[j]} \Gradient\Kernel[i][j].
\end{align}
Given discrete representations of higher-dimensional functions, \eg $\VectorAuxiliaryFunction : \mathbb{R}^d \to \mathbb{R}^n$, even more complex first-order spatial differential operators can be directly discretized, \eg
\begin{align}
	\Gradient\VectorAuxiliaryFunction[i] &\approx \sum_j \frac{\Mass[j]}{\Density[j]} \VectorAuxiliaryFunction[j] \otimes \Gradient\Kernel[i][j] \label{eq:vec_direct_gradient} \\
	\Divergence\VectorAuxiliaryFunction[i] &\approx \sum_j \frac{\Mass[j]}{\Density[j]} \VectorAuxiliaryFunction[j] \cdot \Gradient\Kernel[i][j] \label{eq:vec_direct_divergence} \\
	\Curl\VectorAuxiliaryFunction[i] &\approx -\sum_j \frac{\Mass[j]}{\Density[j]} \VectorAuxiliaryFunction[j] \times \Gradient\Kernel[i][j] \mathcomma \label{eq:vec_direct_curl}
\end{align}
where $\vec a \otimes \vec b = \vec a \vec b^T$ denotes the dyadic product.
Unfortunately, these "direct" derivatives lead to a poor approximation quality and unstable simulations.
For this reason many discrete differential operators have emerged over time.

In this tutorial, we will cover the two most widely used formulations for first order derivatives, \ie the \emph{difference formula} and the \emph{symmetric formula}.

\subsubsection*{Difference Formula}

Analyzing the error in the gradient based on Taylor series expansion (similar to the one carried out in Eq.~\eqref{eq:sph_discrete_taylor}) reveals that the gradient estimate is only \nth{0}-order (\nth{1}-order) accurate if the first (both) of the following constraints are fulfilled:
\begin{equation}
	\SPHDiscrete{\Gradient 1} = \sum_j \frac{\Mass[j]}{\Density[j]} \Gradient_i \Kernel[i][j] = \vec 0 \quad \text{and} \quad \sum_j \frac{\Mass[j]}{\Density[j]} (\Position[j] - \Position[i]) \otimes \Gradient_i \Kernel[i][j] = \Identity \mathpoint \label{eq:gradient_accuracy_costraints}
\end{equation}
In order to recover \nth{0}-order accuracy we can simply subtract the first error term of the Taylor series resulting in the improved approximation
\begin{equation}
	\Gradient \ScalarAuxiliaryFunction[i] \approx \SPHDiscrete{\Gradient\ScalarAuxiliaryFunction[i]} - \ScalarAuxiliaryFunction[i]\SPHDiscrete{\nabla 1} = \sum_j \frac{\Mass[j]}{\Density[j]} (\ScalarAuxiliaryFunction[j] - \ScalarAuxiliaryFunction[i]) \Gradient_i \Kernel[i][j] \mathpoint
\end{equation}
In the rest of this thesis we will refer to this gradient estimate as \emph{difference formula}.
The same formula can be straightforwardly applied to the higher-dimensional first-order differential operators presented in Eqs.~\eqref{eq:vec_direct_gradient} to \eqref{eq:vec_direct_curl}.
This gradient estimate finally results in a more accurate discretization but keep in mind that we still expect a linear error.
However, linear accuracy is sometimes required and can be restored at the cost of solving a small linear equation system per evaluation, \ie
\begin{equation}
	\begin{split}
		\Gradient \ScalarAuxiliaryFunction[i] &\approx \vec L_i \left( \sum_j \frac{\Mass[j]}{\Density[j]} (\ScalarAuxiliaryFunction[j] - \ScalarAuxiliaryFunction[i]) \Gradient_i \Kernel[i][j] \right) \\
		\vec L_i &= \left( \sum_j \frac{\Mass[j]}{\Density[j]} \Gradient_i \Kernel[i][j] \otimes (\Position[j] - \Position[i]) \right)^{-1} \mathpoint
	\end{split}
\end{equation}

\subsubsection*{Symmetric Formula}

Motivated from classical mechanics for hydrodynamical systems, a discrete formula for the pressure force/gradient, starting from the discrete Lagrangian and the density estimate, can be derived.
This results in the following approximation
\begin{equation}
	\begin{split}
		\Gradient\ScalarAuxiliaryFunction[i] &\approx \Density[i] \left( \frac {\ScalarAuxiliaryFunction[i]} {\Density[i]^2} \SPHDiscrete{\Gradient \Density} + \SPHDiscrete{\Gradient \left( \frac{\ScalarAuxiliaryFunction[i]} {\Density[i]}\right)} \right) \\
		&= \Density[i] \sum_j \Mass[j] \left(\frac{\ScalarAuxiliaryFunction[i]}{\Density[i]^2} + \frac{\ScalarAuxiliaryFunction[j]}{\Density[j]^2}\right) \Gradient_i \Kernel[i][j] \mathpoint \label{eq:symmetric_formula}
	\end{split}
\end{equation}
Please note, that we did not include the lengthy derivation as this is out of the scope of this tutorial but kindly refer the reader to the report of Price~\cite{Pri12a}.

Since this formula does not satisfy the constraints in Eq.~\eqref{eq:gradient_accuracy_costraints}, it is clear that it is not able to exactly reproduce constant or linear gradient functions.
However, the massive advantage of this is that discrete physical forces using this particular gradient estimate exactly conserve linear and angular momentum which is an essential criterion for robust simulations.

Deriving the criterion using Taylor series expansion of Eq.~\eqref{eq:symmetric_formula} reveals that the constant error of the symmetric gradient is governed by how much
\begin{equation}
	\sum_j \Mass[j] \left(\frac 1 {\Density[i]^2} + \frac 1 {\Density[j]^2}\right) \Gradient_i\Kernel[i][j] \approx \vec 0 \label{eq:symmetric_constant_error}
\end{equation}
deviates from $0$.
As noted by Price~\cite{Pri12a}, the symmetric formulation "cares" about the particle ordering and the discrete physical forces will try to reorder the particle configuration until Eq.~\eqref{eq:symmetric_constant_error} is fulfilled.
This is in contrast to forces formulated using the difference formula.

To summarize, the difference formula does indeed lead to a more accurate gradient estimate than the symmetric formula.
In the context of physical forces the higher accuracy comes at the cost of a loss in momentum conservation and can therefore lead to unstable simulations.
For the stated reasons, we recommend to use gradient estimates of the symmetric type when quantities are discretized that directly affect particle trajectories, \eg physical forces, impulses, and to use the difference formula when \nth{1}-order differentials are estimated for indirect use, \eg the velocity divergence during pressure solves.

\subsubsection{Discretization of Laplace Operator}

Similar to the direct \nth{1}-order derivatives (Eqs.~\eqref{eq:vec_direct_gradient}-\eqref{eq:vec_direct_curl}) the Laplace operator can be directly discretized, \ie
\begin{equation}
	\Laplace\ScalarAuxiliaryFunction[i] \approx \sum_j \frac{\Mass[j]}{\Density[j]} \ScalarAuxiliaryFunction[j] \Laplace_i \Kernel[i][j] \mathpoint
\end{equation}
This, however, leads again to a very poor estimate of the \nth{2}-order differential.
A improved discrete operator for the Laplacian was presented by Brookshaw~\cite{Bro85}:
\begin{equation}
	\Laplace \ScalarAuxiliaryFunction[i] \approx -\sum_j \frac{\Mass[j]}{\Density[j]} \ScalarAuxiliaryFunction[ij] \frac{2 \Vert \Gradient_i \Kernel[i][j] \Vert}{\Vert \DistanceVector[ij] \Vert} \mathpoint
\end{equation}
The main idea leading to this particular formulation is to effectively use solely a \nth{1}-order derivative of the kernel function and to realize the second derivative using a finite-difference-like operation, \ie dividing by the particle distance.

\nth{2}-order derivatives of vectorial field quantities are realized analogously resulting in
\begin{align}
	\Laplace\VectorAuxiliaryFunction[i] &= -\sum_j \frac{\Mass[j]}{\Density[j]} \VectorAuxiliaryFunction[ij] \frac{2 \Vert \Gradient_i \Kernel[i][j] \Vert}{\Vert \DistanceVector[ij] \Vert} \label{eq:discrete_vector_laplace} \\
	\Gradient(\Divergence\VectorAuxiliaryFunction[i]) &= \sum_j \frac{\Mass[j]}{\Density[j]} \left[ (d + 2) (\VectorAuxiliaryFunction[ij] \cdot \NormalizeQuantity{\DistanceVector}_{ij}) \NormalizeQuantity{\DistanceVector}_{ij} - \VectorAuxiliaryFunction[ij] \right] \frac{\Vert \Gradient_i \Kernel[i][j] \Vert}{\Vert \DistanceVector[ij] \Vert} \mathcomma \label{eq:discrete_vector_grad_div}
\end{align}
where $\SpatialDimension$ and $\NormalizeQuantity{\DistanceVector} = \frac{\DistanceVector[ij]}{\Vert \DistanceVector[ij] \Vert}$ denote the spatial dimension and the normalized distance vector between particles $i$ and $j$, respectively.
A problem of the discrete Laplace operator defined in Eq.~\eqref{eq:discrete_vector_laplace} in the context of physics simulations is that forces derived using this operator, \eg viscosity forces, are not momentum conserving.
Fortunately, we get the following expression by adding together Eqs.~\eqref{eq:discrete_vector_laplace} and \eqref{eq:discrete_vector_grad_div}:
\begin{equation}
	\sum_j \frac{\Mass[j]}{\Density[j]} (\VectorAuxiliaryFunction[ij] \cdot \NormalizeQuantity{\DistanceVector}_{ij}) \NormalizeQuantity{\DistanceVector}_{ij} \frac{\Vert \Gradient_i \Kernel[i][j] \Vert}{\Vert \DistanceVector[ij] \Vert} \approx \frac {\Gradient(\Divergence\VectorAuxiliaryFunction[i])} {\SpatialDimension + 2} - \frac{\Laplace\VectorAuxiliaryFunction[i]}{2(\SpatialDimension + 2)} \mathpoint
\end{equation}
This identity has the important consequence that in the case of a divergence-free vector field, \ie $\Divergence\VectorAuxiliaryFunction = 0$, the Laplace operator can be discretized using
\begin{equation}
	\Laplace\VectorAuxiliaryFunction[i] \approx 2(\SpatialDimension + 2) \sum_j \frac{\Mass[j]}{\Density[j]} \frac{\VectorAuxiliaryFunction[ij] \cdot \DistanceVector[ij]}{\Vert \DistanceVector[ij] \Vert^2} \Gradient_i \Kernel[i][j] \mathcomma \label{eq:discrete_laplace}
\end{equation}
resulting in forces composed of terms that solely act along the "line of sight" between two interacting particles $i$ and $j$.
This particular choice has the advantage that derived physical forces recover momentum conservation \cite{Pri12a}.
Therefore, we recommend to use Eq.~\eqref{eq:discrete_laplace} as discrete Laplace operator for divergence-free vector fields.
In order to improve readability, we will drop the differentiation index for differential operators in the remainder of this tutorial.
We will use the convention that the spatial operators always differentiate with respect to the variable according to the first index such that \eg $\Gradient \Kernel[i][j] \equiv \Gradient_i \Kernel[i][j]$.

\subsection{Governing Equations for Fluids and Solids}
\label{sec:governing_equations}

In order to simulate the dynamic behavior of fluids and solids, a mathematical model that describes physical phenomena and motion of the matter is required.
In computer graphics related research, we are generally interested in the appearance of objects and fluids in motion on humanly perceivable scales which is dominantly governed by the matter's macroscopic behavior.
An important class of mathematical models that describe the macroscopic mechanical behavior of fluids and solids is based on continuum theory.
Unfortunately, we can not cover an introduction to continuum mechanics as this is out of the scope of this tutorial.
For a thorough introduction we would like to refer the reader to the works of Abeyaratne~\cite{Abe12} and Lai et al.~\cite{LKR10}.
Nevertheless, we would like to informally describe the basic idea of continuum theoretical models in the following.

Physics teaches us that all matter is formed out of discrete particles such as atoms, molecules, \etc* 
Therefore, we know that the distribution of mass within matter is not continuous but can rather be interpreted as a system of discrete mass points.
Nonetheless, the vast majority of macroscopic mechanical phenomena can be accurately described when the corresponding matter is idealized as a continuum, \ie a region of continuously distributed mass.
This idealization then implies that a portion of matter can always be divided into smaller portions independent of the size of the regions.
This in turn confirms the theoretical existence of a \emph{material particle}, \ie a portion of matter contained in an infinitesimal volume.
Continuum theory then aims to model macroscopic physical phenomena and neglects effects that can be observed on microscales.
In the following, we will summarize the most important local conservation laws required for the numerical simulation of (in)compressible fluids and solids.

\subsubsection*{Continuity Equation}

The continuity equation describes the evolution of an object's mass density $\Density$ over time, \ie
\begin{equation}
	\frac{D \Density}{D \Time} = -\Density \left(\Divergence\Velocity\right) \mathcomma
\end{equation}
where $\frac{D (\cdot)}{D \Time}$ denotes the \emph{material derivative}.
This relation is especially important when incompressible materials are modelled.
In this particular case the constraint
\begin{equation}
	\frac{D \Density}{D \Time} = 0 \quad \Leftrightarrow \quad \Divergence\Velocity = 0 \label{eq:incompressibility_constraint}
\end{equation}
has to be fulfilled at every material point and at all times within the described matter.

\paragraph*{A note on the material derivative:}
The material derivative describes the time rate of change of a field quantity at a material point.
It is important to understand that the explicit form of the material derivative is dependent on the type of coordinates that are used to the describe the system.
\emph{Eulerian coordinates} describe a field quantity at spatially fixed points in space, observing the motion of the continuum as time passes.
This type of coordinates is usually employed for mesh-based simulation techniques for fluids.
In contrast, \emph{Lagrangian coordinates} "track" the individual material particles as they move through space and time.
Lagrangian coordinates are commonly employed for the particle based simulation of fluids, such as SPH, or the mesh-based simulation of elastic solids.
Given the same field quantity once described in Eulerian coordinates $\ScalarAuxiliaryFunction[][][E](\Time, \Position)$ and Lagrangian coordinates $\ScalarAuxiliaryFunction[][][L](\Time)$ the material derivative has the following explicit forms
\begin{equation}
	\frac{D\ScalarAuxiliaryFunction[][][E]}{Dt} = \frac{\partial \ScalarAuxiliaryFunction[][][E]}{\partial t} + \Velocity \cdot \Gradient_{\Position} \ScalarAuxiliaryFunction[][][E] \quad \text{and} \quad \frac{D\ScalarAuxiliaryFunction[][][L]}{Dt} = \frac{\partial \ScalarAuxiliaryFunction[][][L]}{\partial t} \mathpoint
\end{equation}
The second term of the material derivative for Eulerian coordinates is referred to as \emph{convection term} or \emph{self-advection} term.
As opposed to some people's beliefs, the convection term is non-existent when a quantity is described in Lagrangian coordinates.
In the remainder of this tutorial, we will exclusively describe quantities using Lagrangian coordinates.

\subsubsection{Conservation Law of Linear Momentum}

The conservation law of linear momentum can be interpreted as a generalization of Newton's second law of motion for continua and is also often called the \emph{equation of motion}.
It states that the rate of change of momentum of a material particle is equal to the sum of all internal and external volume forces acting on the particle, \ie
\begin{equation}
	\Density \frac{D^2 \Position}{D t^2} = \Divergence\CauchyStress + \BodyForce[\text{ext}] \mathcomma \label{eq:momentum_conservation}
\end{equation}
where $\CauchyStress$ denotes the stress tensor and $\BodyForce[\text{ext}]$ body forces -- we understand a body force as a force per unit volume.
This equation is independent of the material of the underlying matter as the material's behavior is "encoded" in the stress tensor and described using so-called constitutive laws.

\paragraph*{Navier-Stokes Equation}
A typical constitutive relation for incompressible flow is
\begin{equation}
	\CauchyStress = -\Pressure \Identity + \DynamicViscosity (\Gradient\Velocity + \Gradient\Velocity^T) \mathcomma \label{eq:nv_constitutive_law}
\end{equation}
where $\Pressure$ and $\DynamicViscosity$ denote the pressure and dynamic viscosity of the fluid.
If the incompressibility is intended to be strongly enforced, the pressure $\Pressure$ can be interpreted as a Lagrange multiplier that has to be chosen such that the Eq.~\eqref{eq:incompressibility_constraint} is fulfilled.
If strong enforcement of incompressibility is not required, the constitutive relation can be instead completed by a so-called \emph{state equation} that relates geometric compression (changes in mass density) with the pressure, \ie $\Pressure = \Pressure[][\Density]$ (see Section~\ref{sub:state_equation}).
A simple example for a state equation is a variation of the ideal gas equation that linearly penalizes deviations from a rest density $\RestDensity$ scaled by a positive stiffness factor $k$ resulting in $\Pressure[][\Density] = k \left (\frac \Density \RestDensity - 1 \right )$.

By plugging Eq.~\eqref{eq:nv_constitutive_law} into Eq.~\eqref{eq:momentum_conservation} we arrive at the incompressible Navier-Stokes equation
\begin{equation}
	\Density \frac{D\Velocity}{D\Time} = -\Gradient \Pressure + \DynamicViscosity \Laplace\Velocity + \BodyForce[\text{ext}] \mathpoint \label{eq:navier_stokes}
\end{equation}

\paragraph*{Elasticity}
The stress tensor of elastic solids is solely dependent on the geometric deformation of an object, \eg \CauchyStress = \CauchyStress[][\DeformationGradient], where $\DeformationGradient[]$ denotes the deformation gradient which will be later introduced in Section~\ref{sec:deformable_solids}.
Obviously, the constitutive model can be augmented accordingly, if viscoelastic, plastic, thermoelastic, or other deformation inducing phenomena have to be modeled. 

\subsection{Mixed Initial-Boundary Value Problem}
The previously introduced linear momentum conservation law (Eq.~\eqref{eq:momentum_conservation}) in combination with a constitutive relation, \eg Eq.~\eqref{eq:nv_constitutive_law}, is a PDE in time and space that describes the motion of any object composed of the material modeled by the constitutive law.
In order to model a specific problem and to ensure a unique solution, initial conditions, \ie the initial shape and velocity of the object at every point, and boundary conditions constraining the position and/or velocity field have to be specified.
As there is, in general, no known analytic solution to the mixed initial-boundary value problem in arbitrary scenarios, numerical solving is inevitable and requires discretization of the associated differential operators.
In the previous sections, we have seen several discrete differential operators based on the SPH formalism that can be employed to discretize the spatial differential operators.
After spatial discretization, we are left with a system of ordinary differential equations (this methodology is often called \emph{method of lines}) that is typically discretized using standard time integration schemes such as the implicit or explicit Euler method, Runge-Kutta schemes, \etc*
In the remainder of this tutorial, we will see several variations of these discretizations tailored to specific problems in physics based simulation.

\subsection{Operator Splitting}

Before we will discuss a simple example of a complete simulation loop, the concept of operator splitting is introduced.
Its importance is emphasized by the fact that the vast majority of today's SPH based simulators follow the concept.
The basic idea is to decompose the underlying PDE, \eg the Navier-Stokes equation in the case of fluids, into several sequential subproblems and to employ individual techniques for solving each subproblem.
This simplifies the complexity of the overall problem and sometimes also decouples field variables such as velocity and pressure in the numerical solver.
It moreover allows us to use stable implicit updates for stiff subproblems while cheap explicit updates for the remaining terms can be used.
A potential operator split for the incompressible Navier-Stokes equation (Eq.~\eqref{eq:navier_stokes}) for low-viscous fluids with strong enforcement of the incompressibility constraint (Eq.~\eqref{eq:incompressibility_constraint}) might look as described in the following.
Given the current geometry of the continuum $\Position[][\Time]$ and its velocity field $\Velocity[][\Time]$ at time $\Time$, we split the problem into a sequence of subproblems in order to obtain $\Position[][\Time + \Delta\Time]$ and $\Velocity[][\Time + \Delta\Time]$:
\begin{enumerate}
	\item Update $\Velocity$ by solving $\frac{D\Velocity}{D\Time} = \KinematicViscosity \Laplace\Velocity + \frac 1 \Density \BodyForce[\text{ext}]$,
	\item determine $\Gradient\Pressure$ by enforcing $\frac{D\Density}{D\Time} = 0$,
	\item update $\Velocity$ by solving $\frac{D\Velocity}{D\Time} = -\frac 1 \Density \Gradient p$ and
	\item update $\Position$ by solving $\frac{D\Position}{D\Time} = \Velocity$,
\end{enumerate}
where $\KinematicViscosity = \frac{\DynamicViscosity}{\Density}$ denotes the kinematic viscosity.
In this way, the "weaker" forces could be handled using explicit time integration while we can solve for the pressure gradients using a more sophisticated implicit solver in order to keep the simulation robust for large time steps.
It should further be noticed that the individual steps are not performed in parallel but the updated fields (in this case $\Velocity$ and $\Gradient\Pressure$) are fed forward into the next substep resulting in a somewhat implicit handling which has demonstrated to improve stability in practice as also discussed by Bridson~\cite{Bri15} for grid-based fluid simulation.

\subsection{Time Integration}

As previously described, an SPH discretization of the underlying PDE leaves us with a system of ordinary differential equations (ODEs) in time following the method of lines.
This, of course, requires us to discretize the ODE in time.
Due the operator splitting approach, as introduced in the previous section, each individual subproblem has to be numerically integrated in time.
Theoretically, a different time integration scheme can be employed for each individual step.
In, practice most methods mainly rely on simple and efficient explicit time integration schemes.
The, by far, most frequently used scheme is the semi-implicit Euler scheme, as \eg employted in \cite{BK17,IAAT12,SB12,ICS+14}).
The integration scheme is often also referred to as symplectic Euler or Euler-Cromer scheme.
Sometimes it is useful to solve some of the individual substeps using implicit time integration schemes to ensure stability in the case of "stiff" forces.
A typical example where this strategy is employed is in the case of simulating highly viscous fluids. 
Here, the viscosity force is often integrated implicitly using the implicit Euler scheme as discussed in Section~\ref{sec:viscosity}.

Naturally, we aim for the best performance of our simulator and, therefore, try to use a very large time step width $\Delta\Time$\footnote{We will later see that larger time step widths not always result in better performance. This is especially true when iterative pressure solvers are employed (see Sec.~\ref{sec:pressure_solver}).}.
However, we also understand that choosing an overly large time step width results in decreased accuracy of the numerical approximation and may lead to a less stable simulation which might ultimately result in a breakdown of the simulation.
In the context of computer graphics research, we care most about carrying out a robust and stable simulation in a resource-efficient manner while the numerical accuracy is often of subordinate importance.
This does not mean that we do not care about accuracy at all, as the realism of the resulting animations often improves with better accuracy of the numerical approximation.
We are simply putting a higher priority on maintaining a robust and stable simulation under extreme conditions and in highly complex scenarios than on achieving the highest possible accuracy.

In order to find a "good" time step width $\Delta\Time$ that is as large as possible to achieve high performance but sufficiently small to maintain stability, the vast majority of approaches adaptively estimate the time step using a heuristic based on the Courant-Friedrichs-Lewy (CFL) condition.
The CFL condition is a necessary condition for the convergence of numerical solvers for differential equations and, as a result, provides an upper bound for the time step width, \ie
\begin{equation}
	\Delta\Time \leq \lambda \frac{\ParticleSize}{\Vert \Velocity[][][\text{max}] \Vert} \mathcomma \label{eq:CFL}
\end{equation}
where $\ParticleSize$, $\Velocity[][][\text{max}]$, and $\lambda$ denote the particle size, the velocity at which the fastest particle travels and a user-defined scaling parameter, respectively.
The intuition behind this condition is that all particles are only allowed to move less than the particle diameter per time step for $\lambda = 1$.
As this is only a necessary but no generally sufficient condition, the scaling parameter is heuristically chosen to keep the simulation stable, \ie $\lambda \approx 0.4$ \cite{Mon92}.
This can not strongly guarantee stability but experience from practice has shown that the condition typically leads to stable simulations \cite{SP09,ICS+14,BK17}.
Although obvious from Eq.~\eqref{eq:CFL}, we would like to stress the fact that the maximally allowed time step decreases with higher velocities and spatial resolution.
We would further like to point out that it is in practice useful to specify global bounds, \ie a lower and upper bound, for the time step as we want to produce a certain number of frames per second and want to avoid that the simulation comes to halt if a single particle moves with very high velocity.

\subsection{Example: Simple Fluid Simulator}

Based on the knowledge that we have acquired up to this point, we are now able to implement a simple state-equation based simulator for weakly compressible fluids with operator splitting using SPH and symplectic Euler integration.
\begin{algorithm}
	\caption{Simulation loop for SPH simulation of weakly compressible fluids.}
	\label{alg:WCSPH}
	\begin{algorithmic}
	\algnotext{EndFor}
	\algnotext{EndWhile}
	\algnotext{EndProcedure}
	\ForAll{$particle~i$}
	\State Reconstruct density $\Density[i]$ at $\Position[i]$ with Eq.~\eqref{eq:density_reconstruction}
	\EndFor
	\ForAll{$particle~i$}
	\State Compute $\Force[i][][\text{viscosity}] = \Mass[i] \KinematicViscosity \Laplace \Velocity[i]$, \eg using Eq.~\eqref{eq:discrete_vector_laplace}
	\State $\Velocity[i][][\Predicted] = \Velocity[i] + \frac{\Delta\Time}{\Mass[i]} (\Force[i][][\text{viscosity}] + \Force[i][][\text{ext}])$
	\EndFor
	\ForAll{$particle~i$}
	\State Compute $\Force[i][][\text{pressure}] = -\frac{1}{\Density} \Gradient\Pressure$ using state eq. and Eq.~\eqref{eq:symmetric_formula}	
	\EndFor
	\ForAll{$particle~i$}
	\State $\Velocity[i](\Time + \Delta\Time) = \Velocity[i][][\Predicted] + \frac{\Delta\Time}{\Mass[i]} \Force[i][][\text{pressure}] $
	\State $\Position[i](\Time + \Delta\Time) = \Position[i] + \Delta\Time \Velocity[i](\Time + \Delta\Time)$
	\EndFor
	\end{algorithmic}
\end{algorithm}

The few lines in Algorithm~\ref{alg:WCSPH} are already enough to implement a simple fluid solver.
However, the algorithm does, unfortunately, not handle boundary conditions.
A practical workaround to model boundaries in the discrete model is to sample the boundary geometry with static (non-moving) fluid particles.
The pressure forces will then "push away" particles that attempt to penetrate the boundary.
A more consistent handling of boundary conditions will be discussed in Section~\ref{sec:boundary_handling}.

	\section{Neighborhood Search}
\label{sec:neighborhood_search}

A major insight that we can gain from Algorithm~\ref{alg:WCSPH} is that evaluating the individual force terms is rather inefficient.
It requires to compute the previously defined discrete differential operators which in turn require to compute a sum over all particles resulting in a runtime complexity of $\Landau{n^2}$, where $n$ is the number of particles.
If we, however, use a smoothing kernel that fulfills the \emph{compact support condition}, most terms of the sums vanish since the kernel function and its derivatives for particles that are further away from $i$ than the kernel support radius $\KernelSupportRadius$ vanish.
Assuming that we have a list of \emph{neighbors} for each particle $i$ that lie within a radius of $\KernelSupportRadius$ around $i$, the algorithmic complexity reduces to $\Landau{mn}$, where $m$ is the maximum number of neighboring particles.
In practice, $m$ is usually bounded by a constant such that we can expect linear runtime complexity, \ie $\Landau{n}$.

\begin{figure}[t]
	\centering
	\begin{overpic}[width=\linewidth]{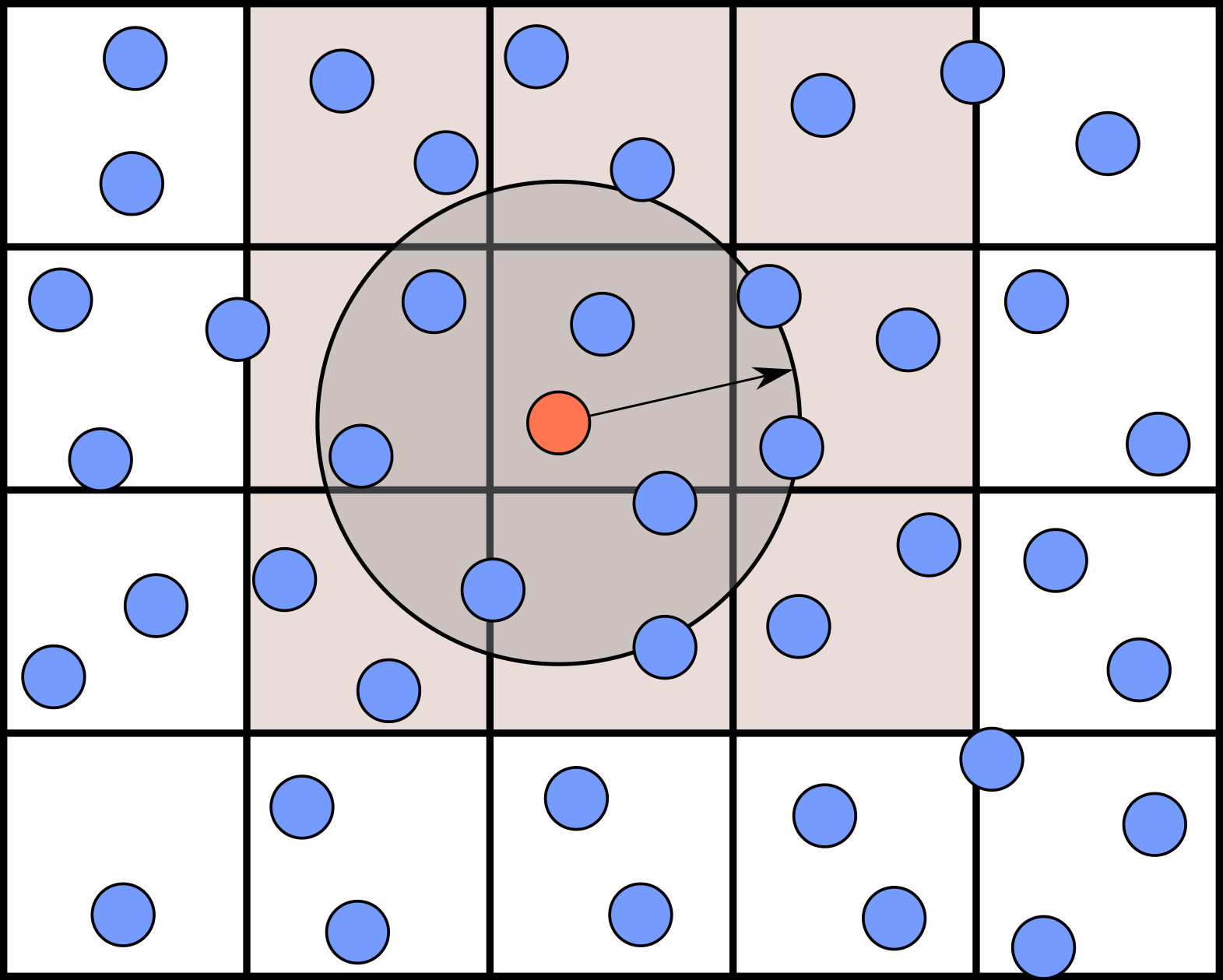}
		\put(55,44){$\KernelSupportRadius$}
		\put(0, 0){\Large $\underbrace{\hphantom{\hspace{1.7cm}}}_{ \KernelSupportRadius}$}
	\end{overpic}
	\vspace*{0.0cm}
	\caption{Uniform grid approach to efficiently find neighboring particles. Particles that are closer to the orange particle than the kernel support radius $\KernelSupportRadius$ must lie within the one ring of the cell containing the orange particle if the grid size is  $\geq \KernelSupportRadius$. }
	\label{fig:uniform_grid}
\end{figure}

The problem of finding the neighbor list is commonly referred to as the \emph{fixed-radius near neighbor problem} and is widely addressed in the computational geometry literature.
The na\"ive approach, \ie brute-force, has a computational complexity of $\Landau{n^2}$ and is therefore not optimal.
In this section, we will present an algorithm to approach the problem in a computationally more efficient way, \ie compact hashing \cite{IABT11}. 
The basic idea of the approach is to place a uniform grid over the domain spanned by the particles with a grid cell size equal to the kernel support radius $\KernelSupportRadius$.
Assuming that a particle is located in the grid cell represented by the tuple $\vec c = (i, j, k)$, where $i$, $j$, and $k$ denote row, column, and depth column of the cell in the grid.
Then, it is obvious that we only have to query for potentially neighboring particles in the cell $c$ itself and its one-ring, \ie $(i - 1, j - 1, k - 1)$, $(i - 1, j - 1, k)$, $(i - 1, j - 1, k + 1)$, $\dots$, $(i + 1, j + 1, k + 1)$.
The strategy then results in an algorithm with a computational complexity of $\Landau{n}$ for construction and $\Landau{1}$ to find the neighbors of a single particle implying $\Landau{n}$ to find the set of all the neighbors of all of the particles.
Obviously, the grid-based approach can easily be generalized to higher dimensions.
Please see Fig.~\ref{fig:uniform_grid} for a graphical illustration.

\subsection{Compact Hashing}

As discussed before, the uniform grid based approach results in a good computational runtime complexity.
However, there is still potential for optimizations in terms of memory consumption, cache efficiency, and parallel processing.
In this regard, the concept of \emph{compact hashing} was proposed by Ihmsen et al.~\cite{IABT11} and will be explained in the following.
An open source \texttt{C++} implementation of a variant of this approach can be found online \cite{CompactNSearch}.

A particular disadvantage of spatial grids is that memory for all cells in the grid has to be allocated although only a small number of cells might be occupied by particles.
Due to the curse of dimensionality the memory requirements increase quickly with increasing domain size.
It would be more memory efficient to only store the populated cells and, hence, employ a sparse representation of the grid.
Therefore, Ihmsen et al. suggest to store the grid cells in a hash map by hashing the index tuple $\vec c = (i, j, k)$ to a scalar index following \cite{THM+03}:
\begin{equation}
	\text{hash}(\vec c) = \left[ (p_1 \, i) \; \mathrm{XOR} \; (p_2 \, j) \; \mathrm{XOR} \; (p_3 \, k) \right] \; \text{mod} \; m,
\end{equation}
where $p_1 = 73856093$, $p_2 = 19349663$, and $p_3 = 83492791$ are large prime numbers and where $m$ is the hash table size.
Please note, that it generally cannot be avoided that several spatial cells are mapped to the same hash value (hash collision).
The effect of overpopulated entries in the hash table might lead to a slow-down of the neighborhood query.
However, as suggested by Teschner et al.~\cite{THM+03} the number of hash collisions can be reduced by increasing the hash table size, \ie trading memory for speed.
As noted by Ihmsen et al.\ the hash table is usually sparsely filled when used in conjuction with SPH discretizations.
Therefore, we would like to avoid to unnecessarily preallocate a large amount of memory.
Moreover, the cache-hit rate of this approach can not expected to be optimal as the cells that are spatially close are not necessarily close in memory.

\begin{figure}[t]
	\centering
	\includegraphics[width=0.8\linewidth]{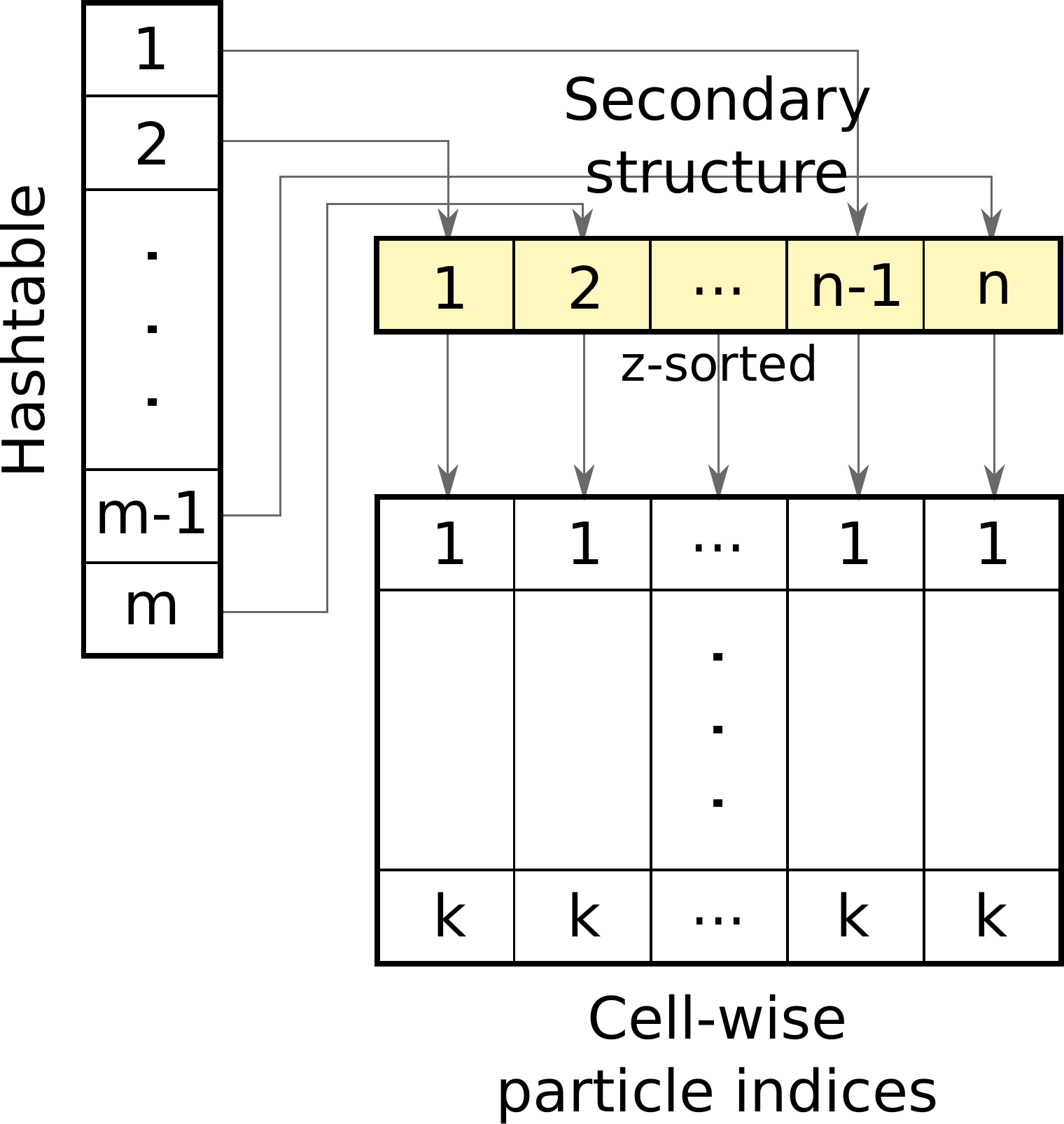}
	\caption{Hash table of size $m$ pointing to secondary data structure (of $n$ non-empty cells) to realize compact hashing. Each array to store the particle index lists in the secondary structure preallocates $k$ entries to minimize the number of memory allocations in the course of the simulation.}
	\label{fig:compact_hashing}
\end{figure}

In order to reduce the frequency of allocations, Ihmsen et al.\ suggest to only store a handle per hash table entry that points to a secondary data structure -- a contiguous array of the populated cells (see Fig.~\ref{fig:compact_hashing}).
Each item of the secondary structure stores a list of the particle indices contained in the respective cell.
In this way memory for a used cell is only allocated if it contains particles and the memory can be (optionally) deallocated if the cell gets empty.
Each storage for the index arrays in the secondary data structure can be further preallocated with the maximally expected number of particles in a cell.
To summarize, the memory consumption now scales linear with the number of particles and not with the volume of the simulation domain.

As it lies in the nature of spatial hash tables to scatter data according to spatially close cells, the indirection to the secondary data structure allows us to optimize for spatial locality in memory.
To realize this, Ihmsen et al.\ suggest to sort the non-empty cells according to a space-filling Z-curve.
The cache hit-rate can further be optimized by analogously sorting the per-particle data in the same way.
However, performing the actual sort ($\Landau{n \log n}$) causes computational overhead and since the particles are constantly moving throughout space during the simulation, it is advised to update the Z-sort in fixed intervals, \eg after every \nth{1000} time step.
This is justified as the order is expected to be roughly maintained over a small number of time steps due to temporal coherence.

Finally, several operations such as the hash table construction, updates and neighborhood queries can be (partially) parallelized to further optimize performance.
For further details on the approach, we would like to refer the reader to the according original paper~\cite{IABT11}.


\section{Pressure Solvers }
\label{sec:pressure_solver}

Incompressibility is an essential aspect in realistic fluid simulations. The fluid volume should not noticeably oscillate or generally grow or shrink over time. Fluid solvers preserve the fluid volume by computing a pressure acceleration $-\frac{1}{\Density}\Gradient \Pressure$ where the pressure $\Pressure$ is proportional to the volume deviation. Then, the term $-\frac{1}{\Density}\Gradient \Pressure$ accelerates particles from high pressure, \ie regions with large volume deviations, to low pressure, \ie regions with small volume deviations. If there would be no volume deviation everywhere in the fluid, the pressure would be zero and the pressure gradient and the pressure acceleration would also be zero. 

Solver implementations typically distinguish pressure acceleration $\PressureAcceleration = -\frac{1}{\Density}\Gradient \Pressure$ and all other non-pressure accelerations $\NonPressureAcceleration$ which improves the intuition of the incompressibility concept. First, a predicted velocity is computed with, \eg $\Velocity[][][\Predicted] = \Velocity[][t] + \Dt \NonPressureAcceleration[][t]$. Then, pressure is computed from the volume deviation after advecting the fluid with $\Velocity[][][\Predicted]$. Finally, the respective pressure acceleration would be applied as, \eg $\Velocity[][t+\Dt] = \Velocity[][][\Predicted] + \Dt \PressureAcceleration[][t]$ to minimize the volume deviation. This final velocity update is often referred to as pressure projection which is related to the fact that the velocity change $\Dt \PressureAcceleration[][t]$ should be minimal. \Ie the pressure acceleration should change the velocity field as little as possible. 

Conceptually, pressure is proportional to the volume deviation. However, there exist various alternatives to actually compute the pressure. First, the volume deviation can be explicitly computed from the density or the velocity divergence can be used to compute a differential update of the volume deviation. Second, pressure can be computed locally with a state equation or it can be computed globally by solving a Pressure Poisson Equation (PPE). The first aspect determines whether the fluid volume oscillates or continuously changes, while the second aspect influences the solver performance.  

\subsection{Explicit Volume Deviation}

The volume deviation is typically deduced from the density deviation. Although SPH solvers can easily handle both formulations, it is probably due to historical reasons that the density formulation is preferred over the volume formulation. The SPH density at a particle $i$ is computed with $\Density[i] = \sum_j \Mass[j] \Kernel[i][j]$ and the deviation to the rest density $\RestDensity$ is considered for the pressure computation.
Note that the density deviation is often clamped, \eg $\max(\Density[i] - \RestDensity,0)$ or $\max\left (\frac{\Density[i]}{\RestDensity}-1,0 \right )$, as a simple solution to the particle deficiency problem at the free surface (see Fig.~\ref{fig:particle_deficiency}).

\subsection{Differential Volume Deviation}

The continuity equation relates the time derivative of the density to the velocity divergence: $\frac{\D \Density}{\D t} = - \Density \Divergence\Velocity$. This fact can be used to predict a particle density from its previous density and, \eg the predicted velocity: $\Density[i][][\Predicted]= \Density[i][t] - \Dt \Density[i][t] \Divergence\Velocity[i][][\Predicted]$. Here, $\Density[i][][\Predicted]$ is a prediction of the particle density after advecting the particles with $\Velocity[i][][\Predicted]$ for time $\Dt$. If it is assumed that the current density equals the rest density, the predicted density is computed as $\Density[i][][\Predicted]= \RestDensity - \Dt \RestDensity \Divergence\Velocity[i][][\Predicted]$ which means that $\Density[i][][\Predicted] - \RestDensity = - \Dt \RestDensity \Divergence\Velocity[i][][\Predicted]$ can be used as a measure for the density deviation. It can be seen that minimizing the density deviation is related to minimizing the velocity divergence. The term $- \Dt \RestDensity \Divergence\Velocity[i][][\Predicted]$ is a density change at a particle if the particles are advected with $\Velocity[i][][\Predicted]$ for time $\Dt$. 

\subsection{Discussion -- Explicit vs. Differential Volume Deviation}

Both forms imply challenges. If pressure accelerations are derived from the explicit form of the volume deviation, the fluid volume oscillates due to an over-correction of the pressure acceleration. These oscillations have to minimized. At least, they should not be perceivable. Using the differential form to compute the volume deviation results in a drift of the fluid volume, typically a volume loss. The differential form assumes that the current density is correct. It minimizes density changes between simulation steps, but potentially existing density deviations are not detected or corrected. Here, the challenge is to minimize the volume drift. Although volume drift often occurs in Eulerian pressure solvers and volume oscillations often occur in Lagrangian solvers, both issues are not related to the Eulerian or Lagrangian perspective. If an SPH solver was using the differential form to compute the density deviation, it would suffer from volume drift. If a Eulerian solver, \eg FLIP, was using the explicit form for the density computation, it would suffer from oscillations.

\subsection{State Equation SPH (SESPH)}
\label{sub:state_equation}

State equations are used to compute pressure from density deviations. The density deviation can be computed explicitly or from a differential form. The deviation can be represented as a quotient or a difference of actual and rest density. One or more stiffness constants are involved. Some examples are: $\Pressure[i] = k \left( \frac{\Density[i]}{\RestDensity}-1\right)$, $\Pressure[i] = k (\Density[i] - \RestDensity)$ or $\Pressure[i] = k_1 \left( \left( \frac{\Density_i}{\RestDensity} \right)^{k_2} - 1\right)$. As $\Density[i] < \RestDensity$ is not considered to solve the particle deficiency problem at the free surface, the computed pressure is always non-negative. SPH fluid simulations that use a state equation to compute pressure are often referred to as compressible or weakly compressible. In contrast, fluid implementations that solve a PPE to compute pressure are known as incompressible. These terms basically indicate that it is more challenging to minimize compression with a state equation than with a PPE.  

It might look confusing that arbitrary pressure values can be computed for a given density $\Density[i]>\RestDensity$ dependent on the state equation and the stiffness constant(s). Here, it is interesting to note that the parameters do not govern the pressure, but the compressibility of the SPH fluid. This can be seen in a simple example with a fluid at rest under gravity. In this case, the pressure acceleration at all particles cancels gravity, \ie $\Gravity -\frac{1}{\Density[i]}\Gradient \Pressure[i] = \vec{0}$. Discretizing the pressure gradient with SPH yields $\Gravity = \sum_j \Mass[j] \left( \frac{\Pressure[i]}{\Density[i]^2} + \frac{\Pressure[j]}{\Density[j]^2}\right) \KernelGradient[i][j]$. Using, \eg $\Pressure[i] = k (\Density[i] - \RestDensity)$, yields $\Gravity = k \sum_j \Mass[j] \left( \frac{\Density[i] - \RestDensity}{\Density[i]^2} + \frac{\Density[i] - \RestDensity}{\Density[j]^2}\right) \KernelGradient[i][j]$. It can be seen that a variation of the stiffness constant $k$ is related to a variation in the density deviation $\Density[i] - \RestDensity$. This relation, however, is not simply $k (\Density[i] - \RestDensity) = \text{const}$ since the erroneous particle sampling for $\Density[i] \neq \RestDensity$ influences the SPH discretization of the pressure gradient. But generally, the stiffness constant in the state equation governs the density deviation. Larger values result in smaller deviations and require smaller time steps. Smaller values lead to larger density deviations, \ie less realistic simulations. Also, the boundary handling fails if the tolerated density deviation is too large.

\subsection{Pressure Poisson Equation (PPE)}

The general idea of the pressure computation is to end up with pressure accelerations that cause velocity changes that in turn cause displacements such that all particles are uncompressed, \ie have their rest density or rest volume. PPE solvers or projection schemes solve a linear system to compute the respective pressure field.

\subsubsection{Derivation}
\label{sec:pressure:derivation}

We consider the predicted velocity after all non-pressure accelerations: $\Velocity[][][\Predicted] = \Velocity[][t] + \Dt \NonPressureAcceleration[][t]$. If the particles would be advected with this velocity, we can use the continuity equation to estimate a predicted density $\Density[][][\Predicted]= \Density[][t] - \Dt \Density[][t] \Divergence\Velocity[][][\Predicted]$. Now, the goal of the pressure computation is a pressure acceleration $- \frac{1}{\Density[][t]} \Gradient \Pressure[][t]$ that corresponds to a velocity change $- \Dt \frac{1}{\Density[][t]} \Gradient \Pressure[][t]$ whose divergence $- \Divergence \left( \Dt \frac{1}{\Density[][t]} \Gradient \Pressure[][t] \right)$ corresponds to a density change per time $- \Density[](t) \Divergence \left( \Dt \frac{1}{\Density[][t]} \Gradient \Pressure[][t] \right)$ that cancels the predicted density deviation per time $\frac{\RestDensity - \Density[][][\Predicted]}{\Dt}$, \ie $\frac{\RestDensity - \Density[][][\Predicted]}{\Dt} - \Density[](t) \Divergence \left( \Dt \frac{1}{\Density[][t]} \Gradient \Pressure[][t] \right) = 0$. This is one form of a PPE, typically written as
\begin{align}
\Dt \Laplace \Pressure[][t]  = \frac{\RestDensity - \Density[][][\Predicted]}{\Dt}.
\label{eq:pressure:PPE1}
\end{align} 
Note that $\Divergence \Gradient \Pressure [][t] = \Laplace \Pressure[][t]$. In this equation, the pressure $\Pressure$ is unknown. We have one equation per particle, resulting in a system with $n$ equations and $n$ unknown pressure values for $n$ particles. Various similar PPE forms can be derived more formally, \eg starting with the continuity equation at time $t+\Dt$: $- \Density[][t+\Dt] \Divergence \Velocity[][t+\Dt] =\frac{\D \Density[][t+\Dt]}{\D t} $. The time derivative of the density is approximated with $\frac{\D \Density[][t+\Dt]}{\D t} = \frac{\Density[][t+\Dt] - \Density[][t]}{\Dt}$. The velocity is written as $\Velocity[][t+\Dt] = \Velocity[][][\Predicted] - \Dt \frac{1}{\Density[][t+\Dt]} \Gradient \Pressure[][t+\Dt]$ using an implicit update with the pressure acceleration at time $t+\Dt$. Imposing the constraint $\Density[][t+\Dt] = \RestDensity$, we get $- \RestDensity \Divergence \Velocity[][][\Predicted] + \Divergence \left( \Dt \Gradient \Pressure[][t+\Dt] \right) = \frac{\RestDensity - \Density[][t]}{\Dt}$ and finally 
\begin{align}
\Dt \Laplace \Pressure[][t+\Dt] = \frac{\RestDensity - (\Density[][t] - \Dt \RestDensity \Divergence \Velocity[][][\Predicted])}{\Dt}.
\label{eq:pressure:PPE2}
\end{align}
If compared carefully, there are very minor differences between Eqs.~\eqref{eq:pressure:PPE1} and~\eqref{eq:pressure:PPE2} in the computation of the predicted density and the pressure acceleration. As $\Density[][t] \approx \RestDensity$, however, these differences are negligible. The biggest difference would actually be the SPH discretizations of the Laplacian. Eq.~\eqref{eq:pressure:PPE1} works with the particle neighborhood at time $t$, while Eq.~\eqref{eq:pressure:PPE2} requires the neighborhood at time $t+\Dt$. As this would require an additional neighbor search per simulation step, it is generally ignored. 

Eqs.~\eqref{eq:pressure:PPE1} and~\eqref{eq:pressure:PPE2} make use of the density invariance as source term in the PPE. Motivated by the continuity equation, the divergence of the predicted velocity could be used alternatively. To derive the respective form, we start with $\Dt \frac{1}{\Density[][t]} \Gradient \Pressure[][t] = \Velocity[][][\Predicted] - \Velocity[][t+\Dt] $. Taking the divergence and imposing the constraint that the velocity field at $t+\Dt$ should be divergence free, \ie $\Divergence \Velocity[][t+\Dt] = 0$, we get $\Divergence \left(\Dt \frac{1}{\Density[][t]} \Gradient \Pressure[][t] \right) = \Divergence \Velocity[][][\Predicted]$ and
\begin{align}
\Dt \Laplace \Pressure[][t] = \Density[][t] \Divergence \Velocity[][][\Predicted].
\label{eq:pressure:PPE3}
\end{align}
We search for pressure $p$ such that the pressure acceleration $- \frac{1}{\Density[][t]} \Gradient \Pressure[][t]$ corresponds to a velocity change $- \Dt \frac{1}{\Density[][t]} \Gradient \Pressure[][t]$ whose divergence $- \Divergence \left( \Dt \frac{1}{\Density[][t]} \Gradient \Pressure[][t] \right)$ cancels the divergence of the predicted velocity, \ie $- \Divergence \left( \Dt \frac{1}{\Density[][t]} \Gradient \Pressure[][t] \right) + \Divergence \Velocity[][][\Predicted]=0$. 

SPH fluid solvers can easily employ any of the PPE forms. If the density invariance is taken as source term, density oscillations occur and have to be minimized. If the predicted velocity divergence is taken as source term, the fluid volume tends to drift. Volume or density oscillations are not a general SPH issue, but related to the pressure computation. In the same way, volume drift is not, \eg a FLIP issue, but only due to the typically used velocity divergence as source term in the PPE.

\subsection{Discretization with Implicit Incompressible SPH (IISPH)}

There exist various alternative discretizations with benefits and drawbacks. Here, we discuss one option referred to as IISPH which has been proposed in \cite{ICS+14}. We consider a slightly rewritten form of the PPE in Eq.~\eqref{eq:pressure:PPE1} for particle $i$: 
\begin{align}
\Dt^2 \Laplace \Pressure[i]  = \RestDensity - \Density[i][][\Predicted].
\label{eq:pressure:PPEIISPH}
\end{align}
If a quantity is considered at time $t$, \eg pressure, the time index is omitted. Using SPH, the source term is computed as
\begin{align}
\RestDensity - \Density[i][][\Predicted] = \RestDensity - \Density[i] - \Dt \sum_j \Mass[j] \left( \Velocity[i][][\Predicted] - \Velocity[j][][\Predicted]\right) \cdot \Gradient \Kernel[i][j]
\label{eq:pressure:sourceTermIISPH}
\end{align}
with $\Velocity[i][][\Predicted] = \Velocity[i] + \Dt \NonPressureAcceleration[i]$. The computation of the Laplacian is realized as the computation of the divergence of the velocity change due to the pressure acceleration, \ie 
\begin{align}
\Dt^2 \Laplace \Pressure[i] & = - \Dt \Density[i] \Divergence \left( \Dt \PressureAcceleration[i] \right) \nonumber \\
& = \Dt^2 \sum_j \Mass[j] \left( \PressureAcceleration[i] - \PressureAcceleration[j] \right) \cdot \Gradient \Kernel[i][j]
\label{eq:pressure:LaplacianIISPH}
\end{align}
with pressure acceleration 
\begin{align}
\PressureAcceleration[i] = -\frac{1}{\Density[i]}\Gradient \Pressure[i] = - \sum_j \Mass_j \left( \frac{\Pressure[i]}{\Density[i]^2} + \frac{\Pressure[j]}{\Density[j]^2} \right) \Gradient \Kernel[i][j].
\label{eq:pressure:accelerationIISPH}
\end{align}
Using Eqs.~\eqref{eq:pressure:sourceTermIISPH} and~\eqref{eq:pressure:LaplacianIISPH}, we can compute the left-hand and right-hand side of Eq.~\eqref{eq:pressure:PPEIISPH} at each particle using SPH sums over adjacent particles. 
The IISPH discretization of Eq.~\eqref{eq:pressure:PPEIISPH} is 
\begin{align}
\Dt^2 \sum_j \Mass[j] \left( \PressureAcceleration[i] - \PressureAcceleration[j] \right) \cdot \Gradient \Kernel[i][j]
= \RestDensity - \Density[i][][\Predicted].
\label{eq:pressure:discretizedPPE}
\end{align} 
Introducing the velocity change due to pressure acceleration $\Velocity[i][][\text{p}] = \Dt \PressureAcceleration[i]$, Eq.~\eqref{eq:pressure:discretizedPPE} can be written as
\begin{align}
\Dt \sum_j \Mass[j] \left( \Velocity[i][][\text{p}] - \Velocity[j][][\text{p}] \right) \cdot \Gradient \Kernel[i][j]
= \Dt \Density[i] \; \Divergence \Velocity[i][][\text{p}] = \RestDensity - \Density[i][][\Predicted].
\label{eq:pressure:discretizedPPE2}
\end{align} 
It can be seen that we search for pressure values such that the pressure acceleration causes a velocity change $\Velocity[i][][\text{p}]$ such that the divergence of $\Velocity[i][][\text{p}]$ causes a density change that corrects from the predicted density $\Density[i][][\Predicted]$ to the rest density $\RestDensity$, \ie $\Density[i][][\Predicted] + \Dt \Density[i]  \; \Divergence \Velocity[i][][\text{p}] = \RestDensity$.
 
\subsubsection{IISPH Solver}

\paragraph*{System} Eq.~\eqref{eq:pressure:discretizedPPE} is considered at all particles. We have $n$ equations with $n$ unknown pressure values $\Pressure[i]$. Each equation does not only contain an unknown pressure value $\Pressure[i]$, but also unknown pressure values at neighbors of $i$ and - due to Eq.~\eqref{eq:pressure:accelerationIISPH} - also unknown pressures at neighbors of neighbors of $i$. The set of Eq.~\eqref{eq:pressure:discretizedPPE} at all particles forms a system $\Matrix \PressureVector = \MatrixSourceTermVector$ and Eq.~\eqref{eq:pressure:discretizedPPE} at particle $i$ can be written as $\MatrixPressureRow[i] = \MatrixSourceTerm[i]$:
\begin{align}
\MatrixPressureRow[i]  & = \Dt^2 \sum_j \Mass[j] \left( \PressureAcceleration[i] - \PressureAcceleration[j] \right) \cdot \Gradient \Kernel[i][j]
\label{eq:pressure:IISPHAp} \\
\MatrixSourceTerm[i] & = \RestDensity - \Density[i][][\Predicted].
\label{eq:pressure:IISPHs}
\end{align} 
$\Matrix$ is not a diagonal matrix, but at least sparse. In row $i$, matrix elements $\MatrixElement[i][i]$ and also elements $\MatrixElement[i][j]$ are non-zero with $j$ being a neighbor or a neighbor of a neighbor of $i$. 

Extracting the matrix elements $\MatrixElement[i][j]$ would not be impossible, but rather tedious and error-prone. Fortunately, element extraction is not required to solve the system as solver implementations typically just require the computation of $\MatrixPressureRow[i]$. The term $\MatrixPressureRow[i]$ can be computed by first evaluating $\PressureAcceleration[i]$ with Eq.~\eqref{eq:pressure:accelerationIISPH}, followed by the computation of $\Dt^2 \Laplace \Pressure[i]$ using Eq.~\eqref{eq:pressure:LaplacianIISPH}. An explicit notion of the elements of $\Matrix$ is typically not required as, \eg for Conjugate Gradients. Some solvers, \eg Jacobi, require the diagonal elements $\MatrixElement[i][i]$.   

\paragraph*{Relaxed Jacobi scheme} In the following, we discuss the implementation of a Jacobi variant to solve $\Matrix \PressureVector = \MatrixSourceTermVector$. The method starts with an initialization of the pressure vector, \eg $\PressureVector[][][(0)] = \vec{0}$. Then, the weighted / damped / relaxed Jacobi scheme iteratively updates all pressure values using
\begin{align}
\Pressure[i][][(l+1)] = (1 - \omega) \Pressure[i][][(l)] + \frac{\omega}{\MatrixElement[i][i]} \left( \MatrixSourceTerm[i] - \sum_{j \neq i} \MatrixElement[i][j] \Pressure[j][][(l)] \right)
\label{eq:pressure:JacobiOriginal} 
\end{align}
with $l$ indicating the iteration and $\omega$ being a relaxation coefficient. The relaxation coefficient is typically set to $\omega=0.5$ in IISPH implementations. Smaller values reduce the convergence, larger values are unstable. The update in Eq.~\eqref{eq:pressure:JacobiOriginal} seems to indicate that the diagonal elements $\MatrixElement[i][i]$, but also elements $\MatrixElement[i][j]$ for neighboring particles are required in an implementation. Interestingly, the update in Eq.~\eqref{eq:pressure:JacobiOriginal} can be rewritten as
\begin{align}
\Pressure[i][][(l+1)] & = (1 - \omega) \Pressure[i][][(l)] + \frac{\omega}{\MatrixElement[i][i]} \left( \MatrixSourceTerm[i] - \MatrixPressureRow[i][][][(l)] + \MatrixElement[i][i] \Pressure[i][][(l)] \right) \nonumber \\
& = \Pressure[i][][(l)] + \frac{\omega}{\MatrixElement[i][i]} \left( \MatrixSourceTerm[i] - \MatrixPressureRow[i][][][(l)] \right).
\label{eq:pressure:JacobiIISPH} 
\end{align}
This update requires the computation of the term $\MatrixPressureRow[i][][][(l)]$ which can be computed from Eqs.~\eqref{eq:pressure:LaplacianIISPH} and~\eqref{eq:pressure:accelerationIISPH} without notion of the elements $\MatrixElement[i][j]$. As IISPH only considers non-negative pressure, the actual update is
\begin{align}
\Pressure[i][][(l+1)] = \max \left( \Pressure[i][][(l)] + \frac{\omega}{\MatrixElement[i][i]} \left( \MatrixSourceTerm[i] - \MatrixPressureRow[i][][][(l)] \right), 0 \right) .
\label{eq:pressure:JacobiIISPHclamped} 
\end{align}
\paragraph*{Pressure clamping} The clamping has been proposed in~\cite{ICS+14} where simulation artifacts are discussed in case of negative pressure values. Negative pressure values are sometimes briefly discussed, but rarely carefully analyzed. Theoretically, arbitrary constant offsets to all pressure values do not change the pressure gradient. Shifting a pressure range from $[0, \Pressure[\text{max}]]$ to, \eg $[-100, \Pressure[\text{max}]-100]$ or $[100, \Pressure[\text{max}]+100]$ by adding constant offsets to all values does not change the pressure gradient. SPH discretizations, however, behave differently for negative and positive pressure values. Reasons have not been investigated yet, but one could speculate that negative and positive pressures result in different pressure gradients in case of incomplete neighborhoods where missing contributions are implicitly assumed to have zero pressure. In IISPH simulations, minimum pressure is generally zero being consistent with the implicit assumption of zero pressure for missing samples at free surfaces.  

\paragraph*{Diagonal element} The update in Eq.~\eqref{eq:pressure:JacobiIISPHclamped} requires the diagonal element $\MatrixElement[i][i]$. The element can be calculated by accumulating all coefficients of $\Pressure[i]$ after inserting Eq.~\eqref{eq:pressure:accelerationIISPH} into Eq.~\eqref{eq:pressure:LaplacianIISPH}. Here, it is important to keep in mind that $i$ is one of the neighbors of each neighbor of $i$. Finally, the diagonal element is 
\begin{align}
\MatrixElement[i][i] = & - \Dt^2 \sum_j \Mass_j 
\left( \sum_j \frac{\Mass[j]}{\Density[j]^2} \KernelGradient[i][j] \right)
\cdot \KernelGradient[i][j] \nonumber \\
& -
\Dt^2 \sum_j \Mass[j] \left( \frac{\Mass[i]}{\Density[i]^2} \KernelGradient[i][j] \right) \cdot \KernelGradient[i][j] .
\label{eq:pressure:diagonalElement}
\end{align} 
\paragraph*{Stop criterion} There is no agreement on when to stop  the Jacobi iterations in Eq.~\eqref{eq:pressure:JacobiIISPHclamped}. The iterations could be stopped after a fixed number. Alternatively, a predicted density deviation is often considered. This is motivated by the fact that $\MatrixPressureRow[i][][][(l)]$ is a predicted density change at particle $i$ due to the pressure field at iteration $l$. \Ie $\Density[i][][\text{err},\Predicted] = (\MatrixPressureRow[i][][][(l)] - \MatrixSourceTerm[i]) / \RestDensity = (\MatrixPressureRow[i][][][(l)] + \Density[i][][\Predicted] - \RestDensity ) / \RestDensity$ is a predicted relative density error at particle $i$, if pressure accelerations according to the pressure field $\PressureVector[][][(l)]$ would be applied. Typically, the average of all $\Density[i][][\text{err},\Predicted]$ is taken as a stop criterion, \eg $\Density[i][][\text{avg\_err},\Predicted] = \frac{1}{n} \sum_i | \Density[i][][\text{err},\Predicted] |$. In \cite{ICS+14}, it is proposed to stop if, \eg $\Density[i][][\text{avg\_err},\Predicted] < 0.1\%$, \ie the oscillation of the overall fluid volume is below $0.1\%$. In addition to the predicted average density error, the maximum of the predicted density errors could also be taken as a stop criterion. 

\paragraph*{Implementation} Alg.~\ref{alg:pressure:IISPH} shows the implementation of IISPH. Source term $\MatrixSourceTerm[i]$ and diagonal element $\MatrixElement[i][i]$ are computed once. In iteration $l$, the pressure Laplacian $\MatrixPressureRow[i][][][(l)] = \Dt^2 \Laplace \Pressure[i]$ is computed in two steps. First, the pressure acceleration $(\PressureAcceleration[i])^{(l)}$ is computed and stored at the particles. Then, the divergence of the velocity change due to the pressure acceleration, \ie $\MatrixPressureRow[i][][][(l)]$, is computed.
Neighborhood search, the computation of the predicted velocity $\Velocity[i][][\Predicted]$ and the advection of particles are omitted in Alg.~\ref{alg:pressure:IISPH}.    

\begin{algorithm}
\caption{Pressure computation with the IISPH PPE solver.}
\label{alg:pressure:IISPH}
\begin{algorithmic}
\algnotext{EndFor}
\algnotext{EndWhile}
\algnotext{EndProcedure}
\ForAll{$particle~i$}
\State compute diagonal element $\MatrixElement[i][i]$ with Eq.~\eqref{eq:pressure:diagonalElement}
\State compute source term $\MatrixSourceTerm[i]$ with Eq.~\eqref{eq:pressure:sourceTermIISPH}
\State initialize pressure $\Pressure[i][][(0)] = 0$
\EndFor
\State $l=0$
\Repeat
\ForAll{$particle~i$}
\State compute pressure acceleration $(\PressureAcceleration[i])^{(l)}$ with Eq.~\eqref{eq:pressure:accelerationIISPH}
\EndFor
\ForAll{$particle~i$}
\State compute Laplacian $\MatrixPressureRow[i][][][(l)]$ with Eq.~\eqref{eq:pressure:LaplacianIISPH}
\State update pressure $\Pressure[i][][(l+1)]$ with Eq.~\eqref{eq:pressure:JacobiIISPHclamped}
\EndFor
\State $l = l+1$
\Until $\Density[i][][\text{avg\_err},\Predicted] < 0.1\%$
\end{algorithmic}
\end{algorithm}

\subsection{Predictive--Corrective Incompressible SPH (PCISPH)}

In the following a pressure solver based on a predictor-corrector approach is introduced~\cite{SP09}.
  	
\paragraph*{Motivation} The density $\Density[i][t+\Dt]$ can be estimated with
\begin{align}
\Density[i][t+\Dt]  = & \sum_j \Mass_j \Kernel[i][j] + \Dt \sum_j \Mass_j (\Velocity[i] - \Velocity[j]) \cdot \KernelGradient[i][j] \nonumber \\
& + \Dt \sum_j \Mass_j (\Dt \NonPressureAcceleration[i] - \Dt \NonPressureAcceleration[j]) \cdot \KernelGradient[i][j] \nonumber \\
& + \Dt \sum_j \Mass_j (\Dt \PressureAcceleration[i] - \Dt \PressureAcceleration[j]) \cdot \KernelGradient[i][j] .
\label{eq:pressure:PCISPHMotivation1}
\end{align}
Again, the time index is omitted for quantities at time $t$. Using the predicted density 
\begin{align}
\Density[i][][\Predicted] = & \sum_j \Mass_j \Kernel[i][j] + \Dt \sum_j \Mass_j (\Velocity[i] - \Velocity[j]) \cdot \KernelGradient[i][j] \nonumber \\
& + \Dt \sum_j \Mass_j (\Dt \NonPressureAcceleration[i] - \Dt \NonPressureAcceleration[j]) \cdot \KernelGradient[i][j]
\label{eq:pressure:PCISPHpredictedDensity}
\end{align}
and the constraint $\Density[i][t+\Dt] = \RestDensity$, we can write
\begin{align}
\RestDensity = \Density[i][][\Predicted] + \Dt \sum_j \Mass_j (\Dt \PressureAcceleration[i] - \Dt \PressureAcceleration[j]) \cdot \KernelGradient[i][j] .
\label{eq:pressure:PCISPHMotivation2}
\end{align}
When using the symmetric SPH formulation for the pressure acceleration, the terms $\PressureAcceleration[i]$ and $\PressureAcceleration[j]$ are computed from unknown pressure values at particle $i$, at neighbors of $i$ and at neighbors of neighbors of $i$. Now, PCISPH introduces approximations and simplifications to end up with only one unknown pressure value $\Pressure[i]$ in Eq.~\eqref{eq:pressure:PCISPHMotivation2} \cite{SP09}. Then, each pressure value $\Pressure[i]$ can be computed from one equation. Solving a linear system is avoided.  

\paragraph*{Simplifications} The pressure acceleration is discretized with the symmetric formulation. It is assumed that the pressure $\Pressure[j]$ at neighboring particles equals the pressure $\Pressure[i]$ at particle $i$. Further, $\Mass_j = \Mass_i$ for all neighbors and $\Density[i]=\Density[j] \approx \RestDensity$. Then, the pressure acceleration can be written as
\begin{align}
\PressureAcceleration[i] & = - \sum_j \Mass_j \left(
\frac{\Pressure[i]}{\Density[i]^2} + \frac{\Pressure[j]}{\Density[j]^2} 
\right) \KernelGradient[i][j] 
\label{eq:pressure:PCISPHpressureAcceleration}
\\
& \approx - \Mass_i \frac{2 \Pressure[i]}{(\RestDensity)^2} \sum_j \KernelGradient[i][j].
\end{align}
Using this approximation, Eq.~\eqref{eq:pressure:PCISPHMotivation2} can be written as
\begin{align}
\RestDensity = \Density[i][][\Predicted] + 2 \Dt^2 \frac{\Mass_i^2}{(\RestDensity)^2} \sum_j \left( 
-  \Pressure[i] \sum_j \KernelGradient[i][j] 
+  \Pressure[j] \sum_k \KernelGradient[j][k]
\right) \cdot \KernelGradient[i][j] .
\end{align}
Using $\Pressure[i] \approx \Pressure[j]$ and approximating $\sum_k \KernelGradient[j][k] \approx \KernelGradient[j][i]$, the equation can further be simplified to 
\begin{align}
\RestDensity & = \Density[i][][\Predicted] + \Pressure[i] \frac{2 \Dt^2 \Mass_i^2}{(\RestDensity)^2} \sum_j \left( 
-  \sum_j \KernelGradient[i][j] 
+  \KernelGradient[j][i]
\right) \cdot \KernelGradient[i][j] \nonumber \\
& = \Density[i][][\Predicted] - \Pressure[i] 
\underbrace{\frac{2 \Dt^2 \Mass_i^2}{(\RestDensity)^2} 
\left(
\sum_j \KernelGradient[i][j] \cdot \sum_j \KernelGradient[i][j] +
\sum_j ( \KernelGradient[i][j] \cdot \KernelGradient[i][j] )  
\right)}_{-\tfrac{1}{k^\text{PCI}}}.
\end{align}
In PCISPH, the coefficient $k^\text{PCI}$ is considered for a template particle with perfect sampling, \ie  $\sum_j \KernelGradient[i][j] \cdot \sum_j \KernelGradient[i][j] +
\sum_j ( \KernelGradient[i][j] \cdot \KernelGradient[i][j])  = \text{const}$. 
Finally, we have the following equation per particle:
\begin{align}
\Pressure[i] = k^\text{PCI} (\RestDensity - \Density[i][][\Predicted])
\label{eq:pressure:PCISPHstateEquation}
\end{align}
with
\begin{align}
k^\text{PCI} = -\frac{0.5 (\RestDensity)^2}{\Dt^2 \Mass_i^2} \cdot
\frac{1}{
\sum_j \KernelGradient[i][j] \cdot \sum_j \KernelGradient[i][j] +
\sum_j ( \KernelGradient[i][j] \cdot \KernelGradient[i][j] )  
} 
\label{eq:pressure:PCISPHstiffness}
\end{align}
for a template particle $i$ with perfect sampling. This is a state equation, where the stiffness constant is not user-defined, but motivated by the fact that Eq.~\eqref{eq:pressure:PCISPHMotivation1} should be satisfied, \ie the pressure should induce pressure accelerations such that the particles have their rest density at time $t+\Dt$. 

\paragraph*{Iterative refinement} The locally optimized state equation is one important property of the PCISPH concept. A second significant characteristics is the iterative refinement of the pressure field. While this sounds expensive, it is motivated by large time steps compared to simple state-equation solvers. PCISPH computes a first estimate of the pressure \Pressure[i][][(1)] with Eq.~\eqref{eq:pressure:PCISPHstateEquation}. This predicted pressure field is used to compute $(\PressureAcceleration[i])^{(1)}$ with Eq.~\eqref{eq:pressure:PCISPHpressureAcceleration}. Then, the pressure field is iteratively refined by
\begin{align}
\Pressure[i][][(l+1)] = & \Pressure[i][][(l)] \nonumber \\
& + k^\text{PCI} (\RestDensity - \Density[i][][\Predicted] - \Dt \sum_j \Mass_j (\Dt (\PressureAcceleration[i])^{(l)} - \Dt (\PressureAcceleration[j])^{(l)}) \cdot \KernelGradient[i][j]) 
\label{eq:pressure:PCISPHpressureUpdate}
\end{align}  
in iteration $l$. The term 
\begin{align}
(\Density[i][][\text{p}])^{(l)} = \Dt \sum_j \Mass_j (\Dt (\PressureAcceleration[i])^{(l)} - \Dt (\PressureAcceleration[j])^{(l)}) \cdot \KernelGradient[i][j]
\label{eq:pressurePCISPHpressureDensity}
\end{align}
is one option to predict the density change due to the pressure accelerations. If this density change does not cancel the predicted density deviation $\RestDensity - \Density[i][][\Predicted]$, the predicted pressure is corrected. Similarly to IISPH, the process is stopped if $(\RestDensity - \Density[i][][\Predicted] + (\Density[i][][\text{p}])^{(l)})/ \RestDensity$ is sufficiently small, \eg smaller than $0.1 \%$. The original PCISPH depiction proposes to compute $\Density[i][][\text{p}]$ from particle displacements due to the pressure accelerations, \ie
\begin{align}
(\Density[i][][\text{p}])^{(l)} = m_i 
\left( 
(\Delta \Position[i])^{(l)} \cdot \sum_j \KernelGradient[i][j] - \sum_j \KernelGradient[i][j] \cdot (\Delta \Position[j])^{(l)} 
\right).
\label{eq:pressurePCISPHpressureDensity2}
\end{align}
If $(\Delta \Position[i])^{(l)} = \Dt^2 (\PressureAcceleration[i])^{(l)}$ and $\Mass_i = \Mass_j$, Eqs.~\eqref{eq:pressurePCISPHpressureDensity} and \eqref{eq:pressurePCISPHpressureDensity2} are identical. We prefer Eq.~\eqref{eq:pressurePCISPHpressureDensity} as it will also be used in the discussion of the relation between PCISPH and IISPH.  

\paragraph*{Implementation} Alg.~\ref{alg:pressure:PCISPH} shows the implementation of PCISPH. The stiffness constant $k^\text{PCI}$ is computed once at the beginning of the simulation. The same coefficient is used for all particles. The pressure field is predicted with the state equation in Eq.~\eqref{eq:pressure:PCISPHstateEquation}. The effect of the respective pressure acceleration onto the density is estimated. Remaining deviations from the rest density are used to compute pressure corrections with Eq.~\eqref{eq:pressure:PCISPHpressureUpdate}.
Neighbor search and the advection of the particles are omitted in Alg.~\ref{alg:pressure:PCISPH}.

\begin{algorithm}
\caption{Pressure computation with the PCISPH solver.}
\label{alg:pressure:PCISPH}
\begin{algorithmic}
\algnotext{EndFor}
\algnotext{EndWhile}
\algnotext{EndProcedure}
\State compute stiffness constant $k^\text{PCI}$ with Eq.~\eqref{eq:pressure:PCISPHstiffness}
\ForAll{$particle~i$}
\State compute predicted density $\Density[i][][\Predicted]$ with Eq.~\eqref{eq:pressure:PCISPHpredictedDensity}
\State initialize pressure $\Pressure[i][][(1)]$ with Eq.~\eqref{eq:pressure:PCISPHstateEquation}
\EndFor
\State $l=1$
\Repeat
\ForAll{$particle~i$}
\State compute pressure acceleration $(\PressureAcceleration[i])^{(l)}$ with Eq.~\eqref{eq:pressure:PCISPHpressureAcceleration}
\EndFor
\ForAll{$particle~i$}
\State compute density change $(\Density[i][][\text{p}])^{(l)}$ with Eq.~\eqref{eq:pressurePCISPHpressureDensity}
\State update pressure $\Pressure[i][][(l+1)]$ with Eq.~\eqref{eq:pressure:PCISPHpressureUpdate}
\EndFor
\State $l = l+1$
\Until $\Density[i][][\text{avg\_err},\Predicted] < 0.1\%$
\end{algorithmic}
\end{algorithm}

\subsection{Relations between SESPH, IISPH and PCISPH}

The PCISPH pressure solver updates the pressure with
\begin{align}
\Pressure[i][][(l+1)] = & \Pressure[i][][(l)] \nonumber \\
& + k^\text{PCI} (\RestDensity - \Density[i][][\Predicted] - \Dt \sum_j \Mass_j (\Dt (\PressureAcceleration[i])^{(l)} - \Dt (\PressureAcceleration[j])^{(l)}) \cdot \KernelGradient[i][j]) 
\label{eq:pressure:PCISPHpressureUpdate2}
\end{align}
which simplifies to the state equation
\begin{align}
\Pressure[i][][(1)] = k^\text{PCI} (\RestDensity - \Density[i][][\Predicted])
\label{eq:pressure:PCISPHstateEquation2}
\end{align}
for the first update if the pressure is initialized with $\Pressure[i][][(0)] = 0$. The same applies to IISPH. The solver updates pressure with 
\begin{align}
\Pressure[i][][(l+1)] = \Pressure[i][][(l)] + \frac{\omega}{\MatrixElement[i][i]} \left( \MatrixSourceTerm[i] - \MatrixPressureRow[i][][][(l)] \right) 
\label{eq:pressure:JacobiIISPH2} 
\end{align}
which simplifies to the state equation
\begin{align}
\Pressure[i][][(1)] =  \frac{\omega}{\MatrixElement[i][i]} \MatrixSourceTerm[i] = \frac{\omega}{\MatrixElement[i][i]} (\RestDensity - \Density[i][][\Predicted])
\label{eq:pressure:JacobiIISPH3} 
\end{align}
for the first update if the pressure is initialized with $\Pressure[i][][(0)] = 0$. \Ie if IISPH or PCISPH stop after one pressure update, they are state-equation solvers.

Another remarkable aspect are the stiffness constants in PCISPH (Eqs.~\eqref{eq:pressure:PCISPHpressureUpdate2} and \eqref{eq:pressure:PCISPHstateEquation2}) and IISPH (Eqs.~\eqref{eq:pressure:JacobiIISPH2} and \eqref{eq:pressure:JacobiIISPH3} ). The PCISPH stiffness constant is
\begin{align}
k^\text{PCI} = -\frac{0.5 (\RestDensity)^2}{\Dt^2 \Mass_i^2} \cdot
\frac{1}{
\sum_j \KernelGradient[i][j] \cdot \sum_j \KernelGradient[i][j] +
\sum_j ( \KernelGradient[i][j] \cdot \KernelGradient[i][j] )  
} .
\label{eq:pressure:PCISPHstiffness2}
\end{align}
The IISPH constant is $\frac{\omega}{\MatrixElement[i][i]}$ with $\omega=0.5$ and 
\begin{align}
\MatrixElement[i][i] = & - \Dt^2 \sum_j \Mass_j 
\left( \sum_j \frac{\Mass[j]}{\Density[j]^2} \KernelGradient[i][j] \right)
\cdot \KernelGradient[i][j] \nonumber \\
& -
\Dt^2 \sum_j \Mass[j] \left( \frac{\Mass[i]}{\Density[i]^2} \KernelGradient[i][j] \right) \cdot \KernelGradient[i][j] .
\label{eq:pressure:diagonalElemen2}
\end{align} 
Applying the same assumptions as for PCISPH, \ie $\Density[i] = \RestDensity$ and $\Mass[i] = \Mass[j]$, the diagonal element simplifies to
\begin{align}
\MatrixElement[i][i] = & - \frac{\Dt^2 \Mass[i]^2}{(\RestDensity)^2} \left( \sum_j \KernelGradient[i][j] \cdot \sum_j \KernelGradient[i][j] +
\sum_j  ( \KernelGradient[i][j]  \cdot \KernelGradient[i][j] ) \right).
\label{eq:pressure:diagonalElemen3}
\end{align} 
Thus, 
\begin{align}
k^\text{PCI} = \frac{\omega}{\MatrixElement[i][i]}.
\end{align}
Further, 
\begin{align}
\Dt \sum_j \Mass_j (\Dt (\PressureAcceleration[i])^{(l)} - \Dt (\PressureAcceleration[j])^{(l)}) \cdot \KernelGradient[i][j])  = \MatrixPressureRow[i][][][(l)]
\end{align}
which means that the pressure update with Eq.~\eqref{eq:pressure:PCISPHpressureUpdate2} in the PCISPH solver  is equal to the pressure update with Eq.~\eqref{eq:pressure:JacobiIISPH2} in the IISPH solver. There might be insignificant differences in the computations of $\Density[i][][\Predicted]$ and $\MatrixPressureRow[i]$ between PCISPH and IISPH, but both solvers are essentially equal. 

Ihmsen et al.~\cite{ICS+14} report significant performance differences between PCISPH and IISPH. These differences are possibly due to the fact that PCISPH computes one global stiffness constant $k^\text{PCI}$ for a template particle, while IISPH computes $\MatrixElement[i][i]$ for each particle. In particular, $ \sum_j \KernelGradient[i][j] \cdot \sum_j \KernelGradient[i][j] +
\sum_j  ( \KernelGradient[i][j]  \cdot \KernelGradient[i][j] ) $ is assumed to be constant in PCISPH, while it is computed per particle in IISPH. The respective difference affects the performance. If $k^\text{PCI} < \frac{\omega}{\MatrixElement[i][i]}$, the convergence of PCISPH is worse than with IISPH. If $k^\text{PCI} > \frac{\omega}{\MatrixElement[i][i]}$, PCISPH could be unstable. Such potential instabilities are difficult to deal with and might result in the requirement of significantly smaller time steps compared to IISPH. There are also other smaller differences between PCISPH and IISPH. \Eg PCISPH computes $\Density[i][][\Predicted]$ from advected samples without updated neighborhood, while IISPH uses the velocity divergence to estimate $\Density[i][][\Predicted]$. Such differences, however, are probably less relevant for stability or convergence differences.

\subsection{PPE Variants}

In addition to PCISPH, there exist various other pressure solvers that are closely related to a PPE solver, \eg Local Poisson SPH \cite{HLL+12}, Constraint Fluids (CF) \cite{BLS12} and Position-based Fluids (PBF) \cite{MM13}. It is beyond the scope of these course notes to provide a detailed analysis of the aforementioned solvers, but the close relation can be derived from three aspects. First, CF and PBF compute constraint values at particles: $C_i = \frac{\Density[i]}{\RestDensity}-1$. The magnitudes of these constraints are equal to pressure values computed with a state equation $\Pressure[i] = k \left (\frac{\Density[i]}{\RestDensity}-1 \right )$ with stiffness constant $k=1$. Second, CF and PBF compute forces or position changes that are proportional to the negative of the constraint gradient. As constraint and pressure are closely related, constraint gradient and pressure gradient are related as well. Third, CF and PBF iteratively update their solution like a Jacobi solver.  

In addition to these variants there exist publications that analyze the discretizations of the pressure Laplacian and the source term in the PPE. The general idea is always the same, \ie to compute pressure such that the resulting pressure accelerations minimize density deviations or the divergence of the velocity field. Nevertheless, the computed velocity field depends on the employed discretizations. \Eg F{\"{u}}rstenau et al.~\cite{FAW17} compare three discretizations of the pressure Laplacian. They consider the following form of the PPE with velocity divergence as source term:
\begin{align}
\Divergence \left( \frac{1}{\Density[i]} \Gradient \Pressure[i] \right)
= \frac{\Divergence \Velocity[i][][\Predicted]}{\Dt}.
\end{align}
Three variants to compute the pressure Laplacian are analyzed. The first variant is 
\begin{align}
\Divergence \left( \frac{1}{\Density[i]} \Gradient \Pressure[i] \right)
= \frac{\Mass[i]}{\Density[i]} \sum_j \frac{\Density[i]+\Density[j]}{\Density[i] \Density[j]} \frac{(\Pressure_j - \Pressure_i)(\Position[i]-\Position[j])}{(\Position[i]-\Position[j])^2 + \epsilon} \cdot \KernelGradient[i][j],
\end{align}
referred to as finite difference scheme, where $\epsilon$ is a small constant which is added to avoid singularities. The second variant is
\begin{align}
\Divergence \left( \frac{1}{\Density[i]} \Gradient \Pressure[i] \right)
= \frac{\Mass[i]}{\Density[i]} \sum_j \left( \frac{\Gradient \Pressure[j]}{\Density[j]} - \frac{\Gradient \Pressure[i]}{\Density[i]} \right) \cdot \KernelGradient[i][j],
\end{align}
referred to as double summation scheme. This approximation is used, \eg in IISPH \cite{ICS+14}. It computes the divergence of the pressure accelerations. The third variant is
\begin{align}
\Divergence \left( \frac{1}{\Density[i]} \Gradient \Pressure[i] \right)
= \frac{\Mass[i]}{\Density[i]} \sum_j \frac{\Density[i] + \Density[j]}{2 \Density[i] \Density[j]} (\Pressure[j]-\Pressure[i]) \KernelLaplace[i][j],
\end{align}
referred to as second derivative scheme. All these options can be used to compute the left-hand side of the PPE that contains the pressure Laplacian. If the PPE is solved with a Jacobi method, the required diagonal element $\MatrixElement[i][i]$ varies accordingly. 

The three discretizations result in different solutions for the velocity field. It is difficult to come up with general conclusions, as all discretizations have benefits and drawbacks. \Eg the double-summation approach used in IISPH seems to have an improved solver convergence compared to the finite-difference scheme and the second-derivative scheme. On the other hand, the computed velocity field suffers from high-frequency noise. This noise, however, is rather low and it depends on the application whether this is an issue or not. In typical free-surface scenario, this noise is not an issue and the double-summation discretization is just faster than the other two options as less solver iterations are required for a specified tolerated density deviation.

Another degree-of-freedom is the form of the source term. As already shown and discussed in Section~\ref{sec:pressure:derivation}, the source term can either represent the divergence of the predicted velocity or the deviation of the predicted density to the rest density. The predicted velocity is the velocity after applying all non-pressure accelerations. The predicted density is the estimated density after advecting all samples with the predicted velocity. The equivalence of both formulations follows from the continuity equation. The PPE with density invariance as source term is
\begin{align}
\Dt^2 \Laplace \Pressure[i] = \RestDensity - \Density[i][][\Predicted]
\label{eq:PPE_density_deviation}
\end{align}
with source term 
\begin{align}
\MatrixSourceTerm[i] = \RestDensity - \Density[i][][\Predicted].
\label{eq:pressure:DFSPH:densitySourceTerm}
\end{align}
The PPE with velocity divergence as source term is
\begin{align}
\Dt^2 \Laplace \Pressure[i] = \Dt \Density[i] \Divergence \Velocity[i][][\Predicted]
\label{eq:PPE_divergence}
\end{align}
with source term
\begin{align}
\MatrixSourceTerm[i] = \Dt \Density[i] \Divergence \Velocity[i][][\Predicted].
\label{eq:pressure:DFSPH:velocitySourceTerm}
\end{align}
The properties of both variants have been analyzed in \cite{CBG+18}. As already discussed, the density invariance results in an oscillating density deviation over time, while the velocity divergence causes a continuous drift of the density, typically a growing density over time. Another interesting aspect that is discussed in \cite{CBG+18} is the artificial viscosity. Using the density invariance as source term causes more artificial viscosity than using the velocity divergence. Further, the velocity-divergence source term causes less high-frequency or short-range noise in the velocity field. 

Considering all properties has an interesting conclusion. What about solving two PPEs, one with the density invariance as source term and one with the velocity divergence as source term? We would get two velocity changes due to pressure accelerations. According to the properties of the source terms, it can make sense to advect the particles with the velocity from the PPE with density invariance as source term. This avoids the density drift. The velocity from the PPE with velocity divergence, however, could be used for the final velocities at the particles as these velocities have less artificial viscosity and less noise. This has been actually done by Bender and Koschier~\cite{BK15} who proposed to solve two PPEs with different source terms and to use one solution to compute the advected particle positions and one solution as the final velocity field. Solving two PPEs is obviously more expensive than a simple IISPH solver. However, each PPE solve typically requires very few iterations, \ie less than ten, and both PPEs share the same matrix which can be exploited to get an efficient solver as shown in the next section. 

\subsection{Divergence-Free SPH (DFSPH)}

DFSPH \cite{BK17} is a variant of the idea to solve two PPEs with different source terms.
The original DFSPH solver does not compute pressure, but some stiffness parameter $\kappa_i$ per particle $i$. 
According to the work of Band et al.~\cite{BGPT18}, however, this stiffness parameter is closely related to pressure with $\Pressure[i] = \kappa_i \Density[i]$. 
In the following we introduce DFSPH using a pressure formulation and replace the stiffness parameter in order to get a formulation which is closer to the ones of PCISPH and IISPH.
In this way it is easier to see the similarities and the differences.

DFSPH conceptually solves two PPEs, one with density invariance as source term, the other one with velocity divergence as source term. 
Instead of using one solution for the position update and one solution as the final velocity field, DFSPH combines both solutions to compute the final velocity field which is used to advect the particles. 
This combination is motivated by the fact that a first PPE solve with density invariance computes particle positions of an incompressible fluid state, but not necessarily a divergence-free velocity field. 
That's why, a second PPE solve with velocity divergence computes a divergence-free velocity field. 

\paragraph*{Divergence-Free Solver}
If we solve the PPE with velocity divergence as source term (see Eq.~\eqref{eq:PPE_divergence}), we can derive the following equation for the corresponding pressure value $\Pressure[i][][v]$ of a particle $i$:
\begin{equation}
	\Pressure[i][][v] = \frac{1}{\Dt} \frac{D \Density[i]}{D \Time} \cdot \underbrace{\frac{\Density[i]^2}{\| \sum_j \Mass[j] \KernelGradient[i][j] \|^2  + \sum_j \| \Mass[j] \KernelGradient[i][j] \|^2}}_{k_i^\text{DFSPH}},
	\label{eq:dfsph_factor}
\end{equation}
where the time derivative of the density is determined as
\begin{equation}
	\frac{D \Density[i]}{D \Time} = \sum_j \Mass[j] (\Velocity[i] - \Velocity[j]) \cdot  \KernelGradient[i][j].
\end{equation}
Note that this formulation is similar to the pressure computation of PCISPH (see Eqs.~\eqref{eq:pressure:PCISPHstateEquation} and \eqref{eq:pressure:PCISPHstiffness}).
But in contrast to PCISPH the source term is the velocity divergence and not the density deviation.
Moreover, DFSPH does not use a global factor $k^\text{PCI}$ that is determined for a template particle but computes the actual factor $k_i^\text{DFSPH}$ for each particle $i$ in each time step.

Algorithm~\ref{alg:dfsph_divergenceSolver} shows the divergence-free solver.
In each iteration first the divergence is updated for all particles. 
Then the pressure values are determined and the predicted velocity is updated accordingly.

\begin{algorithm}
\caption{Divergence-free solver}\label{alg:dfsph_divergenceSolver}
\begin{algorithmic}[1]
\algnotext{EndFor}
\algnotext{EndWhile}
\algnotext{EndProcedure}
\While {$\left (\left (\frac{D \Density}{D \Time} \right )^{\text{avg}} > \eta^{\text{div}} \right ) \vee \left (\text{iter} < 1 \right )$} \label{alg:velCorrBegin}
	\ForAll{particles $i$}				
		\State $\frac{D \Density[i]}{D \Time} = -\Density[i] \Divergence \Velocity[i][][\Predicted]$
	\EndFor
	\ForAll{particles $i$}			
		\State $\Pressure[i][][v]  = \frac{1}{\Dt} \frac{D \Density[i]}{D \Time} k_i^\text{DFSPH}$,\ \ \ \ \ \ $\Pressure[j][][v] = \frac{1}{\Dt} \frac{D \Density[j]}{D \Time} k_j^\text{DFSPH}$
		\State $\Velocity[i][][\Predicted] := \Velocity[i][][\Predicted] - \Dt \sum_j \Mass[j] \left (\frac{\Pressure[i][][v]}{\Density[i]^2} + \frac{\Pressure[j][][v]}{\Density[j]^2} \right ) \KernelGradient[i][j]$				\label{alg:velCorr}
	\EndFor
\EndWhile								\label{alg:velCorrEnd}
\end{algorithmic}
\end{algorithm}

\paragraph*{Constant Density Solver}
The constant density solver uses the PPE with density deviation as source term (see Eq.~\eqref{eq:PPE_density_deviation}).
For this PPE we get a pressure of
\begin{equation}
	\Pressure[i][] = \frac{1}{\Dt^2} (\Density[i][][\Predicted] - \Density[0]) k_i^\text{DFSPH}, 
\end{equation}
where the predicted density is determined by
\begin{equation}
	\Density[i][][\Predicted] = \Density[i] + \Dt \frac{D \Density[i]}{D \Time} = \Density[i] + \Dt \sum_j \Mass[j] (\Velocity[i][][\Predicted] - \Velocity[j][][\Predicted]) \cdot \KernelGradient[i][j].
\end{equation}
Note that the factor $k_i^\text{DFSPH}$ is used for both, the divergence-free and the constant density solver.
Therefore, it has to be computed only once per simulation step.
Algorithm~\ref{alg:dfsph_pressureSolver} demonstrates an implementation of the constant density solver.

\begin{algorithm}
\caption{Constant density solver}\label{alg:dfsph_pressureSolver}
\begin{algorithmic}[1]
\algnotext{EndFor}
\algnotext{EndWhile}
\algnotext{EndProcedure}
\While {$(\Density[][][\text{avg}] -\Density[0] > \eta) \vee (\text{iter} < 2)$} 		\label{alg:pressureSolveBegin}
	\ForAll{particles $i$}	
		\State compute $\Density[i][][\Predicted]$
	\EndFor
	\ForAll{particles $i$}						
		\State $\Pressure[i] = \frac{\Density[i][][\Predicted] - \Density[0]}{\Dt^2} k_i^\text{DFSPH}$,\ \ \ \ \ \ $\Pressure[j] = \frac{\Density[j][][\Predicted] - \Density[0]}{\Dt^2} k_j^\text{DFSPH}$
		\State $\Velocity[i][][\Predicted] := \Velocity[i][][\Predicted] - \Dt \sum_j \Mass[j] \left (\frac{\Pressure[i]}{\Density[i]^2} + \frac{\Pressure[j]}{\Density[j]^2} \right ) \KernelGradient[i][j]$
	\EndFor
\EndWhile										\label{alg:pressureSolveEnd}
\end{algorithmic}
\end{algorithm}

\paragraph*{DFSPH Simulation Step}
Algorithm~\ref{alg:dfsph} shows a simulation step with DFSPH and how both solvers are integrated in the time step.
Note that the neighborhoods, the particle densities and the factor $k_i^\text{DFSPH}$ are computed once at the beginning of the simulation for the initial state and then updated once per time step.
The algorithm first computes predicted velocities by integrating all non-pressure accelerations. 
Then the density deviation is corrected using the constant density solver which yields new particle positions.
Hence, the neighborhoods, the density values and the factors must be updated. 
After correcting the density deviation the velocity field is typically not divergence-free. 
This is corrected in the last step by the divergence-free solver which gives us the final velocities.
Note that the order of the steps is a bit different than the order of other solvers but in this way it is guaranteed that the density deviations and the divergence error are both corrected at the end of a time step. 
Moreover, in this way we have to update the factor $k_i^\text{DFSPH}$ only once per time step but are able to use it twice: for the constant density solver and for the divergence-free solver. 

\begin{algorithm}
\caption{Simulation step with DFSPH}\label{alg:dfsph}
\begin{algorithmic}[1]
\algnotext{EndFor}
\algnotext{EndWhile}
\algnotext{EndProcedure}
\ForAll{particles $i$}
	\State compute non-pressure accelerations $\NonPressureAcceleration[i]$
\EndFor
\State adapt time step size $\Dt$ according to CFL condition
\ForAll{particles $i$}		
	\State predict velocity $\Velocity[i][][\Predicted] = \Velocity[i] + \Dt \NonPressureAcceleration[i] $		\label{alg:predictVel}		
\EndFor
\State correct density error using algorithm~\ref{alg:dfsph_pressureSolver}
\ForAll{particles $i$}			
	\State update position $\Position[i][][\Time+\Dt] = \Position[i] + \Dt \Velocity[i][][\Predicted]$		\label{alg:posIntegration}
\EndFor
\State update neighborhoods 
\ForAll{particles $i$}			
	\State update density $\Density[i]$
	\State update factor $k_i^\text{DFSPH}$ using Eq.~\eqref{eq:dfsph_factor}
\EndFor
\State correct divergence error using algorithm~\ref{alg:dfsph_divergenceSolver}
\ForAll{particles $i$}			
	\State update velocity $\Velocity[i][][\Time+\Dt] = \Velocity[i][][\Predicted]$
\EndFor
\end{algorithmic}
\end{algorithm}

DFSPH solves two PPEs which is more expensive than solving just one like PCISPH or IISPH. 
However, the second solve is not that expensive since the costly computation of the factor $k_i^\text{DFSPH}$ has to be performed only once per step. 
Experiments have shown that solving both PPEs leads to a better stability which enables larger time steps and therefore a faster simulation~\cite{BK17}.
The performance can be further improved by using a warm start. 
More details about this can be found in~\cite{BK17}.

\subsection{The Best Pressure Solver}

Iterative PPE solvers are more expensive to compute than EOS solvers. Their utility, however, is motivated by the fact that PPE solvers work with significantly larger time steps compared to EOS solvers. Solenthaler and Pajarola~\cite{SP09} show an improved overall performance of PCISPH compared to the EOS solver in \cite{BT07}. Ihmsen et al.~\cite{ICS+14} show an improved performance of IISPH compared to PCISPH and Bender and Koschier~\cite{BK17} show a performance gain of DFSPH compared to IISPH. So, DFSPH has the best overall performance of all discussed PPE variants. 

Although PPE solvers work with big time steps, they do not reach their best overall performance for the largest possible time step as discussed in \cite{ICS+14} and \cite{IOS+14}. Although the number of neighborhood searches decreases for larger time steps, the solver iterations increase for larger time steps. 

The reported performance gains of PPE solvers compared to EOS solvers have been estimated for so-called complex scenarios. The term complex refers basically to the height of a simulated fluid body under gravity. The higher the simulated fluid column, the more complex the scenario, the bigger the performance gain of a PPE solver.  If a scenario is simple, \eg one layer of fluid particles on a planar boundary, EOS solvers are faster. The overall number of particles does not necessarily influence the solver performance. An EOS solver is more efficient than a PPE solver for one billion fluid particles in one layer on a plane, while a PPE solver is more efficient than an EOS solver for one hundred particles on top of each other in one column under gravity.

Independent from whether there is a performance gain of a PPE solver, they are more simple to handle than EOS solvers. In an EOS solver, the stiffness constant has to be found to realize a desired density deviation and the time step has to be found to get a stable simulation. In a PPE solver, the desired density deviation is explicitly specified. The time step is also easier to estimate as it is typically rather larger, corresponding to CFL numbers close to one.


\section{Boundary Handling}
\label{sec:boundary_handling}

In order to complete the discretization of a mixed initial-boundary value problem (see Section~\ref{sec:foundations}) the boundary of the simulation domain has to be discretized and the corresponding boundary conditions must be enforced.
In recent years, a wide variety of approaches to represent boundary geometries and to enforce boundary conditions has been presented.
The approaches can be roughly categorized into particle-based approaches, \eg \cite{AIA+12,IAGT10,BT07,BGPT18,BGI+18,GPB+19}, and implicit approaches, \eg \cite{KB17,HKK07,HKK07a,BLS12}.

The particle based strategy is probably the most popular representation type.
The main idea is to sample the boundary geometry using an additional set of so-called \emph{boundary particles} equipped with a (sometimes specialized) kernel function.
The advantages here are that the representation is consistent with the discretization of the fluid or solid.
Modeling, either explicit/implicit boundary forces for weak satisfaction of boundary conditions or algorithms to strongly satisfy the constraints is probably more straightforward than using implicit or mesh-based techniques.
Most methods, however, have the constraint that the particle size used to sample the boundary has to be the same as the particle size of the continuum discretization.
The disadvantages are that even the representation of simple geometries, such as a plane, requires a large number of boundary particles that have to be accounted for during the neighborhood search and in the evaluation of field quantities, \eg Eq.~\eqref{eq:density_reconstruction}.
Moreover, determining "good" samplings is generally non-trivial.
Too sparse samplings might not sufficiently cover the surface of the boundary leading to the issue of SPH particles penetrating the boundary.
Too dense samplings lead to an increased computational effort and to higher memory requirements.
Also a somewhat "bumpy" sampling might lead to a bias in the particle trajectories as the discretized surfaces' smoothness suffers from sampling noise leading to unwanted perturbations in the simulation (\cf \cite{KB17}) if no additional considerations are made, such as proposed by Band et al.~\cite{BGPT18}.

Implicit boundaries use an implicit function -- typically a signed distance field (SDF) -- to represent the boundary geometry.
Advantages of this type of approaches are that the boundary representation is decoupled from the particle size.
As a result, more flexible data structures, \eg adaptive octrees with higher-order approximations~\cite{KDBB17}, can be used to memory-efficiently and accurately represent the boundary geometry.
This circumstance also avoids the problem of noisy boundary samplings, the resulting bias, and unwanted perturbations in the particle trajectories.
Typical disadvantages are, that implicit representations do not directly integrate with particle based continuum discretizations.
In order to couple both discretization types, special considerations have to be made and the corresponding implementation is rather involved.

In the remainder of this section, we will discuss approaches to handle non-penetration of rigid boundaries using particle sampling approaches.
We gradually develop a formulation starting with a simple dense, uniform multilayer sampling of the boundary and show how the method can be simplified to a uniform single layer sampling and consequently even to a robust and consistent formulation using non-uniformly sampled boundaries.
We moreover discuss how the fluid-boundary coupling can be improved using pressure mirroring or pressure extrapolation and how these techniques can be incorporated into the previously discussed pressure solvers. 
For implicit or mesh-based boundary handling techniques we would like to refer the reader to the corresponding literature, \eg \cite{KB17,HKK07,HKK07a,BLS12,MFK+15,FLR+13,FM15}

\subsection{Particle-based Boundary Handling}

\begin{figure}[tb]
	\centering
	\includegraphics[trim = 0mm 40mm 0mm 40mm, clip, width=\linewidth]{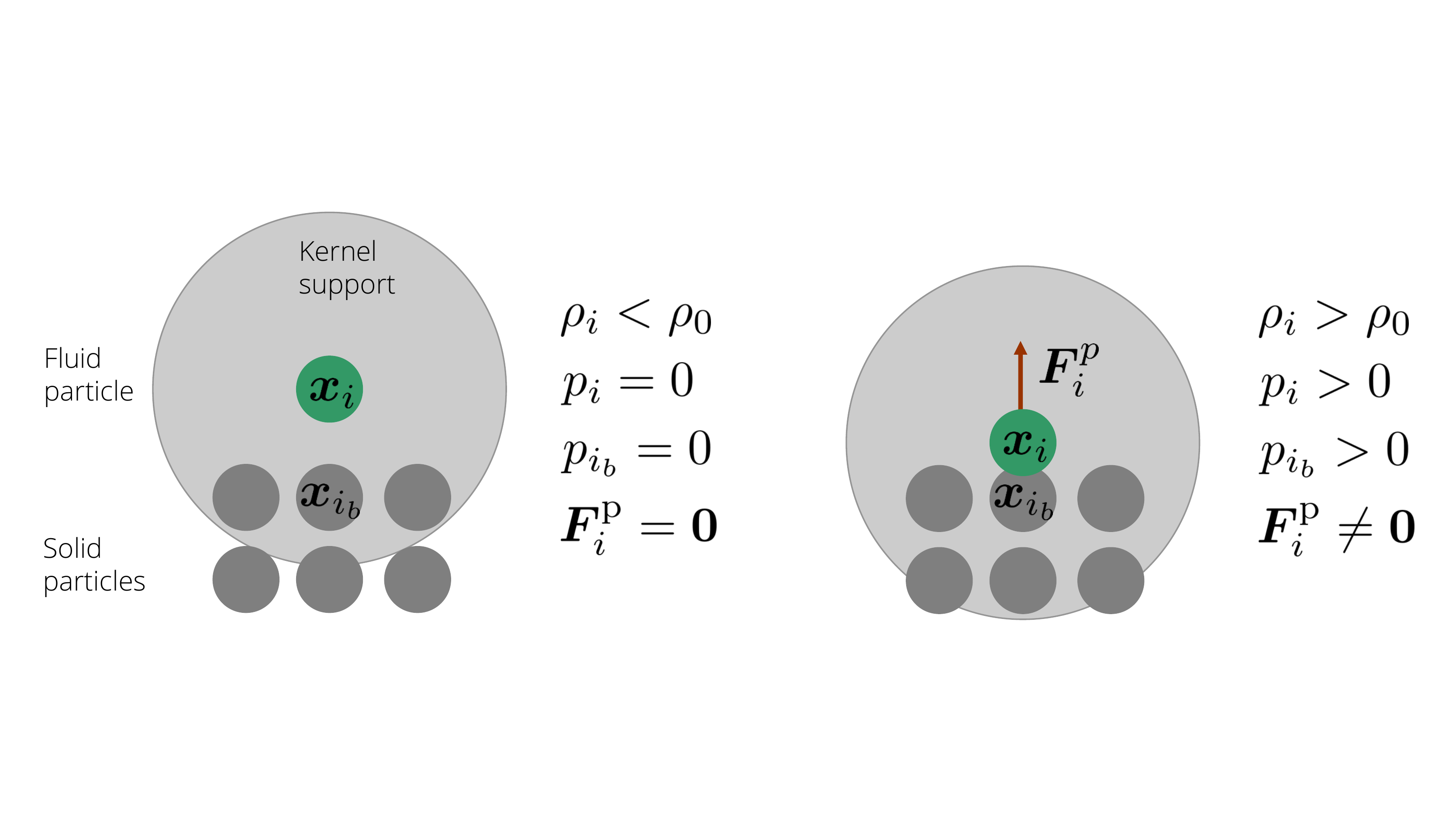} 
	\caption{Idea of the particle-based boundary handling. The boundary is represented with particles $\Position[i_b]$. These particles are considered in the computation of density $\Density[i]$ , pressure $\Pressure[i]$, and pressure force $\PressureForce[i]$ of nearby fluid particles $\Position[i]$. If a fluid particle moves closer to the boundary, its density increases. If a fluid particle is too close, its density is larger than the rest density, \ie $\Density[i]>\RestDensity$, which causes pressure $\Pressure[i]>0$ which in turn causes a pressure force $\PressureForce[i] \neq \vec{0}$ that accelerates the fluid particle away from boundary particles. Please note that $i_b$ denote boundary neighbors of a fluid particle $i$. Accordingly, $i_f$ refer to indices of fluid neighbors of a fluid particle $i$.}
	\label{fig:boundaryHandling:concept}
\end{figure}

In this concept, the boundary is represented with particles and these boundary particles are incorporated into the computation of density $\Density[i]$, pressure $\Pressure[i]$, and $\PressureForce[i]$ at nearby fluid particles $\Position[i]$. This is illustrated in Fig.~\ref{fig:boundaryHandling:concept}. 
There are two main aspects to discuss. First: The boundary can be sampled in different ways. \Eg, the boundary can be sampled with particles of uniform size corresponding to the size of a fluid particle. Or the boundary can be sampled with particles of non-uniform size. Also, the boundary can be sampled with several layers of boundary particles as indicated in Fig.~\ref{fig:boundaryHandling:concept} or the boundary can be sampled with just one layer of particles. The second aspect is the actual computation of density $\Density[i]$ , pressure $\Pressure[i]$, and pressure force $\PressureForce[i]$ of nearby fluid particles $\Position[i]$. Such computations require information from boundary samples, \eg pressure. Such pressure at boundary samples can be estimated in different ways. Here, typical examples are pressure mirroring, \ie it is assumed that the pressure at the boundary particles equals the pressure at adjacent fluid particles. Alternatively, pressure can be extrapolated from the fluid into the boundary. While the pressure extrapolation is theoretically the correct choice, this concept is challenging to realize due to the fact that the computation of the pressure gradient at a fluid particle close to the boundary is error-prone. In the following, we discuss different combinations of the aforementioned variants.

\subsubsection{Different Types of Boundary Samplings}

\begin{figure}[tb]
	\centering
	\includegraphics[trim = 0mm 40mm 0mm 40mm, clip, width=\linewidth]{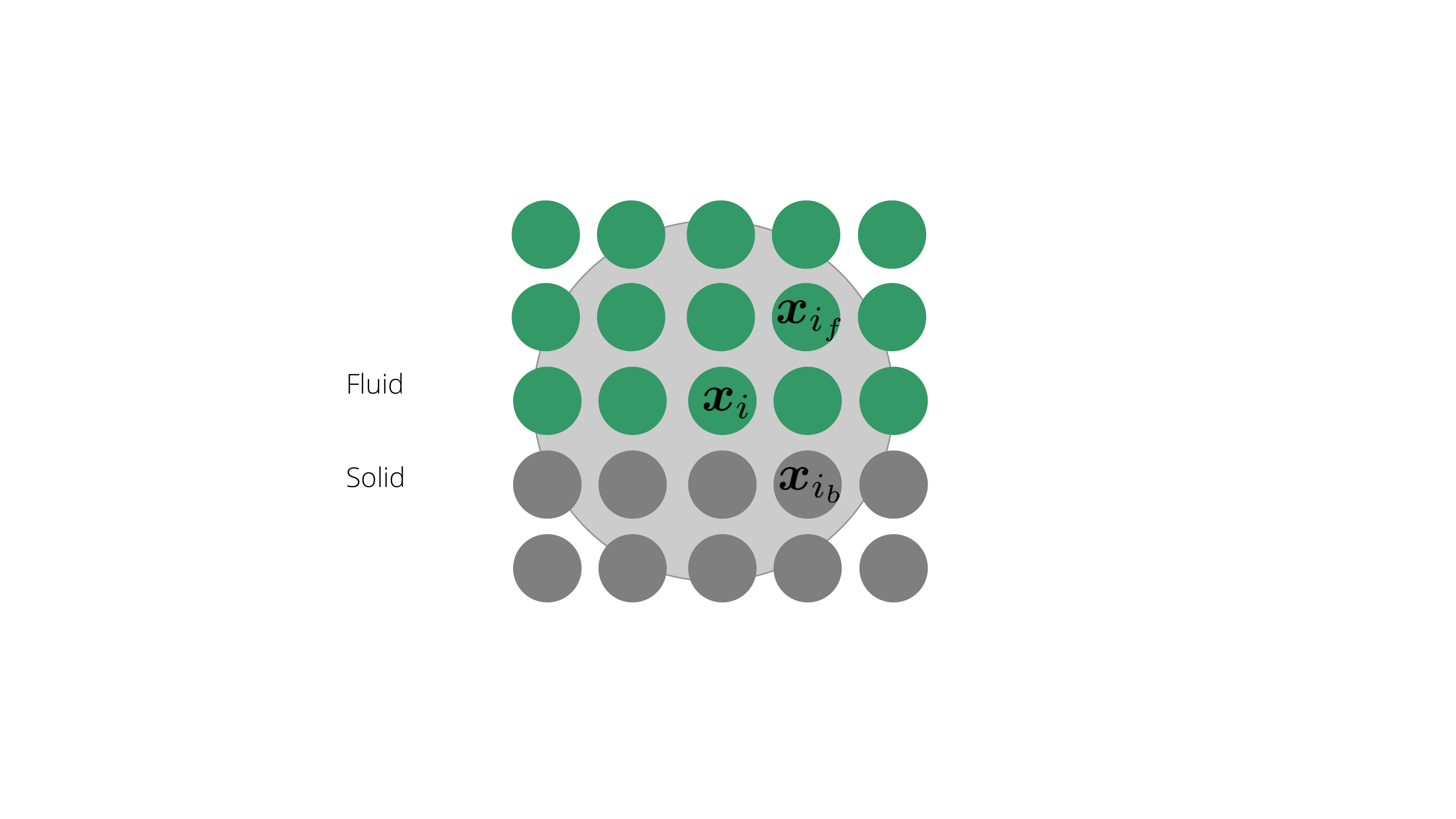} 
	\caption{Sample-based boundary representation with several layers. The usage of several boundary layers avoids incomplete neighborhoods for fluid particles $\Position[i]$ close to the boundary.  }
	\label{fig:boundaryHandling:severalLayers}
\end{figure}

\paragraph*{Several layers with boundary samples of uniform size:} 
If a fluid particle $\Position[i]$ is close to a boundary, it generally has fluid neighbors $\Position[i_f]$ and boundary neighbors $\Position[i_b]$ as illustrated in Fig.~\ref{fig:boundaryHandling:severalLayers}. All these neighbors contribute to the density computation, \ie
\begin{align}
\Density[i] = \sum_{i_f} \Mass[i_f] \Kernel[i][i_f] + \sum_{i_b} \Mass[i_b] \Kernel[i][i_b] \mathpoint
\label{eq:boundary:density1}
\end{align} 
All samples $\Position[i]$, $\Position[i_f]$, $\Position[i_b]$ have the same size and we take the boundary samples as static fluid samples resulting in the same rest density $\RestDensity$ for all fluid and boundary particles. This also means that all masses are equal: $\Mass[i] = \Mass[i_f] = \Mass[i_b]$. Thus, the density computation could also be written as $\Density[i] = \Mass[i] (\sum_{i_f} \Kernel[i][i_f] + \sum_{i_b} \Kernel[i][i_b])$. If the boundary samples belong to a rigid body, it is not perfectly intuitive to represent its boundary with particles that have mass and rest density of a fluid particle. This issue results from the fact that SPH formulations often prefer to weight the contribution of a particle $i$ with $\frac{\Mass[i]}{\Density[i]}$ instead of using its volume $\Volume[i] = \frac{\Mass[i]}{\Density[i]}$. Now, in the boundary handling, one actually works with the volume of boundary particles. Nevertheless, this volume is often represented with some rest density - typically the fluid rest density - and the respective mass.

If the density $\Density[i]$ in Eq.~\eqref{eq:boundary:density1} is larger than the rest density $\RestDensity$, a pressure $\Pressure[i]$ is computed at fluid particles. We have seen that $\Pressure[i]$ can be computed from a state equation, \eg $\Pressure[i] = k \left (\frac{\Density[i]}{\RestDensity} -1 \right )$ or from solving a PPE.      

Now, the pressure force at the fluid particle can be computed as
\begin{align}
\PressureForce[i] = &- \Mass[i] \Mass[i] \sum_{i_f} \left( \frac{\Pressure[i]}{\Density[i]^2} + \frac{\Pressure[i_f]}{\Density[i_f]^2} \right) \KernelGradient[i][i_f] \nonumber \\
&- \Mass[i] \Mass[i] \sum_{i_b} \left( \frac{\Pressure[i]}{\Density[i]^2} + \frac{\Pressure[i_b]}{\Density[i_b]^2} \right) \KernelGradient[i][i_b] \mathpoint
\label{eq:boundary:force1} 
\end{align}
It can be seen in Eq.~\eqref{eq:boundary:force1} that we basically compute a pressure force component with respect to fluid neighbors and a pressure force component with respect to boundary neighbors. Using several layers of boundary particles guarantees that the neighborhood of a fluid particle is completely sampled, even if this particle is very close to the boundary. Fully filled neighborhoods keep the errors due to missing samples in Eqs.~\eqref{eq:boundary:density1} and~\eqref{eq:boundary:force1} small.

\begin{figure}[tb]
	\centering
	\includegraphics[trim = 0mm 30mm 0mm 35mm, clip, width=\linewidth]{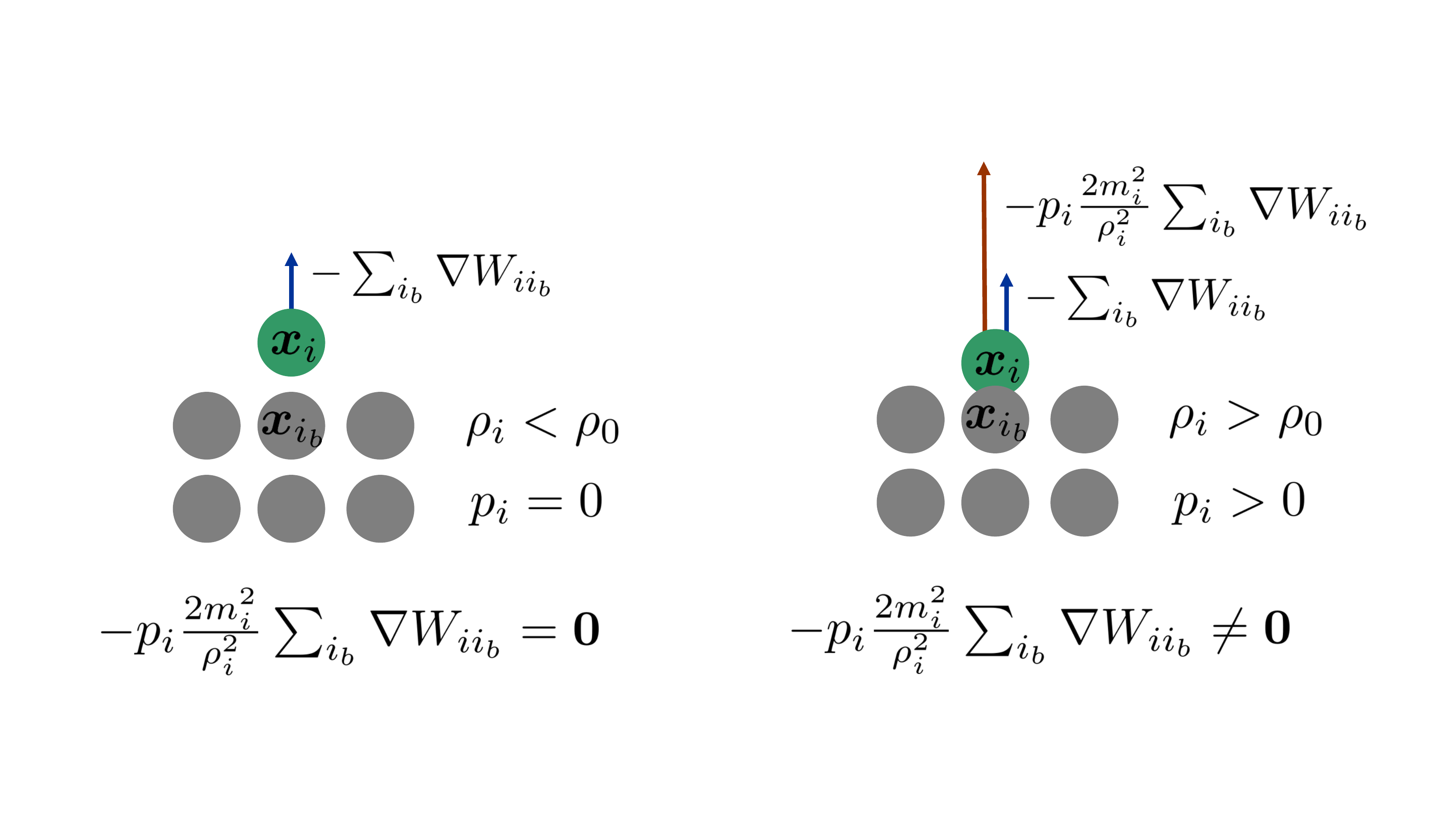} 
	\caption{Working with equal pressure $\Pressure[i]$ at a fluid particle $i$ and at its adjacent boundary samples results in a repulsion force in case of $\Pressure[i] > 0$.  }
	\label{fig:boundaryHandling:severalLayers2}
\end{figure}

The computation in Eq.~\eqref{eq:boundary:force1} requires positions, densities and pressures of adjacent fluid and boundary particles. While these quantities are known for fluid particles, density and pressure at boundary samples are still unknown. As the boundary samples have a fixed volume and their mass is set with respect to the rest density of the fluid, it is appropriate to set the density of a boundary particle to the rest density of the fluid, \ie $\Density[i_b] = \RestDensity$. Regarding the pressure $\Pressure[i_b]$, we have already briefly mentioned pressure mirroring and pressure extrapolation. Here, the simplest idea is to mirror the pressure from a fluid particle to an adjacent boundary particle, \ie $\Pressure[i_b] = \Pressure[i]$. These assumptions result in an adapted form of Eq.~\eqref{eq:boundary:force1} for the pressure force where all required quantities at fluid and boundary neighbors are known:
\begin{align}
\PressureForce[i] = &- \Mass[i] \Mass[i] \sum_{i_f} \left( \frac{\Pressure[i]}{\Density[i]^2} + \frac{\Pressure[i_f]}{\Density[i_f]^2} \right) \KernelGradient[i][i_f] \nonumber \\
&- \Mass[i] \Mass[i] \sum_{i_b} \left( \frac{\Pressure[i]}{\Density[i]^2} + \frac{\Pressure[i]}{(\RestDensity)^2} \right) \KernelGradient[i][i_b] \mathpoint
\label{eq:boundary:force2} 
\end{align}
Working with equal pressure at a fluid sample and at its neighboring boundary samples theoretically corresponds to a pressure gradient of zero which in turn results in a pressure acceleration of zero. In practice, however, the SPH derivative approximation always results in the desired gradient with the respective repulsion force. If the neighborhood of a fluid particle is completely filled, as shown in Fig.~\ref{fig:boundaryHandling:severalLayers}, the true gradient is slightly underestimated by SPH. This leads to a small, practically not relevant amount of penetration of the fluid into the boundary. In the other case, where a single fluid particle without fluid neighbors is close to the boundary as depicted in Fig.~\ref{fig:boundaryHandling:severalLayers2}, the contributions from missing fluid neighbors are implicitly assumed to be zero. Having a fluid neighbor with zero pressure or not having this fluid neighbor has the same effect on the SPH approximation.

For a single fluid particle at a boundary, Eq.~\eqref{eq:boundary:force2} simplifies to $\PressureForce[i] =
- \Pressure[i] \frac{2 \Mass[i]^2}{\Density[i]^2} \sum_{i_b} \KernelGradient[i][i_b]$ for $\Density[i] = \RestDensity$. The term $-\sum_{i_b} \KernelGradient[i][i_b]$ can be interpreted as the surface normal of the boundary close to position $\Position[i]$. The pressure force at particle $\Position[i]$ corresponds to this vector scaled with the fluid particle pressure $\Pressure[i]$. So, if the density of the fluid particle is larger than its rest density, we get a positive pressure and a repulsion force from the boundary into normal direction.

In summary, computing the fluid density $\Density[i]$ with Eq.~\eqref{eq:boundary:density1}, the pressure $\Pressure[i]$ with a state equation or a PPE, and the pressure force with Eq.~\eqref{eq:boundary:force2} realizes a boundary handling with pressure mirroring in case of a uniformly sampled boundary with several layers. The boundary handling works for fluid particles with a complete or incomplete neighborhood.

\begin{figure}[tb]
	\centering
	\includegraphics[trim = 0mm 45mm 0mm 45mm, clip, width=\linewidth]{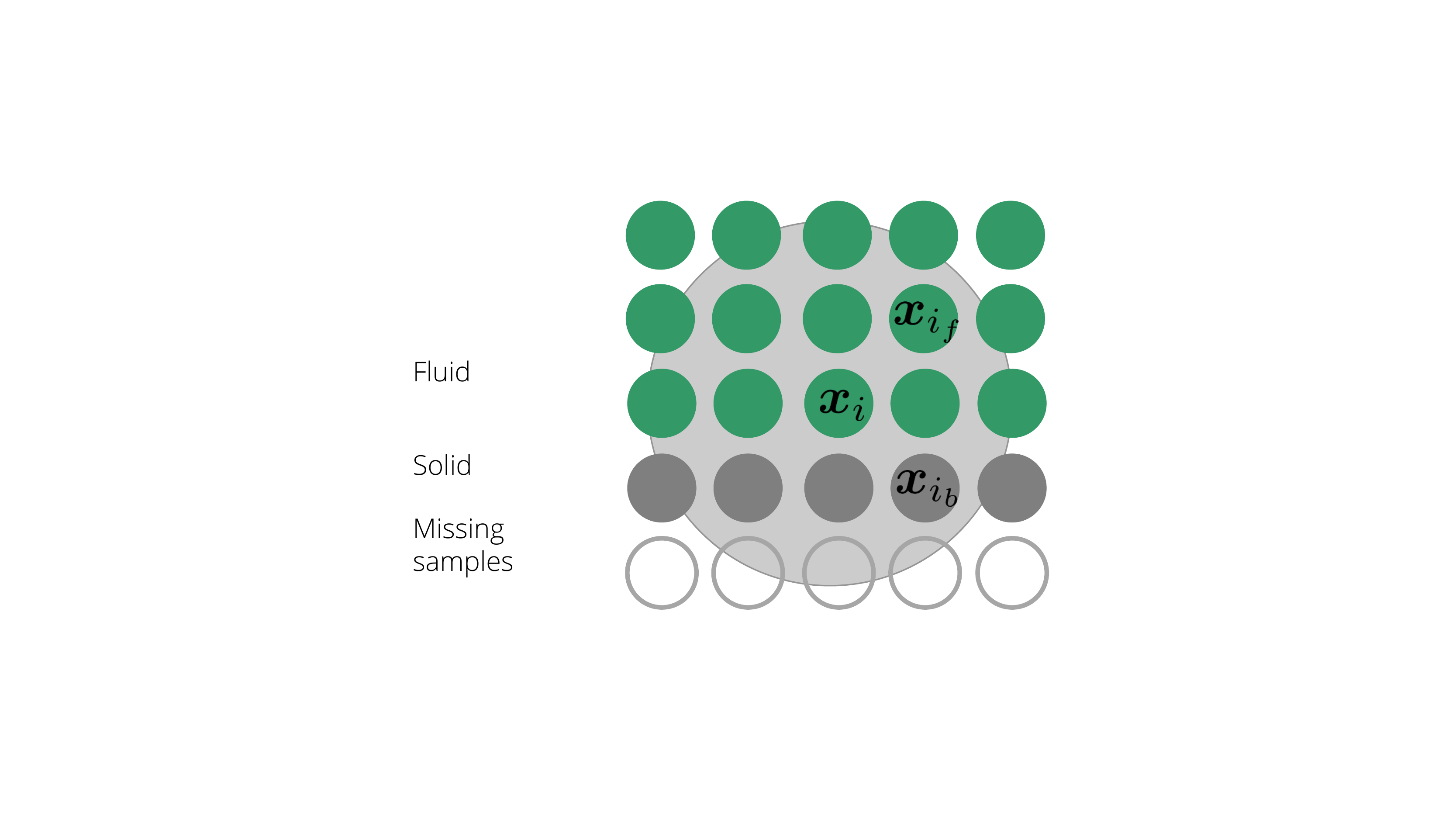} 
	\caption{One layer of uniform boundary samples is easier to generate than several layers. The boundary layer is used to naturally detect the proximity of a fluid sample to a boundary. Due to the size of the kernel support, however, more than one layer of samples might be required in SPH computations. The contributions of these missing samples can be analytically estimated.  }
	\label{fig:boundaryHandling:severalLayers3}
\end{figure}

\paragraph*{One layer of uniform boundary samples:}
It can be difficult to generate multiple layers of uniform boundary samples for arbitrarily shaped geometries. Also, if the relative position of a fluid sample to the boundary can be determined, it is not necessarily required to explicitly represent the boundary particles. Instead, their contributions can  analytically be estimated. Fig.~\ref{fig:boundaryHandling:severalLayers3} shows a setting with one layer of uniform boundary samples. These samples are used to estimate that a fluid particle is close to a boundary. For the computation of SPH approximations, however, more than one layer of samples might be required as indicated in Fig.~\ref{fig:boundaryHandling:severalLayers3}.   

For a given position $\Position[i]$ of a fluid particle close to a one-layer boundary, the density can be written as
\begin{align}
\Density[i] = \Mass[i] \sum_{i_f} \Kernel[i][i_f] + \Mass[i] \sum_{i_b} \Kernel[i][i_b] + \Mass[i] \sum_{i_m} \Kernel[i][i_m]. 
\label{eq:boundary:density2}
\end{align} 
The index $i_m$ refers to missing samples in the neighborhood of particle $i$. While it is a natural way to encode the contributions of missing samples as an offset, these contributions are typically encoded as a correcting factor of the contributions from boundary samples instead, \ie
\begin{align}
\Density[i] = \Mass[i] \sum_{i_f} \Kernel[i][i_f] + \gamma_1 \Mass[i] \sum_{i_b} \Kernel[i][i_b] . 
\label{eq:boundary:density3}
\end{align} 
Correcting coefficients for contributions from missing boundary samples have been proposed in \cite{AIA+12}. They are commonly used. Benefits and drawbacks to correcting offsets as in Eq.~\eqref{eq:boundary:density2} are unclear and have not been analyzed yet. 

The correcting coefficient $\gamma_1$ in Eq.~\eqref{eq:boundary:density3} depends on various aspects, \eg kernel function, kernel support, dimensionality. In particular, however, it depends on the position of a fluid particle relative to the boundary. In practice, the correcting coefficient is determined for a template particle in a perfect sampling pattern as shown in Fig.~\ref{fig:boundaryHandling:severalLayers3}. If the neighborhood of a fluid particle at the boundary is fully filled, the kernel sum over all neighbors gives one over the particle volume according to the general kernel properties, \ie $\sum_{i_f} \Kernel[i][i_f] + \sum_{i_b} \Kernel[i][i_b] = \frac{1}{\Volume[i]}$. If there are samples missing, this equation does not hold and we introduce the correcting coefficient $\gamma_1$ to obtain the desired result:
\begin{align}
\sum_{i_f} \Kernel[i][i_f] + \gamma_1 \sum_{i_b} \Kernel[i][i_b] = \frac{1}{\Volume[i]}.
\end{align}
Solving this equation gives the desired value for the correcting coefficient:
\begin{align}
\gamma_1 = \frac{\frac{1}{\Volume[i]} - \sum_{i_f} \Kernel[i][i_f]}{\sum_{i_b} \Kernel[i][i_b]}.
\end{align}

The corrected density computation is the basis for the pressure computation which is typically not affected by missing boundary samples. A state equation just works with the density of the fluid particle itself. A PPE typically uses the densities of adjacent fluid particles. So, boundary samples are not required in the pressure computation.

In the subsequent computation of the pressure forces, however, the missing samples have to be accounted for. Similar to the density computation, a correcting coefficient $\gamma_2$ is introduced:
\begin{align}
\PressureForce[i] = - \Mass[i] \Mass[i] \sum_{i_f} \left( \frac{\Pressure[i]}{\Density[i]^2} + \frac{\Pressure[i_f]}{\Density[i_f]^2} \right) \KernelGradient[i][i_f] 
- \gamma_2 \Pressure[i] \frac{2 \Mass[i]^2}{\Density[i]^2} \sum_{i_b} \KernelGradient[i][i_b].
\end{align}
This coefficient is derived from a property of the kernel gradient. If the neighborhood of a particle is perfectly sampled, the sum of the kernel gradient over the neighbors is zero: $\sum_{i_f} \KernelGradient[i][i_f] + \sum_{i_b} \KernelGradient[i][i_b] = \vec{0}$. In case of missing boundary samples, the sum is not zero and a correcting coefficient for the boundary contributions is introduced to meet the constraint:
\begin{equation}
\sum_{i_f} \KernelGradient[i][i_f] + \gamma_2 \sum_{i_b} \KernelGradient[i][i_b] = \vec{0}.
\end{equation}
Solving this equation results in 
\begin{equation}
\gamma_2 = \frac{\sum_{i_f} \KernelGradient[i][i_f] \cdot \sum_{i_b} \KernelGradient[i][i_b]}{\sum_{i_b} \KernelGradient[i][i_b] \cdot \sum_{i_b} \KernelGradient[i][i_b]}.
\end{equation} 
Similar to the coefficient $\gamma_1$, $\gamma_2$ also depends on dimenionality, kernel function and support. It also depends on the position $\Position[i]$ of fluid particle $i$ relative to the boundary. In practice, however, the coefficient is determined for a template setting with perfect sampling as depicted in Fig.~\ref{fig:boundaryHandling:severalLayers3}.

In summary: One-layer boundary representations are more simple to generate than multi-layer boundaries. The contributions of missing samples in the computation of the density and the pressure force at a fluid particle close to the boundary can be approximated with correcting coefficients. 

\begin{figure}[tb]
	\centering
	\includegraphics[trim = 0mm 45mm 0mm 45mm, clip, width=\linewidth]{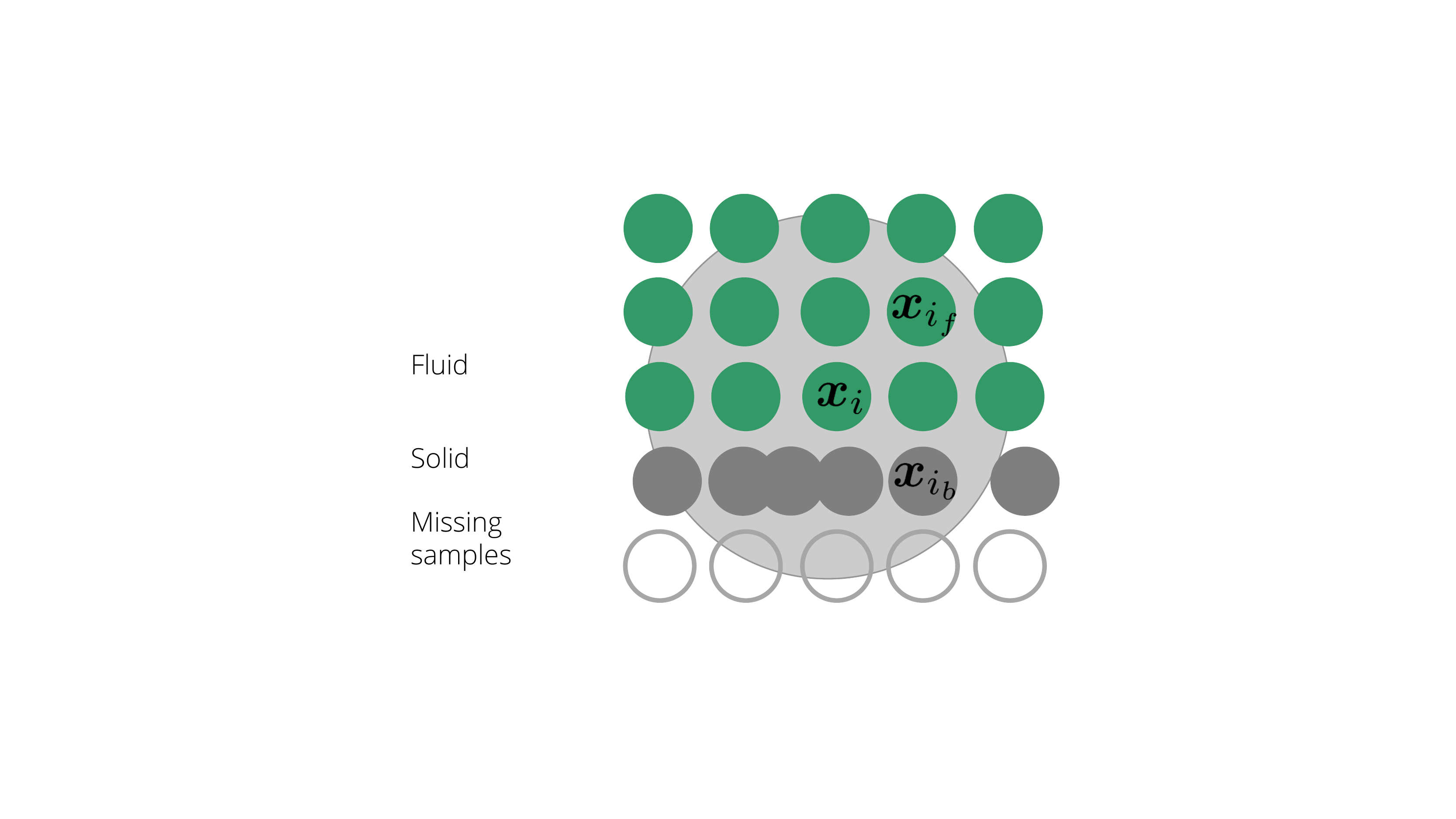} 
	\caption{The boundary geometry is sampled with one layer of particles of non-uniform size. The sampling should be sufficiently dense, \ie each boundary particle should be equal or smaller than a fluid particle. Otherwise, leakage could occur. }
	\label{fig:boundaryHandling:nonuniformSampling1}
\end{figure}

\paragraph*{One layer of non-uniform boundary samples:}
The next step to an even more flexible boundary representation is to work with samples of arbitrary size as proposed in \cite{AIA+12} and illustrated in Fig.~\ref{fig:boundaryHandling:nonuniformSampling1}.
This sampling is motivated by the fact that its generation is really simple. Particles can be arbitrarily placed on a boundary geometry, as long as each boundary particle is equal or smaller than a fluid particle. Even more than one boundary particle at the same position can be handled. This extreme case is certainly not optimal in terms of performance, but the boundary handling works.

\begin{figure}[tb]
	\centering
	\includegraphics[trim = 0mm 20mm 0mm 20mm, clip, width=\linewidth]{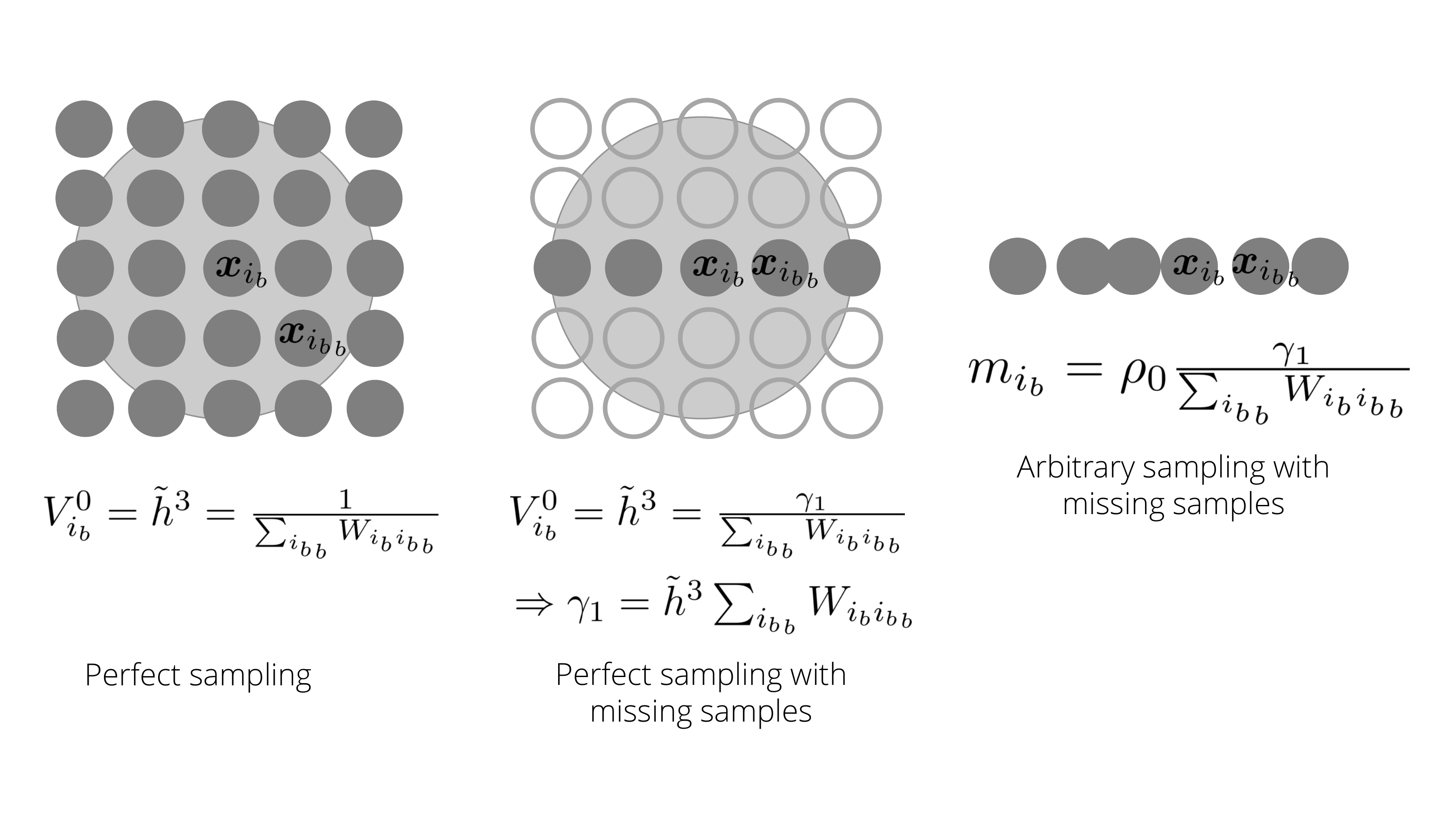} 
	\caption{Derivation of the mass $\Mass[i_b]$ of a boundary sample of arbitrary size. The artificial mass is computed from the volume of the sample and the rest density of the fluid. The left image shows the volume for a filled neighborhood and uniform sample size. The volume is $\ParticleSize^3$. It can also be computed as one over the sum of the kernel values from neighboring particles. In the middle image, we have removed all samples except the samples in a plane which represents the boundary. We still know that the size of $\Position[i_b]$ should be $\ParticleSize^3$, but the kernel sum would not provide the correct result. That's why the correcting coefficient $\gamma_1$ is introduced to get the expected result $\ParticleSize^3$. \Ie $\gamma_1$ accounts for missing neighbors in the volume computation. The volume computation $\Volume[i_b][][0] = \frac{\gamma_1}{\sum_{{i_b}_b} \Kernel[i_b][{i_b}_b]}$ also works for arbitrary sample sizes within the plane as shown in the image on the right. The artificial mass of a boundary sample is finally computed from its volume and the rest density of a fluid particle.  }
	\label{fig:boundaryHandling:nonuniformSampling2}
\end{figure}

The basic idea of the boundary handling with non-uniform boundary samples is the consideration of the actual contribution, \ie the actual volume of each boundary sample. The density of a fluid particle near the boundary is computed with
\begin{align}
\Density[i] = \Mass[i] \sum_{i_f} \Kernel[i][i_f] + \sum_{i_b} \Mass[i_b] \Kernel[i][i_b],
\label{eq:boundary:density4}
\end{align} 
where $\Mass[i_b]$ represents the contribution of boundary sample $i_b$. If a boundary sample is bigger, its contribution is bigger and this contribution is encoded in the mass $\Mass[i_b]$. The contribution, \ie the artificial mass of a boundary sample is deduced from its volume and the rest density of the fluid as explained in Fig.~\ref{fig:boundaryHandling:nonuniformSampling2}.
So, the mass $\Mass[i_b]$ of a boundary sample is computed as 
\begin{align}
\Mass[i_b] = \RestDensity \frac{\gamma_1}{\sum_{{i_b}_b} \Kernel[i_b][{i_b}_b]}.
\label{eq:boundaryHandling:mass1}
\end{align}
The correcting coefficient $\gamma_1$ actually accounts for missing contributions in two cases. It cancels missing contributions in the computation of the mass $\Mass[i_b]$ of a boundary sample. In parallel, it accounts for missing contributions in the density computation of a nearby fluid particle using Eq.~\ref{eq:boundary:density4}. The pressure force is now computed with
\begin{align}
\PressureForce[i] = - \Mass[i] \Mass[i] \sum_{i_f} \left( \frac{\Pressure[i]}{\Density[i]^2} + \frac{\Pressure[i_f]}{\Density[i_f]^2} \right) \KernelGradient[i][i_f] 
- \gamma_2 \Pressure[i] \frac{2 \Mass[i]}{\Density[i]^2} \sum_{i_b} \Mass[i_b] \KernelGradient[i][i_b],
\end{align}
which is very similar to the pressure force for uniform samples. The only difference is the consideration of the individual masses $\Mass[i_b]$ of the boundary particles instead of the standard mass $\Mass[i]$ for samples of uniform size. The correcting factor $\gamma_2$ is 
\begin{equation}
\gamma_2 = \frac{\sum_{i_f} \KernelGradient[i][i_f] \cdot \sum_{i_b} \KernelGradient[i][i_b]}{\sum_{i_b} \KernelGradient[i][i_b] \cdot \sum_{i_b} \KernelGradient[i][i_b]}.
\end{equation} 
The same factor has already been used, motivated, and derived in the case of a one-layer boundary with uniform samples.


\section{Viscosity}
\label{sec:viscosity}

\begin{figure*}[t]
	\centering
	\includegraphics[height=3.22cm]{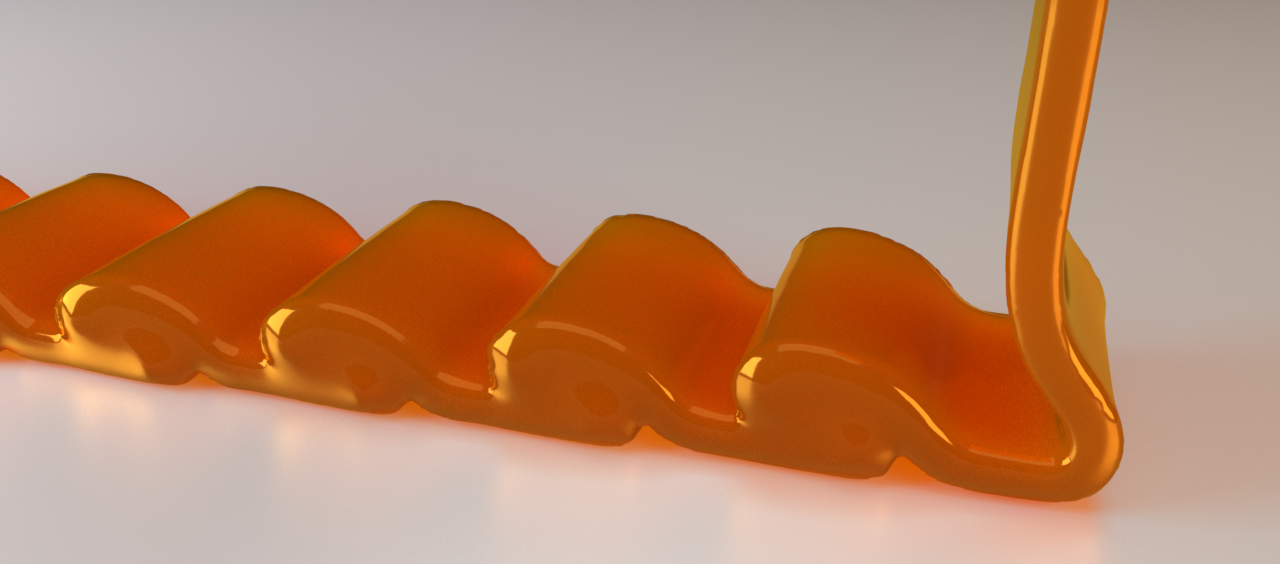} \hfill
	\includegraphics[height=3.22cm]{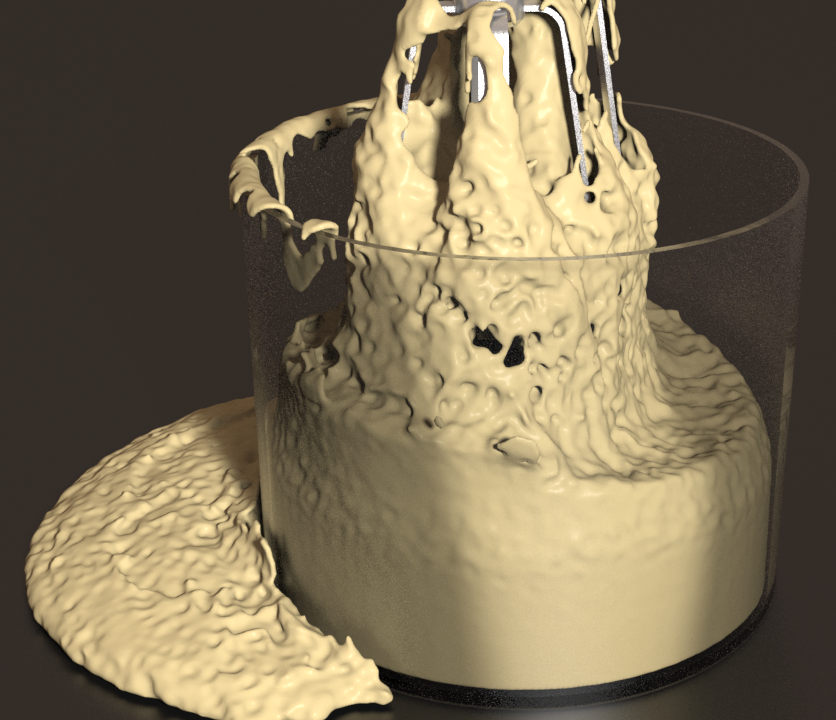} \hfill
	\includegraphics[height=3.22cm]{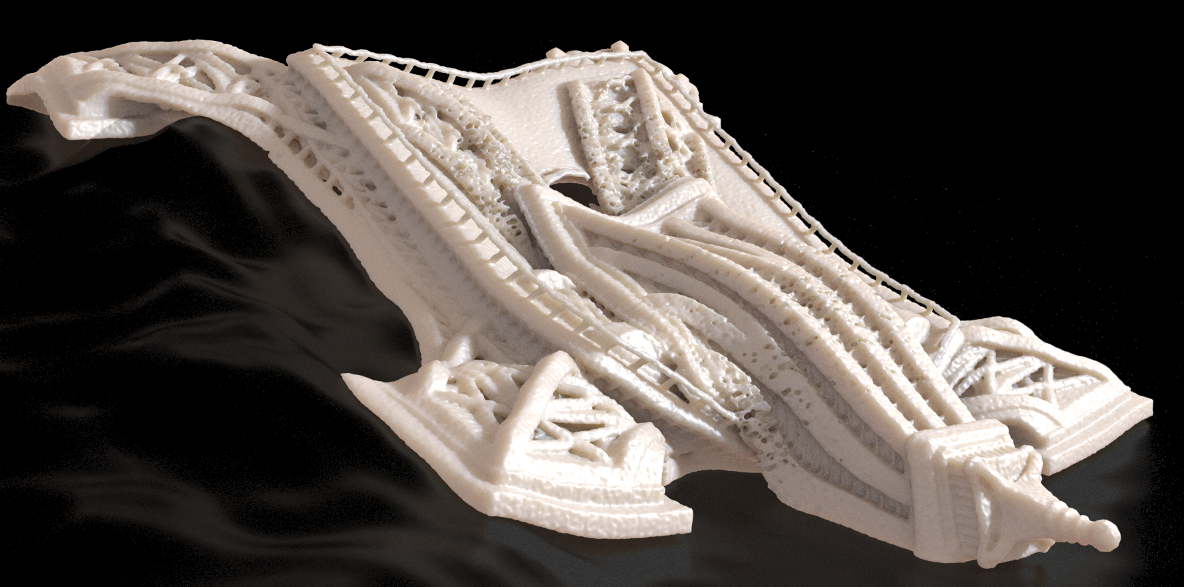}
	\caption{
	Different examples for the simulation of viscous behavior.
	Left: The simulation of highly viscous fluids enables realistic buckling effects~\cite{WKBB18}. Center: A viscous dough interacts with a fast moving solid~\cite{PICT15}. Right: A melting Eiffel tower is simulated (the model is courtesy of Pranav Panchal)~\cite{PT16}.
		}		
	\label{fig:visco_teaser}
\end{figure*}

Modeling and simulating viscosity is often vital for physics simulations as the phenomenon is responsible for various visually appealing effects, such as buckling and coiling but also general energy dissipation.
In recent years, various methods for the realistic simulation of low viscous fluids like water as well as highly viscous fluids like honey, mud, or dough were proposed.
Fig.~\ref{fig:visco_teaser} shows different examples of highly viscous materials.

In this section we first introduce the term for the viscous force in the Navier-Stokes equations.
Then we discuss important methods for the simulation of low viscous flow and highly viscous materials. 
We present explicit approaches which are typically used for low viscous fluids and we discuss implicit methods for highly viscous materials.

\subsection{Viscous Force}

In the Navier-Stokes equations for incompressible fluids, the viscosity term is determined by a material parameter $\DynamicViscosity$ and the Laplacian of the velocity field $\Laplace\Velocity$ (see Eq.~\eqref{eq:navier_stokes}). 
Before we discuss different approaches to compute this viscosity term, we first want to show how this term is derived. 

The stress tensor of a Newtonian fluid is defined as
\begin{equation}
\label{eq:stress_tensor}
 \CauchyStress = -\Pressure \Identity + 2 \DynamicViscosity \StrainRate, 
\end{equation}
where $\StrainRate$ is the strain rate tensor which is determined as
\begin{equation}
	\StrainRate = \frac12 (\Gradient \Velocity + (\Gradient \Velocity)^T ).
	\label{eq:strain_rate}
\end{equation}
When substituting the stress tensor in the conservation law of linear momentum (see Eq.~\eqref{eq:momentum_conservation}) and considering the incompressibility constraint $\Divergence\Velocity = 0$ (see Eq.~\eqref{eq:incompressibility_constraint}), we end up with the Navier-Stokes equations
\begin{equation}
\label{eq:navier_stokes_strain_rate}
    \Density \frac{\D \Velocity}{\D \Time} = -\Gradient \Pressure + \DynamicViscosity \underbrace{\Divergence \Gradient \Velocity}_{\Laplace \Velocity} + \DynamicViscosity \underbrace{\Divergence (\Gradient \Velocity)^T}_{\Gradient(\Divergence \Velocity) = \ZeroVector} + \BodyForce[\text{ext}], 
\end{equation}
where the right term of the strain rate tensor vanishes and the final viscosity force is determined as 
\begin{equation}
	\label{eq:viscous_force}
	\BodyForce[\text{visco}]= \DynamicViscosity \Laplace\Velocity. 
\end{equation}

Recent viscosity solvers either use a strain rate based formulation to compute viscous forces or they directly determine the Laplacian of the velocity field.
While the Laplacian formulation ensures that the right term of the strain rate tensor vanishes by definition, approaches based on the strain rate have to enforce a divergence-free velocity field, otherwise $\Divergence (\Gradient \Velocity)^T \neq \ZeroVector$ which leads to undesired bulk viscosity~\cite{lautrup2011physics,PICT15}.
In the following we introduce both approaches and discuss the advantages and disadvantages. 

\subsection{Explicit Viscosity}
\label{sub:explicit_viscosity}

In the Navier-Stokes equations for incompressible fluids the viscous force is defined by the Laplacian of the velocity field (see Eq.~\eqref{eq:viscous_force}).
The standard SPH discretization of this Laplacian is determined as
\begin{equation}
	\Laplace \Velocity[i]=\sum_{j}\frac{\Mass[j]}{\Density[j]}\Velocity[j] \KernelLaplace[i][j].
	\label{eq:naive_laplace}
\end{equation}
However, this formulation has two major disadvantages. 
First, it is sensitive to particle disorder~\cite{Mon05,Pri12a}.
Second, typically Gaussian-like kernel functions are used in SPH and their second derivatives changes the sign inside the support radius (see Fig.~\ref{fig:kernel}).

Different approaches were proposed to avoid this problem. 
First, instead of computing the second derivative directly, it can also be determined taking two first SPH derivatives~\cite{FMH+94,WBF+96,TDF+15}.
However, this method increases the computation time and memory consumption and introduces additional smoothing. 
Another approach, which was introduced by Brookshaw~\cite{Bro85}, is to determine one derivative using SPH and the second one using finite differences. 
This method is very popular and has been used for scalar quantities~\cite{Mon92,CM99,IOS+14} and for vector quantities~\cite{ER03,JSD04,Mon05,Pri12a,WKBB18}.
In the following we will discuss this approach in more detail.
 
In order to approximate the Laplacian of the velocity field we combine an SPH derivative with a finite difference derivative which yields the following equation~\cite{Mon05}:
\begin{equation}
		\Laplace \Velocity[i] = 2(d+2) \sum_j \frac{\Mass[j]}{\Density[j]}  \frac{\Velocity[ij] \cdot \Position[ij]}{\| \Position[ij] \|^2 + 0.01 \SmoothingLength^2} \KernelGradient[i][j],
		\label{eq:laplace_v}
\end{equation}
where $\Position[ij] = \Position[i] - \Position[j]$, $\Velocity[ij] = \Velocity[i] - \Velocity[j]$ and $d$ is the number of spatial dimensions. 
Note that a term $0.01 h^2$ is introduced in the denominator to avoid singularities.
The approximation of the Laplacian in Eq.~\eqref{eq:laplace_v} has some nice features. 
First, it is Galilean invariant.
Moreover, it vanishes for rigid body rotation. 
This is an important property since there is no friction if all particles rotate uniformly.
Finally, the introduced formulation conserves linear and angular momentum~\cite{Mon92}.

Instead of computing the Laplacian of the velocity field in order to simulate viscosity, in some works XSPH is used as artificial viscosity model (\eg \cite{SB12}).
XSPH determines the smoothed velocity $\hat{\Velocity}_i$ of a particle $i$ as
\begin{equation}
\label{eq:visco_XSPH}
	\hat{\Velocity}_i = \Velocity[i] + \alpha \sum_j \frac{\Mass[j]}{\Density[j]} (\Velocity[j] - \Velocity[i]) \Kernel[i][j], 
\end{equation}
where $0 \leq \alpha < 1$ is a user-defined parameter. 
The core idea of smoothing the velocity field in this way is to reduce the particle disorder by reducing the velocity difference between a particle and its neighborhood. 
The advantage of this formulation is that no kernel derivative is required. 
The disadvantage is that $\alpha$ is not physically meaningful.

\subsection{Implicit Viscosity}

Simulating the behavior of highly viscous materials implies that the viscosity coefficient is large.
However, in this case explicit viscosity solvers tend to get unstable.
Therefore, it is recommended to use an implicit method in order to simulate highly viscous fluids. 
In this subsection we introduce some of the most important implicit viscosity solvers~\cite{TDF+15,PICT15,PT16,WKBB18}.
We first present the concepts of these solvers in chronological order and then compare the different approaches.

\paragraph*{Takahashi et al.}\textbf{\cite{TDF+15}}\ \ 
As discussed in the previous subsection, one way to compute the second derivative of the velocity field is to take two first SPH derivatives. 
This approach is used by Takahashi et al.~to formulate an implicit integration scheme for the viscosity term in the Navier-Stokes equations. 
The implicit integration enables a stable simulation of highly viscous fluids. 
In each simulation step Takahashi et al.~first determine the strain rate $\StrainRate[i]$ for each particle $i$ using Eq.~\eqref{eq:strain_rate}.
Following the Navier-Stokes equations (see Eq.~\eqref{eq:navier_stokes_strain_rate}) the authors then compute the divergence of the strain rate as
\begin{equation}
	\Divergence \left (\Gradient \Velocity[i] + (\Gradient \Velocity[i])^T \right ) =  \sum_j \Mass[j] \left ( \frac{2 \StrainRate[i]}{\rho_i^2} + \frac{2  \StrainRate[j]}{\rho_j^2} \right ) \KernelGradient[i][j].
\label{eq:divergence_of_strainrate}
\end{equation}
Using this formulation the implicit integration scheme can be derived as
 \begin{equation}
	\Velocity[][\Time+\Dt] = \Velocity[][][\Predicted] + \frac{\Dt}{\Density} \DynamicViscosity \Divergence \left (\Gradient \Velocity[][t+\Dt] + \left (\Gradient \Velocity[][t+\Dt] \right )^T \right ),
\label{eq:Takahashi_integration}
\end{equation}
where $\Velocity[][][\Predicted]$ is the predicted velocity which is determined by integrating all non-pressure forces except viscosity.
Takahashi et al.~substitute Eq.~\eqref{eq:divergence_of_strainrate} in Eq.~\eqref{eq:Takahashi_integration} and solve the resulting formula to get the new velocities of the particles. 
The advantage of this implicit scheme is that highly viscous fluids can be simulated in a stable way while the viscosity is independent of the temporal and spatial resolution.
However, in this formulation all second-ring neighbors of a particle have to be considered in order to compute one first-order SPH derivative after the other.
This leads to many non-zero elements in the system matrix which decreases the performance significantly.

\paragraph*{Peer et al.}\textbf{\cite{PICT15,PT16}}\ \ 
Instead of using a classical implicit time integration scheme, Peer et al.~propose to decompose and modify the velocity gradient $\Gradient \Velocity$.
The goal of the authors is to modify only the shear rate in order to simulate a viscous behavior. 
Hence, they exploit the fact that the velocity gradient can be decomposed as
\begin{equation}
\label{eq:peer}
	\Gradient \Velocity = \SpinRate + \ExpansionRate + \ShearRate, 
\end{equation}
where $\SpinRate=\frac12 (\Gradient \Velocity - (\Gradient \Velocity)^T )$ is the spin rate tensor, $\ExpansionRate=\frac13 (\Divergence \Velocity) \Identity$ the expansion rate tensor and $\ShearRate = \StrainRate - \ExpansionRate$ the traceless shear rate tensor.
This decomposition enables to modify the traceless shear rate tensor without influencing the other components of the velocity gradient. 
Therefore, the authors define a target velocity gradient 
\begin{equation}
\label{eq:target_velocity_gradient}
	\Gradient \Velocity[][][\text{target}] = \SpinRate + \ExpansionRate + \xi \ShearRate
\end{equation}
which reduces the shear rate by a user-defined factor $0 \leq \xi \leq 1$.
This modified velocity gradient can be used to determine new particle velocities by a Taylor approximation of first order 
\begin{equation}
\label{eq:Peer_linearization}
	\Velocity[i][t+\Dt] = \frac{1}{\Density[i]} \sum_j \Mass[j] \left ( \Velocity[j][t + \Dt] + \frac{\Gradient \Velocity[i][][\text{target}] + \Gradient \Velocity[j][][\text{target}]}{2} \Position[ij] \right ) \Kernel[i][j].
\end{equation}
This yields a linear system $\Matrix \Velocity(\Time + \Dt) = \vec b$, where the matrix entries and the right hand side vector are defined as
\begin{align}
	\Matrix[ij] &= -\Mass[j] \Kernel[i][j], \\
	\Matrix[ii] &= \Density[i] - \Mass[i] \Kernel[i][i], \\
	\vec b_i &= \sum_j \Mass[j] \frac{\Gradient \Velocity[i][][\text{target}] + \Gradient \Velocity[j][][\text{target}]}{2} \Position[ij] \Kernel[i][j].
	\label{eq:Peer_linear_system}
\end{align}
This system can be decomposed to get three smaller linear systems for the x-, y- and z-component of the velocity.
Finally, the authors propose to solve the three systems using a conjugate gradient method.

Later, Peer and Teschner~\cite{PT16} extended this method by simulating vorticity diffusion in order to improve the rotational motion.
The diffusion process in a viscous fluid is described by
$\frac{D \AngularVelocity}{D t} = \KinematicViscosity \Laplace \AngularVelocity$.
The authors determine  
$\AngularVelocity = ( \AngularVelocity[x], \AngularVelocity[y], \AngularVelocity[z])^T$
from the spin rate tensor as
\begin{equation}
	\SpinRate = \frac12
	\begin{pmatrix} 
	0 & -\AngularVelocity[z] & \AngularVelocity[y]  \\
	\AngularVelocity[z] & 0 & -\AngularVelocity[x] \\
	-\AngularVelocity[y] & \AngularVelocity[x] & 0
	\end{pmatrix}.
\end{equation}
Analogous to Eq.~\eqref{eq:target_velocity_gradient} the vorticity is reduced by solving the system
\begin{equation}
	\Laplace \AngularVelocity[i][][\text{target}] = \xi \Laplace \AngularVelocity.
\end{equation}
The resulting vector $\AngularVelocity[i][][\text{target}]$ is used to determine a target spin rate tensor $\SpinRate[i][][\text{target}]$ which is substituted in Eq.~\eqref{eq:target_velocity_gradient} before reconstructing the velocity field using Eq.~\eqref{eq:Peer_linearization}.

The proposed methods are very efficient and enable a stable simulation of highly viscous materials. 
However, these methods have also some disadvantages.
The reconstruction of the velocity field using SPH is problematic as discussed in \cite{BGAO17} and introduces a significant damping. 
When simulating highly viscous fluids, this damping effect is not that crucial but this approach is not recommended for the simulation of low viscous flow. 
Another disadvantage of the methods is that the viscosity parameter $\xi$ is not physically meaningful and depends on the temporal and spatial resolution.

\paragraph*{Bender and Koschier}\textbf{\cite{BK17}}\ \ 
The authors of this work also reduce the strain rate by introducing a user-defined coefficient which is similar to the core idea of Peer et al. 
However, instead of modifying the velocity gradient and reconstructing the velocity field, Bender and Koschier define a velocity constraint function $\vec C_i(\Velocity) = \StrainRate[i] - \gamma \StrainRate[i]$ with the user-defined coefficient $0 \leq \gamma \leq 1$.
The constraint is defined as six-dimensional vector function where the vector contains the elements of the upper triangular part of the symmetric strain rate tensor.
Finally, the constraint is enforced by first solving the linear system 
\begin{equation}
	\left (\frac{1}{\Density[i]} \frac{\partial \StrainRate[i]}{\partial \Velocity[i]} \left ( \frac{\partial \StrainRate[i]}{\partial \Velocity[i]} \right )^T +  \sum_j \frac{1}{\Density[i]} \frac{\partial \StrainRate[i]}{\partial \Velocity[j]} \left (\frac{\partial \StrainRate[i]}{\partial \Velocity[j]} \right )^T\right ) \boldsymbol{\mu}_i =   
	 \StrainRate[i] - \gamma \StrainRate[i]
\end{equation}
for the Lagrange multiplier $\boldsymbol{\mu}$ by Jacobi iterations.
The final velocities are then determined as 
\begin{equation}
	\Velocity[i][\Time+\Dt] = \Velocity[i][][\Predicted] + \frac{1}{m_i} \left (\frac{m_i}{\rho_i} \left ( \frac{\partial \StrainRate[i]}{\partial \Velocity[i]}\right )^T \boldsymbol{\mu}_i  + \sum_j \frac{m_j}{\rho_j} \left (\frac{\partial \StrainRate[j]}{\partial \Velocity[i]} \right )^T \boldsymbol{\mu}_j  \right ).
\end{equation}
Details about the computation of $\partial \StrainRate / \partial \Velocity$ can be found in~\cite{BK17}.
 
The advantage of solving a constraint function instead of using the velocity field reconstruction approach of Peer et al.~ is that also low viscous fluids can be simulated. 
The disadvantages of the method are that solving six-dimensional constraints using Jacobi iterations is computationally expensive and the introduced viscosity coefficient depends on the temporal and spatial resolution.  

\paragraph*{Weiler et al.}\textbf{\cite{WKBB18}}\ \ 
All implicit viscosity methods introduced so far, use a formulation based on the strain rate tensor $\StrainRate$.
The strain rate is determined by Eq.~\eqref{eq:strain_rate}, where the velocity gradient $\Gradient \Velocity$ is computed using the following SPH discretization (see Section~\ref{sub:discretizatin_differential_operators}):
\begin{equation}
\label{eq:nabla_v}
	\Gradient \Velocity[i] = \frac{1}{\Density[i]} \sum_j \Mass[j] (\Velocity[j] - \Velocity[i]) \KernelGradient[i][j]^T. 
\end{equation}

\begin{figure}[tb]
	\centering
	\includegraphics[width=\linewidth]{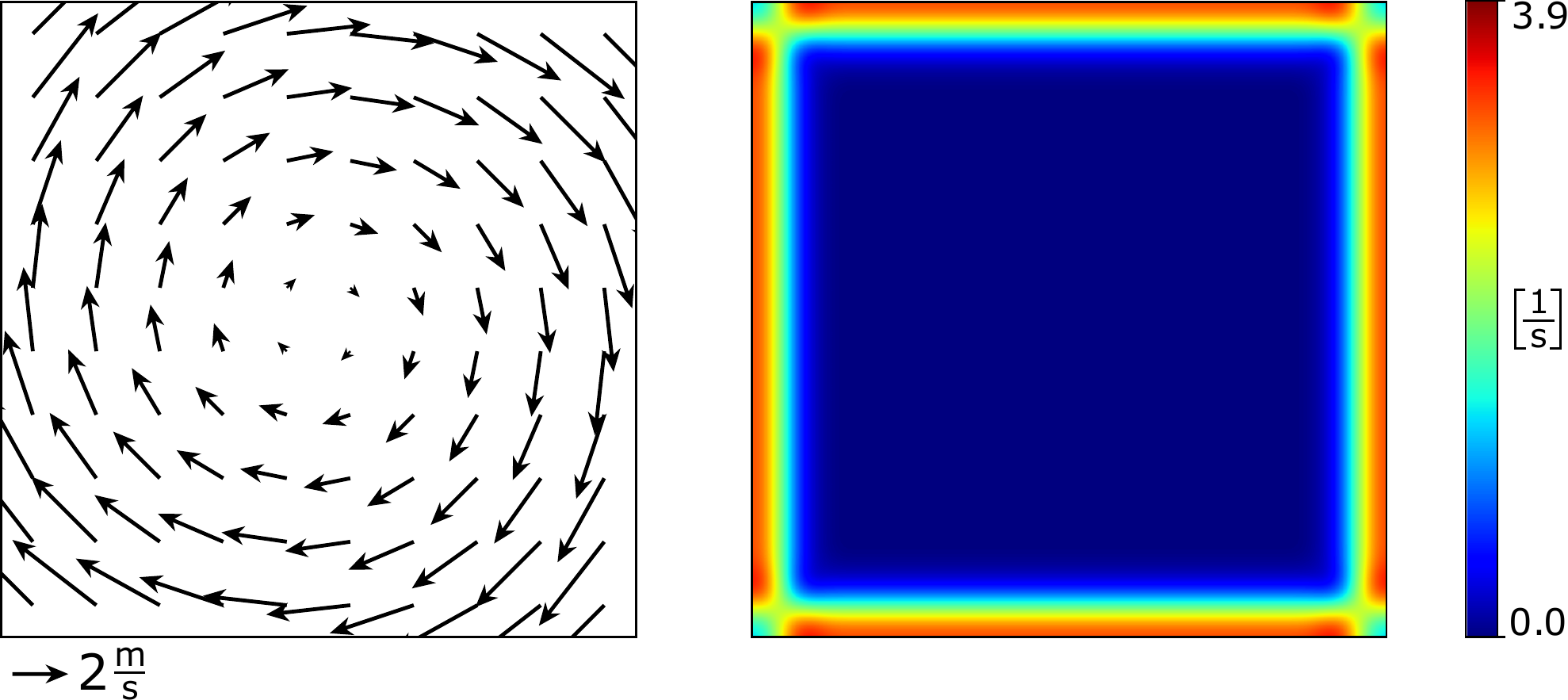} 
	\caption{When computing the strain rate tensor using Eq.~\eqref{eq:nabla_v}, errors occur at the free surface due to particle deficiency~\cite{WKBB18}. Left: Velocity field of a rotational motion. Right: The corresponding strain rate should be zero for all particles. However, the plot of the Frobenius norm of the tensors shows an error at the free surface.}
	\label{fig:strain_rate_error}
\end{figure}

Weiler et al.~found out that this SPH discretization is negatively affected by the particle deficiency problem at the free surface of a fluid.
For a rotational velocity field (see Fig.~\ref{fig:strain_rate_error}, left) the strain rate should be $\StrainRate = \vec 0$ since a rotation is a rigid body motion which does not deform the body.
However, when using the SPH discretization in Eq.~\eqref{eq:nabla_v}, the strain rate is not zero at the free surface (see Fig.~\ref{fig:strain_rate_error}, right).
In this experiment we can observe a significant error at the boundary. 
The main problem is that the viscosity solver tries to counteract this erroneous strain rate which leads to ghost forces.
These forces causes severe visual artifacts and a loss of angular momentum which is discussed later in more detail.

To solve this problem, Weiler et al.~developed an implicit viscosity solver which directly determines the Laplacian of the velocity field instead of using the strain rate. 
Their approach is based on the implicit integration scheme 
\begin{equation}
	\Velocity[][t+\Dt] = \Velocity[][][\Predicted] + \frac{\Dt}{\Density} \DynamicViscosity \Laplace \Velocity[][t+\Dt].
	\label{eq:visco_implicit_update}
\end{equation}
This is similar to the one of Takahashi et al.~\cite{TDF+15} but uses the Laplacian of the velocity field instead of the divergence of the strain rate.
To compute the Laplacian, the approximation in Eq.~\eqref{eq:laplace_v} is used. 
Since this approximation vanishes for rigid body rotations and conserves linear and angular momentum~\cite{Mon92}, the proposed approach solves the problems of the methods above. 

Eq.~\eqref{eq:visco_implicit_update} is a linear system which has to be solved to get the unknown new velocities $\Velocity[][t+\Dt]$.
Using the SPH discretization of the Laplacian in Eq.~\eqref{eq:laplace_v}, we can rewrite this linear system as
\begin{equation}
(\Identity - \Dt \Matrix) \Velocity[][t+\Dt]=\Velocity[][][\Predicted],
\label{eq:visco_linear_system}
\end{equation}
where the matrix $\Matrix$ contains a $3\times 3$ block $\Matrix[ij]$ for each pair of neighboring particles $i$ and $j$:
\begin{equation}
\Matrix[ij] = - 2(d+2)\frac{\DynamicViscosity \, \AverageQuantity{\Mass}_{ij}}{\Density[i]\Density[j]}\frac{\KernelGradient[i][j] \Position[ij]^T}{\left\|\Position[ij]\right\|^2+0.01 \SmoothingLength^2}, \quad\ \
\Matrix[ii]= -\sum_{j}\Matrix[ij].
\label{eq:matrixA}
\end{equation}
Note that the average mass $\AverageQuantity{\Mass}_{ij}=0.5 (\Mass[i]+\Mass[j])$ is used in order to obtain a symmetric system.
The resulting system can be solved efficiently by a matrix-free conjugate gradient method. 
The convergence can be improved by a block Jacobi preconditioner where the preconditioner matrix is block diagonal with the $3\times 3$ blocks $\Identity - \Dt \Matrix[ii]$.
Moreover, starting the conjugate gradient solver with an initial guess of $\Velocity[][][\Predicted] + \Delta \Velocity$ using the velocity difference of the last step $\Delta \Velocity = \Velocity[][t]-\Velocity[][t-\Dt][\Predicted]$ further improves the performance.

Weiler et al.~propose to extend this viscosity formulation also for the boundary in order to simulate materials that stick to solid objects.
In SPH often a particle-based surface representation of the boundary is used. 
For such a surface representation the diagonal matrix blocks in Eq.~\eqref{eq:matrixA} and the right hand side of the system in Eq.~\eqref{eq:visco_linear_system} have to be adapted as
\begin{align}
\Matrix[ii] &= -\sum_{j}\Matrix[ij]
 +2(d+2) \sum_k \frac{\DynamicViscosity_b \Mass[k]}{\Density[i]^2}\frac{\KernelGradient[i][k] \Position[ik]^T}{\left\|\Position[ik]\right\|^2+0.01 \SmoothingLength^2}, \\
	\vec{b}_i &= \Velocity[i][][\Predicted]  - 2(d+2) \Dt \sum_k \frac{\DynamicViscosity_b \Mass[k]}{\Density[i]^2}\frac{\Velocity[k] \cdot \Position[ik]}{\left\| \Position[ik]\right\|^2+0.01 \SmoothingLength^2} \KernelGradient[i][k],
\end{align}
where $\Mass[k]$ is the mass of the boundary particle $k$ (see Section~\ref{sec:boundary_handling}).
Note that a reaction force has to be applied to the boundary particles to get a consistent two-way coupling~\cite{AIA+12}. 
Using the proposed extension sticky and separating boundaries can be simulated.

\paragraph*{Comparison}
\begin{figure*}[t]
	\centering
	\subfloat[Takahashi et al.~\cite{TDF+15}\label{fig:grid_model_Takahashi}]{\includegraphics[width=0.245\linewidth]{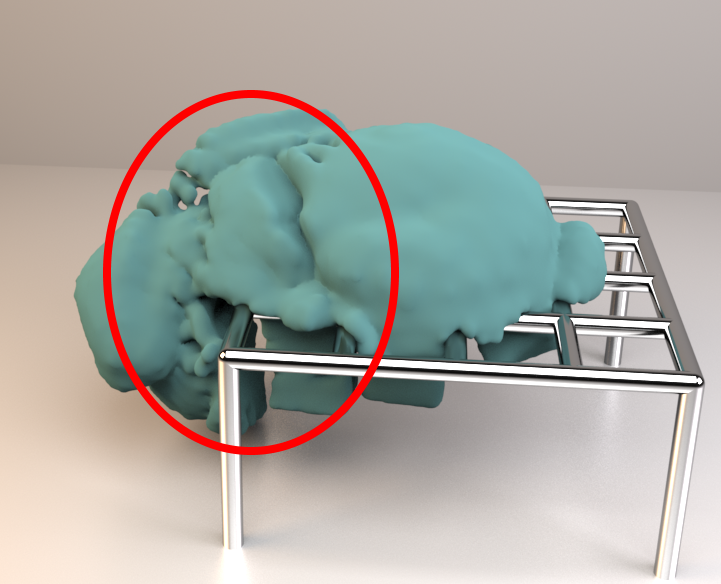}} \hfill
	\subfloat[Peer \& Teschner~\cite{PT16}\label{fig:grid_model_Peer}]{\includegraphics[width=0.245\linewidth]{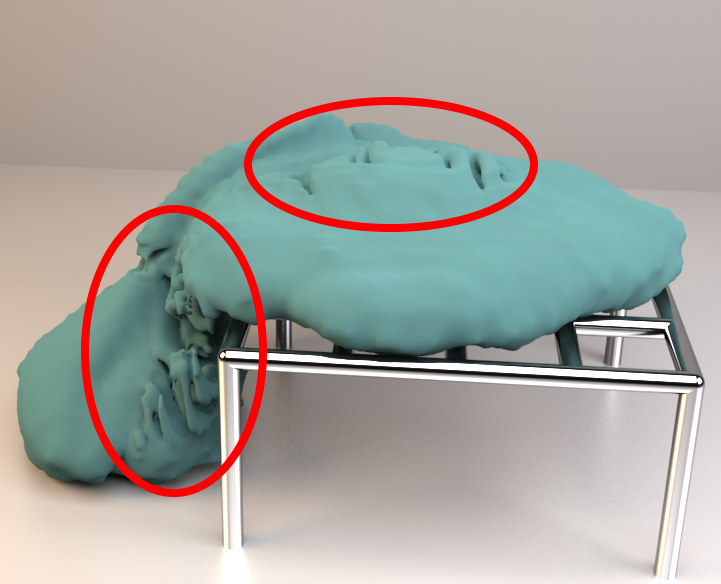}} \hfill
	\subfloat[Bender \& Koschier~\cite{BK17}\label{fig:grid_model_Bender}]{\includegraphics[width=0.245\linewidth]{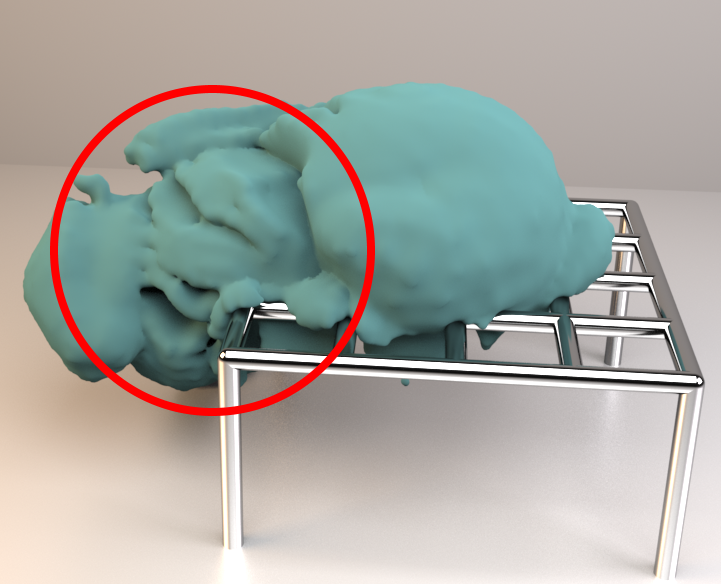}} \hfill	
	\subfloat[Weiler et al.~\cite{WKBB18}\label{fig:grid_model_IL}]{\includegraphics[width=0.245\linewidth]{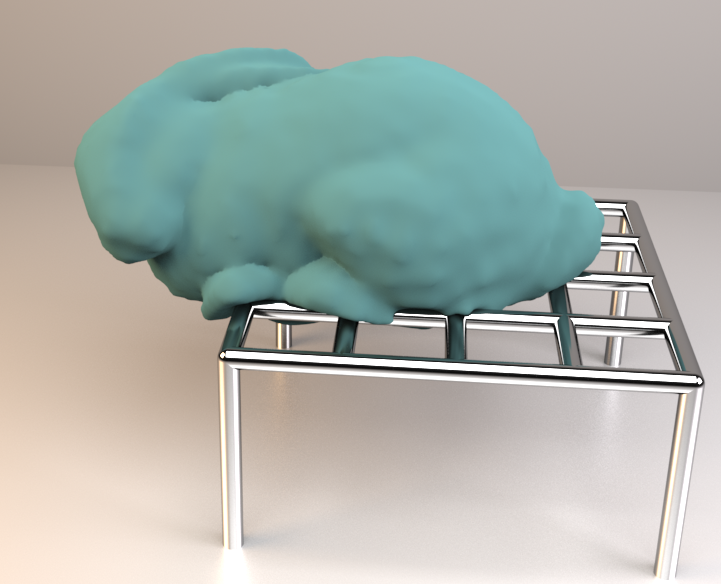}}	
	\caption{The SPH approximation of the strain rate tensor is erroneous at the free surface due to particle deficiency which leads to artifacts (a,b,c). The formulation of Weiler et al.~avoids this strain rate computation by using the Laplacian of the velocity field which solves this problem (d)~\cite{WKBB18}.
	}
	\label{fig:GridModel}
\end{figure*}
\begin{figure*}[t]
	\centering
	\subfloat[Takahashi et al.~\cite{TDF+15}\label{fig:Coiling_Takahashi}]{\includegraphics[width=0.245\linewidth]{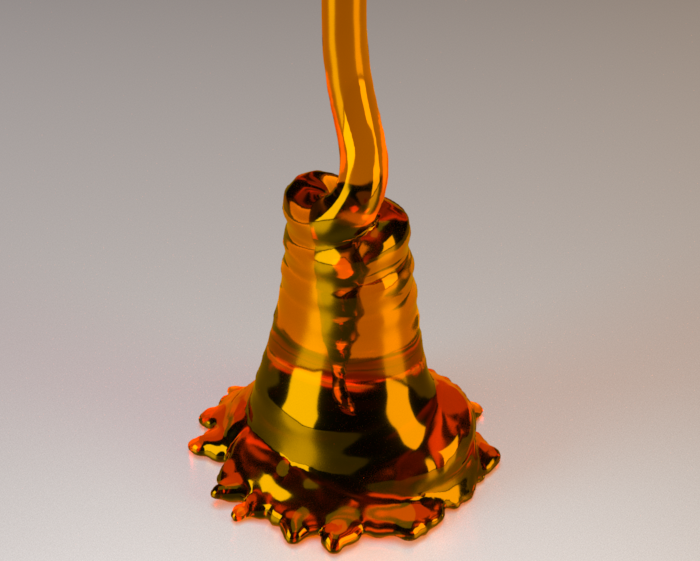}} \hfill
	\subfloat[Peer \& Teschner~\cite{PT16}\label{fig:Coiling_Peer}]{\includegraphics[width=0.245\linewidth]{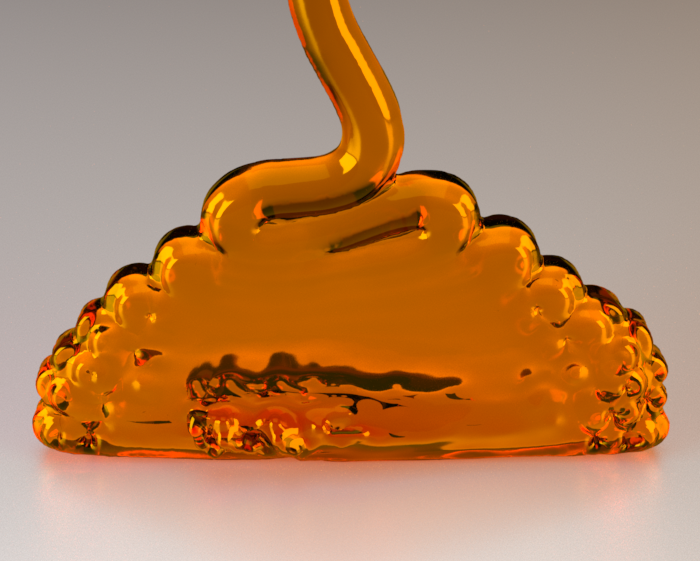}} \hfill
	\subfloat[Bender \& Koschier~\cite{BK17}\label{fig:Coiling_Bender}]{\includegraphics[width=0.245\linewidth]{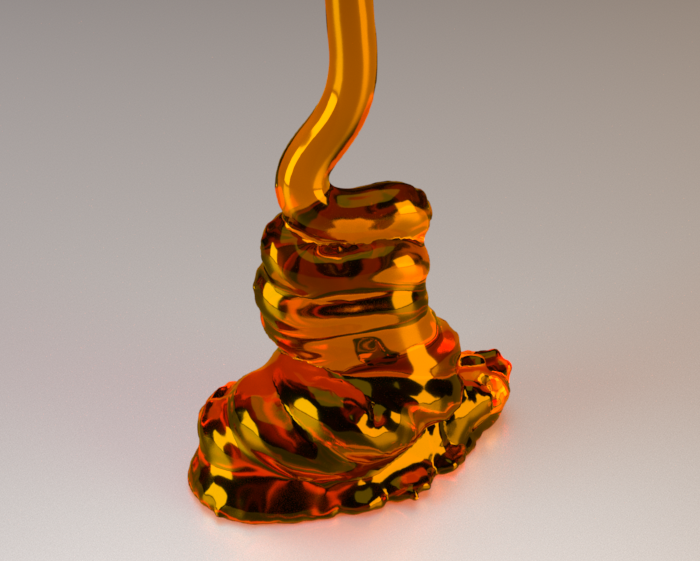}} \hfill	
	\subfloat[Weiler et al.~\cite{WKBB18}\label{fig:Coiling_IL}]{\includegraphics[width=0.245\linewidth]{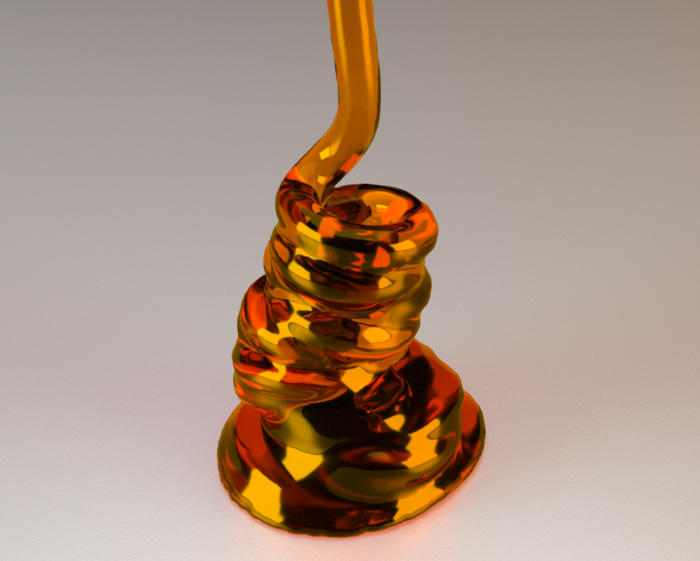}}	
	\vspace{-0.2cm}
	\caption{When simulating the rope coiling effect with the introduced implicit viscosity solvers, we can observe a loss of angular momentum caused by errors in the SPH strain rate approximation (a,b,c). Using a formulation based on the Laplacian of the velocity field instead, solves this problem and enables a uniform coiling (d)~\cite{WKBB18}.
	}
	\label{fig:Coiling}
\end{figure*}
In this part we want to compare all implicit viscosity solvers introduced above.
All approaches except the one of Weiler et al.~\cite{WKBB18} are based on a strain rate formulation. 
As discussed above and shown in Fig.~\ref{fig:strain_rate_error}, this leads to errors at the free surface due to particle deficiency.
In practice this can lead to artifacts at the surface (see Figs.~\ref{fig:grid_model_Takahashi}, \ref{fig:grid_model_Peer} and \ref{fig:grid_model_Bender}).
Moreover, the strain rate error at the free surface causes a significant loss of angular momentum which leads to a damped rotational motion (see Figs.~\ref{fig:Coiling_Takahashi}, \ref{fig:Coiling_Peer} and \ref{fig:Coiling_Bender}).

Weiler et al.~analyzed these problems and proposed a new method which computes the Laplacian of the velocity field instead of using the divergence of the strain rate.
Moreover, they use an SPH approximation of the Laplacian that vanishes for rigid body rotation and conserves linear and angular momentum.
In this way the problems at the free surface can be solved (see Figs.~\ref{fig:grid_model_IL} and \ref{fig:Coiling_IL}).

While the methods of Peer et al.~\cite{PICT15,PT16} and Bender and Koschier~\cite{BK17} use a viscosity parameter that depends on the temporal and spatial resolution, Takahashi et al.~\cite{TDF+15} and Weiler et al.~\cite{WKBB18} solve this problem by using a consistent implicit time integration.

Finally, when comparing the performance, the approach of Peer et al.~\cite{PICT15} is the fastest method.
This is due to the fact that they can decompose their linear system in three smaller ones while this cannot be done for the approach of Weiler et al.
Bender and Koschier use a Jacobi solver which converges slower and Takahashi et al.~have to consider the second-ring neighbors which results in large computational overhead.

\subsection{Conclusion}

For the simulation of low viscous fluids an explicit viscosity formulation should be used since explicit methods are computationally less expensive. 
We recommend to compute the viscous force in Eq.~\eqref{eq:viscous_force} by approximating the Laplacian of the velocity field using Eq.~\eqref{eq:laplace_v}.
An alternative, which is computationally less expensive but also less accurate, is to use XSPH as artificial viscosity. 

An implicit viscosity solver is recommended for the simulation of highly viscous fluids due to stability reasons.
As discussed above, the implicit strain rate based formulations suffer from an error in the SPH approximation that causes visual artifacts and leads to a loss of angular momentum.
The method of Weiler et al.~avoids the problem and therefore generates more realistic results.
Finally, we think that it would be an interesting open problem for future research to find a better SPH approximation of the strain rate without problems at the free surface.
Such an approximation would solve the problems of the strain rate based formulations.

Note that all viscosity methods that were discussed in this section are implemented in our open-source framework SPlisHSPlasH~\cite{SPlisHSPlasH}.


\section{Surface Tension}
\label{sec:surface_tension}

Surface tension is an important physical phenomenon which is a ubiquitous effect in daily life.
For example, surface tension forces keep liquid molecules together when pouring water into a glass. 
The surface forces are the result of intermolecular attractive forces at microscopic scales.
The molecules attract each other inside of a fluid while the molecules at the surface are pulled inwards.
Therefore, surface tension minimizes the surface area which causes droplets of water to form a sphere when external forces are excluded.
We typically speak of cohesion if molecules of the same type attract each other while adhesion describes the attractive forces between molecules of different types.
Cohesion and adhesion are important effects when simulating surface tension.

\begin{figure*}[t]
	\centering
	\includegraphics[height=3.92cm]{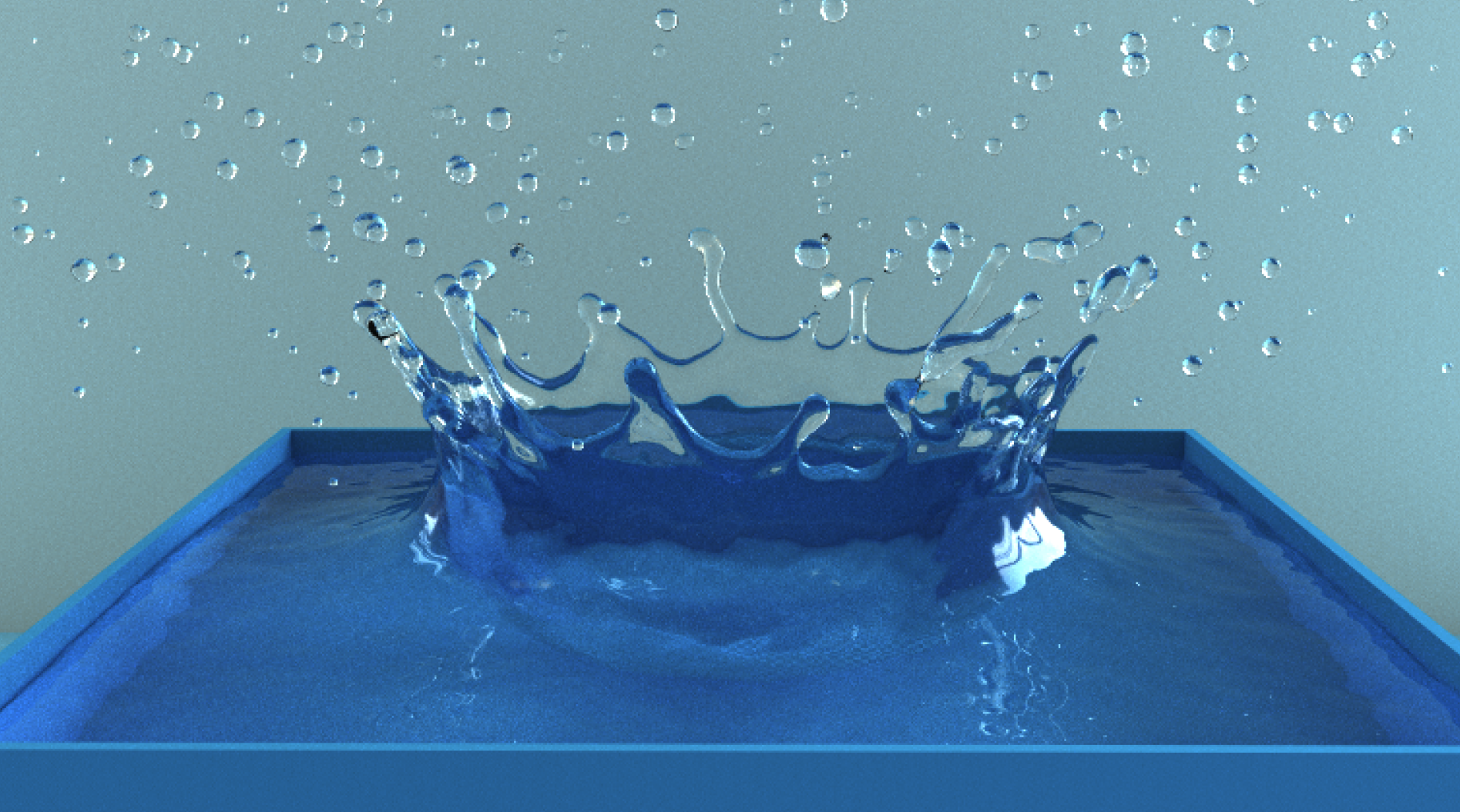} \hfill
	\includegraphics[height=3.92cm]{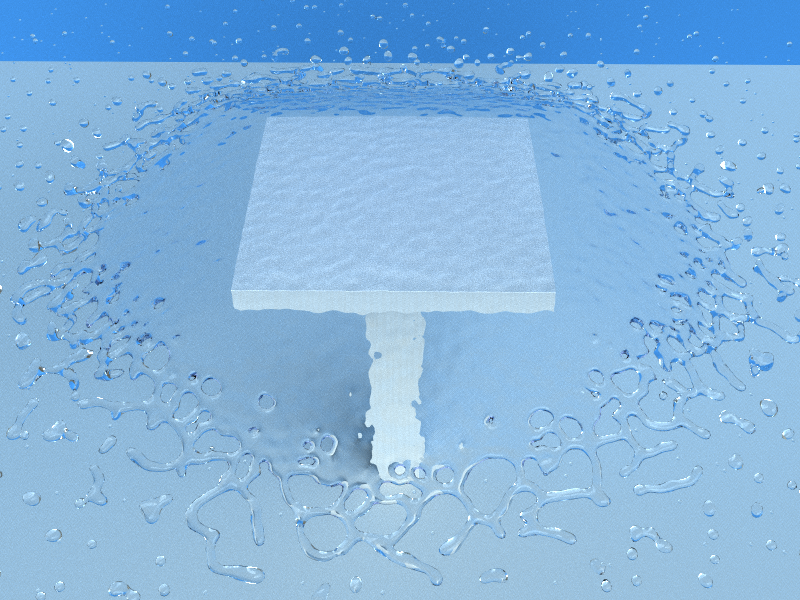} \hfill
	\includegraphics[height=3.92cm]{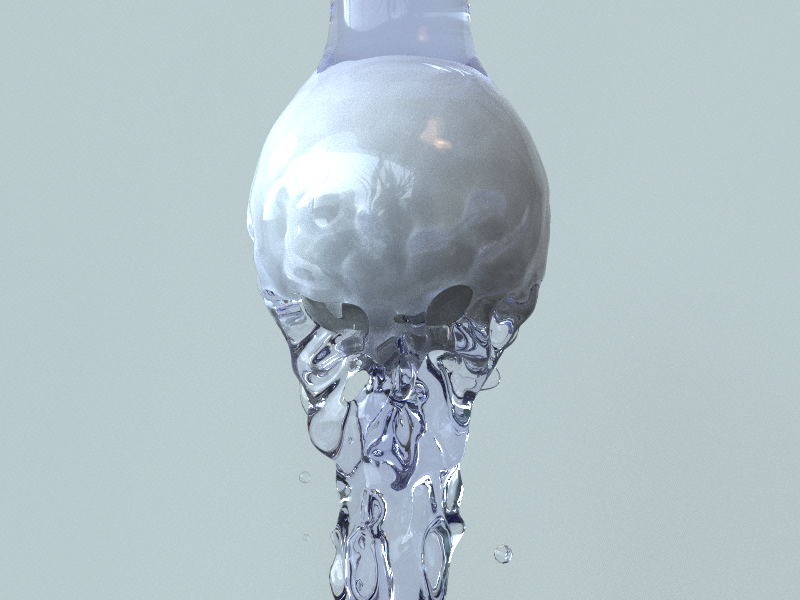}
	\caption{Different surface tension effects are realized by approximating intermolecular attractive forces in the fluid~\cite{AAT13}.
		}		
	\label{fig:surface_tension_teaser}
\end{figure*}

In recent years, various methods were proposed to simulate surface tension effects in SPH-based fluid simulations (\eg \cite{BT07,AAT13,HWZ+14}).
Fig.~\ref{fig:surface_tension_teaser} shows SPH simulation examples of surface tension.
Typically we can differentiate between surface tension approaches that are inspired by a microscopic point of view and approaches that compute the forces on a macroscopic level.
In the following, we will present one microscopic method and one macroscopic method in order to introduce the core idea of both approaches.

\subsection{Microscopic Approach}

Surface tension is the result of attracting forces between molecules. 
Methods that are based on a microscopic point of view aim to simulate the intermolecular cohesive forces.
However, since the smallest element in an SPH simulation is a particle, these forces are determined at the particle-level.

Becker and Teschner~\cite{BT07} propose to compute the particle acceleration due to cohesion by
\begin{equation}
	\frac{\D\Velocity[i]}{\D\Time} = - \frac{\alpha}{\Mass[i]} \sum_j \Mass[j] \left (\Position[i] - \Position[j] \right ) \Kernel[i][j], 
\label{eq:surface_tension_Becker07}
\end{equation}
where $\alpha$ is a coefficient to control the surface tension of the fluid. 
This equation interpolates the position differences in the neighborhood of a particle $i$ and computes an acceleration to attract the neighboring particles.
Note that adhesion can also be simulated by Eq.~\eqref{eq:surface_tension_Becker07} when using a sum over all neighboring boundary particles.

\subsection{Macroscopic Approach}

\paragraph*{Cohesion}
Akinci et al.~\cite{AAT13} present a method based on a macroscopic point of view.
Instead of only considering cohesive forces, also forces to minimize the surface area are determined.
Their method computes the cohesive force of a particle as
\begin{equation}
	\Force[i \leftarrow j][][\text{cohesion}] = -\alpha \Mass[i] \Mass[j] \frac{\Position[i] - \Position[j]}{\| \Position[i] - \Position[j] \|} \Kernel[][][\|\Position[i] - \Position[j]\|][\text{cohesion}],
\end{equation}
where $i$ and $j$ are neighboring particles and $\Kernel[][][][\text{cohesion}]$ is a special cohesion kernel which is defined by
\begin{equation}
	\Kernel[][][r][\text{cohesion}] = \frac{32}{\pi h^9} \begin{cases}
		(h-r)^3 r^3 & 2r > h \wedge r \leq h \\
		2(h-r)^3 r^3 - \frac{h^6}{64} & r > 0 \wedge 2r \leq h \\
		0 & \text{otherwise.}
	\end{cases}
\end{equation}
In contrast to the model of Becker and Teschner~\cite{BT07} the cohesive forces can become positive and negative. 
In this way repulsion forces for close particles are generated which prevents undesired particle clustering at the free surface.

Additionally, Akinci et al.~compute a force in order to minimize the surface area. 
This additional force counteracts the surface curvature which requires the computation of the surface normals. 
The normals can be determined by a so-called color field. 
The idea is to set the color of a particle to 1 while it is 0 everywhere else.  
Then the gradient of the smoothed color field
\begin{equation}
	\vec n_i = \SmoothingLength \sum_j \frac{\Mass[j]}{\Density[j]} \KernelGradient[i][j]
\end{equation}
yields a surface normal pointing into the fluid.
Note that the factor $\SmoothingLength$ is used to make the normal scale independent.
The magnitude of the resulting vector is close to zero in the interior of the fluid and proportional to the curvature at the free surface. 
Hence, a symmetric force that counteracts the curvature can be defined as
\begin{equation}
	\Force[i \leftarrow j][][\text{curvature}] = -\alpha \Mass[i] (\vec n_i - \vec n_j).
\end{equation}
This force is used to minimize the surface area.

Finally, both forces are combined as
\begin{equation}
	\Force[i \leftarrow j][][\text{surface tension}] = K_{ij} (\Force[i][][\text{cohesion}] + \Force[i][][\text{curvature}]), 
\end{equation}
where $K_{ij} = \frac{2 \Density[0]}{\Density[i] + \Density[j]}$ is a symmetric factor that amplifies the surface tension forces at the free surface.
At the surface $\Density[i]$ and $\Density[j]$ are underestimated due to particle deficiency and $K_{ij} > 1$ while for a particle with a full neighborhood $K_{ij} \approx 1$.

\paragraph*{Adhesion}
To simulate the attractive forces between fluid particles and the boundary, an adhesion force is introduced as
\begin{equation}
	\Force[i \leftarrow k][][\text{adhesion}] = -\beta \Mass[i] \Mass[k] \Kernel[i][k][][\text{adhesion}] \frac{\Position[i]-\Position[k]}{\|\Position[i]-\Position[k]\|},
\end{equation}
where $\beta$ is the adhesion coefficient and $k$ denotes a neighboring boundary particle.
The computation of the mass $\Mass[k]$ of a boundary particle is described in detail in Section~\ref{sec:boundary_handling}. 
Akinci et al.~propose to use another specialized kernel function for the computation of the adhesion forces which is defined as
\begin{equation}
	\Kernel[][][r][\text{adhesion}] = \frac{0.007}{h^{3.25}} \begin{cases}
		\sqrt[4]{-\frac{4 r^2}{h} + 6r -2h} & 2r > h \wedge r \leq h \\
		0 & \text{otherwise.}
	\end{cases}
\end{equation}
Note that only fluid particles with a distance between $h/2$ and $h$ are attracted by the boundary. 


\section{Vorticity}
\label{sec:vorticity}

One of the most visually appealing phenomena in dynamic fluids is the evolution of chaotic structures due to turbulences.
Turbulent motions are largely caused by the interaction of many unsteady vortices on various scales.
A vortex is, moreover, defined as a local spinning motion in the fluid -- mathematically spoken the vorticity is a vector field 
\begin{equation}
	\AngularVelocity = \Curl\Velocity\mathpoint \label{eq:vorticity}
\end{equation}

In SPH fluid simulations turbulent details quickly get lost due to numerical diffusion \cite{dGWH+15} or due to coarse sampling of the velocity field \cite{IOS+14,CIPT14} which negatively influences the visual liveliness of the flow.
In order to facilitate the formation of vortices in the simulation and to counteract numerical diffusion, a range of approaches has been proposed in the past.
Most of these approaches originate from research concerning Eulerian, grid-based discretizations and can roughly be categorized into vorticity confinement techniques, Lagrangian vortex methods, fluid up-sampling, and more recently micropolar models.
In this section we will discuss a simple vorticity confinement approach following Macklin and M\"uller~\cite{MM13} and an SPH discretization of a micropolar model that facilitates the formation of vortices as proposed by Bender et al.~\cite{BKKW18}.

\subsection{Vorticity Confinement}

As already discussed, SPH discretizations tend to overly dissipate energy in turbulent flow.
Therefore, Macklin and M\"uller~\cite{MM13} employ a method based on vorticity confinement in order to counteract the dissipation by amplifying existing vortices.
The technique consists of three steps:
\begin{enumerate}
	\item The vorticity $\AngularVelocity[i]$ for each particle $i$ is computed using a discrete curl operator, \eg
	\begin{equation}
		\AngularVelocity[i] = \Curl\Velocity[i] = -\sum_j \frac{\Mass[j]}{\Density[j]} \Velocity[ij] \times \Gradient_i\Kernel[i][j] \mathpoint
	\end{equation}
	\item A corrective force is computed and applied that amplifies the already existing vortical motion, \ie
	\begin{align}
		\Force[i][][\text{vorticity}] &= \epsilon^{\text{vorticity}} (\frac{\vec\eta}{\Vert \vec\eta \Vert} \times \AngularVelocity[i]) \\
		\vec\eta &= \sum_j \frac{\Mass[j]}{\Density[j]} \Vert \AngularVelocity[j] \Vert \Gradient_i\Kernel[i][j] \mathcomma
	\end{align}
	where $\epsilon^{\text{vorticity}}$ denotes a small constant used to steer the amount of amplification.
	\item The velocity field is smoothed using XSPH (see Eq.~\eqref{eq:visco_XSPH}) in order to enforce a coherent particle motion.
\end{enumerate}
Algorithmically, this correction is applied just before the advection of the SPH particle positions.

While this approach is very simple and effectively amplifies existing vortices it has some drawbacks.
It is hard to choose the control parameter $\epsilon^{\text{vorticity}}$ such that overamplification is avoided.
Moreover, the method can, in the best case, only conserve existing vortices but does not facilitate the formation of new ones.

\subsection{Micropolar Model}

The most prominent mathematical model describing the dynamics of Newtonian fluids is the Navier-Stokes model.
As described in Section~\ref{sec:foundations} the model can be derived from the conservation law of linear momentum (see Eq.~\eqref{eq:momentum_conservation}) and by presuming that the mechanical stress is composed of isotropic pressure and a diffusing
viscous term.
An important assumption of the model is that the infinitesimally small particles which compose a fluid continuum are not subject to rotational motion.
This also implies that the law of angular momentum conservation is identically fulfilled if and only if the stress tensor is symmetric.

In this section we introduce the concept of micropolar fluids and present a material model that generalizes the Navier-Stokes equations for the simulation of incompressible, inviscid turbulent flow as proposed by Bender at al.~\cite{BKKW18}.
Following the definition of {\L}ukaszewicz~\cite{L99}, a micropolar fluid follows constitutive laws modeled using a generally non-symmetric stress tensor.
Moreover, the definition includes that the fluid consists of rigid, spherical (and therefore rotationally invariant) particles.
Based on the non-symmetric stress measures, the micropolar model additionally models rotating motions of the infinitesimal spherical particles using an angular velocity field.
Due to the additional rotational degrees of freedom, the generation of vortices is facilitated and a wider range of potential dynamic effects are captured by the model.
Please note, that for this section we will neglect any dissipation terms, such as viscosity, as the main goal is to generate undamped, highly turbulent flows.
For the complete model, we would like to kindly refer the reader to the original paper~\cite{BKKW18}.

\subsubsection*{Balance Law for Angular Momentum Conservation}
Similar to the conservation law of linear momentum (see Eq.~\eqref{eq:momentum_conservation}) a balance law for angular momentum can be derived, \ie
\begin{equation}
	\Density\MicroInertia \frac{D\AngularVelocity}{D\Time} = \CauchyStress[\times] + \BodyTorque \mathcomma \label{eq:angular_momentum_conservation}
\end{equation}
with $\left[\CauchyStress_{\times}\right]_i = \sum_j \sum_k \epsilon_{ijk} T_{jk}$, where $\epsilon_{ijk}$ and $\BodyTorque$ denote the Levi-Civita tensor and external body torque.
Further, $\MicroInertia$ represents a scalar, isotropic microinertia coefficient.
A physical interpretation for this quantity is that each infinitesimal fluid particle has a certain inertial resistance against rotational accelerations.
Bender et al. suggest to choose $\MicroInertia = 2 \si{\square\meter\per\second}$ based on experimentation.
We would further like to stress the fact that $\MicroInertia$ is not at all related to the spatial extents of an SPH particle as it is defined in the continuous setting.

\subsubsection*{Constitutive Model}

It is essential to understand that the classical model actually also respects angular momentum conservation (see Eq.~\eqref{eq:angular_momentum_conservation}).
It is in this context, however, rarely explicitly mentioned as the balance law is usually identically fulfilled based on two assumptions.
Firstly, the classical approach does not model external torques, \ie $\BodyTorque \equiv \vec 0$.
Secondly, stress tensor $\CauchyStress$ is usually chosen as a symmetric tensor and such that $\CauchyStress_\times \equiv 0$ and, hence, $\frac{D\AngularVelocity}{D\Time} \equiv 0$.
For this reason, the balance law of angular momentum is not particularly useful for symmetric stress measures.

In order to account for the microstructured particles and to utilize the balance law of angular momentum, Bender et al.\ propose to use the following constitutive relation:
\begin{equation}
	\CauchyStress = -\Pressure \Identity - \DynamicViscosity_t \Gradient\Velocity + \DynamicViscosity_t \AngularVelocity[][][\times],\label{eq:micropolar_conservation_law}
\end{equation}
with $\left[\AngularVelocity[][][\times]\right] = \sum_i \epsilon_{jik}\omega_i$, where $\DynamicViscosity_t$ denotes the "transfer coefficient".
We will later discuss a physical interpretation of the term in the final PDE that is controlled using $\DynamicViscosity_t$.
In order to ensure consistency with the second law of thermodynamics $\DynamicViscosity_t \geq 0$ must be satisfied.
Please note that this constitutive model allows for a non-symmetric stress resulting in the fact that we have to explicitly account for the angular balance law in our simulation.

\subsubsection*{Equations of Motion}

As the conservation laws and constitutive equation are now established, we can finally derive the augmented equations of motion that build the basis for the numerical simulation.
By plugging the constitutive relation~\eqref{eq:micropolar_conservation_law} into the conservation laws \eqref{eq:momentum_conservation} and \eqref{eq:angular_momentum_conservation} and by applying the incompressibility condition \eqref{eq:incompressibility_constraint}, we arrive at the following representation:
\begin{align}
	\frac{D\Velocity}{D\Time} &= - \frac 1 \Density \Gradient \Pressure + \KinematicViscosity_t \Curl\AngularVelocity + \frac{\BodyForce[\text{ext}]}{\Density} \label{eq:linear_micropolar_momentum}\\
	\MicroInertia \frac{D\AngularVelocity}{D\Time} &= \KinematicViscosity_t (\Curl\Velocity - 2\AngularVelocity) + \frac{\BodyTorque}{\Density} \label{eq:angular_micropolar_momentum} \mathcomma
\end{align}
where $\KinematicViscosity_t$ denotes the kinematic transfer coefficient.
Looking at Eq.~\eqref{eq:linear_micropolar_momentum}, we quickly notice that it is identical to the inviscid Navier-Stokes equation (Euler equation) but augmented by the term $\KinematicViscosity_t \Curl\AngularVelocity$.
A complementary term also governed by the transfer coefficient resides in Eq.~\eqref{eq:angular_micropolar_momentum}, \ie $\KinematicViscosity_t (\Curl\Velocity - 2\AngularVelocity)$.
The terms effectively convert angular accelerations into linear accelerations and vice versa.
Physically, they can be interpreted as dissipation-free friction or as a dissipation-free viscosity coupling linear and rotational motion.

In order to realize the model in the implementation, it is required to discretize Eqs.~\eqref{eq:linear_micropolar_momentum} and \eqref{eq:angular_micropolar_momentum}.
This means that additional to a discrete representation of $\Position$ and $\Velocity$ it is necessary to discretize the vorticity $\AngularVelocity$.
Please note, that due to the assumption in the micropolar model, that the microstructure of the material particles is spherical we do not have to discretize and track the rotational field (only the angular velocity field) which makes the implementation less complicated and delivers better performance.
The transfer forces and torques can then, following the splitting approach, simply be applied in line with the non-pressure forces and integrated explicitly as they are considerably less stiff than the pressure forces.
It is further advised to "filter" the resulting velocity field using XSPH (see Eq.~\eqref{eq:visco_XSPH}) in order to ensure coherent particle motion.
Algorithm~\ref{alg:micropolar} shows an exemplary pseudocode of the resulting method.

\begin{algorithm}
	\caption{Simulation loop for SPH simulation turbulent micropolar fluids.}
	\label{alg:micropolar}
	\begin{algorithmic}
	\algnotext{EndFor}
	\algnotext{EndWhile}
	\algnotext{EndProcedure}
	\ForAll{$particle~i$}
		\State Find neighbors
	\EndFor
	\ForAll{$particle~i$}
		\State Compute density $\Density_i$
	\EndFor
	\ForAll{$particle~i$}
		\State Compute non-pressure forces $\Force[i][][\text{non-pressure}]$
		\State Compute transfer forces $\Force[i][][\text{transfer}] = \Mass[i] \KinematicViscosity_t \Curl\AngularVelocity[i]$
		\State Compute transfer torque $\Torque[i][][\text{transfer}] = \Mass[i] \KinematicViscosity_t (\Curl\Velocity[i] - 2 \AngularVelocity[i])$
	\EndFor
	\State Compute time step size $\Delta\Time$ according to CFL
	
	\ForAll{$particle~i$}
		\State $\Velocity[i][][\Predicted] = \Velocity[i] + \frac{\Delta\Time}{\Mass[i]} (\Force[i][][\text{non-pressure}] + \Force[i][][\text{transfer}] + \Force[i][][\text{ext}])$
	\EndFor
	\ForAll{$particle~i$}
		\State Enforce incompressibility using pressure solver
		\State Update $\Velocity[i][][\Predicted]$
	\EndFor
	\ForAll{$particle~i$}
		\State $\Velocity[i](\Time + \Delta\Time) = \Velocity[i][][\Predicted]$
		\State $\Position[i](\Time + \Delta\Time) = \Position[i] + \Delta\Time \Velocity[i](\Time + \Delta\Time)$
		\State $\AngularVelocity[i](\Time + \Delta\Time) = \AngularVelocity[i][\Time] + \frac{\Delta\Time}{\Mass[i] \MicroInertia[i]} (\Torque[i][][\text{transfer}] + \Torque[i][][\text{ext}])$
	\EndFor
	\end{algorithmic}
\end{algorithm}

\begin{figure}[t]
	\centering
	\includegraphics[width=\linewidth]{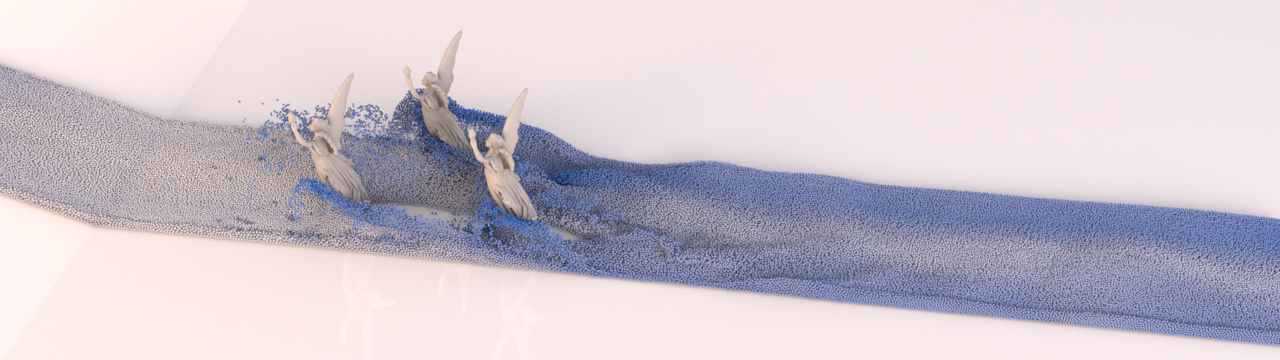}
	\includegraphics[width=\linewidth]{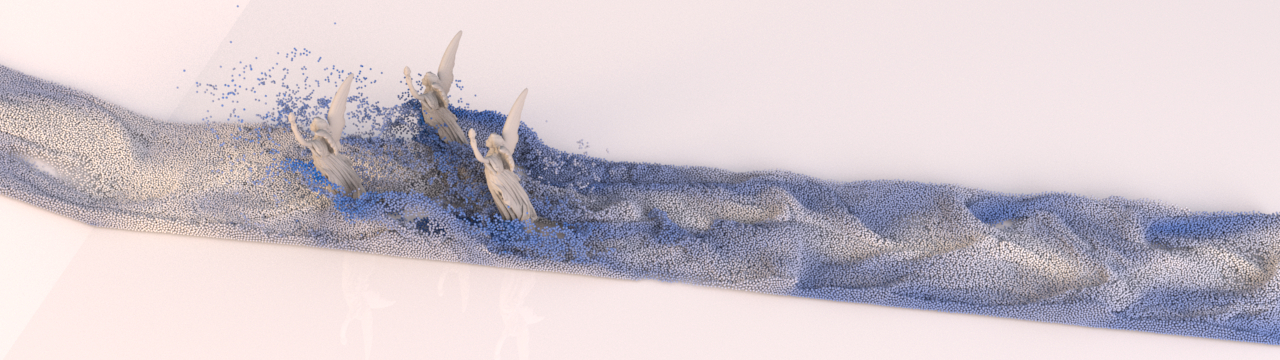}
	\includegraphics[width=\linewidth]{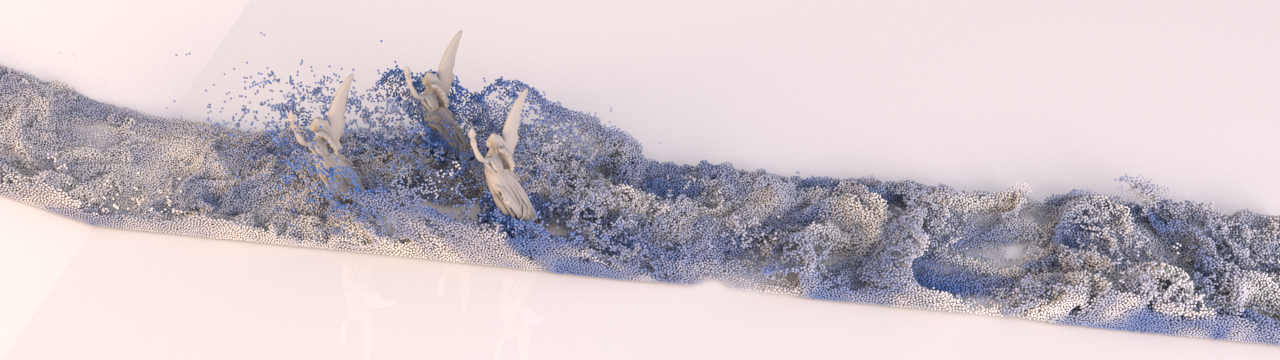}
	\caption{Simulation of $4.7$M turbulent fluid particles with three obstacles and increasing transfer coefficient $\KinematicViscosity_t$. Top-down: $\KinematicViscosity_t = 0.2\si{\square\meter\per\second}$, $\KinematicViscosity_t = 0.3\si{\square\meter\per\second}$, $\KinematicViscosity_t = 0.4\si{\square\meter\per\second}$.}
	\label{fig:increasing_vorticity}
\end{figure}
\begin{figure}[t]
	\centering
	\includegraphics[width=\linewidth]{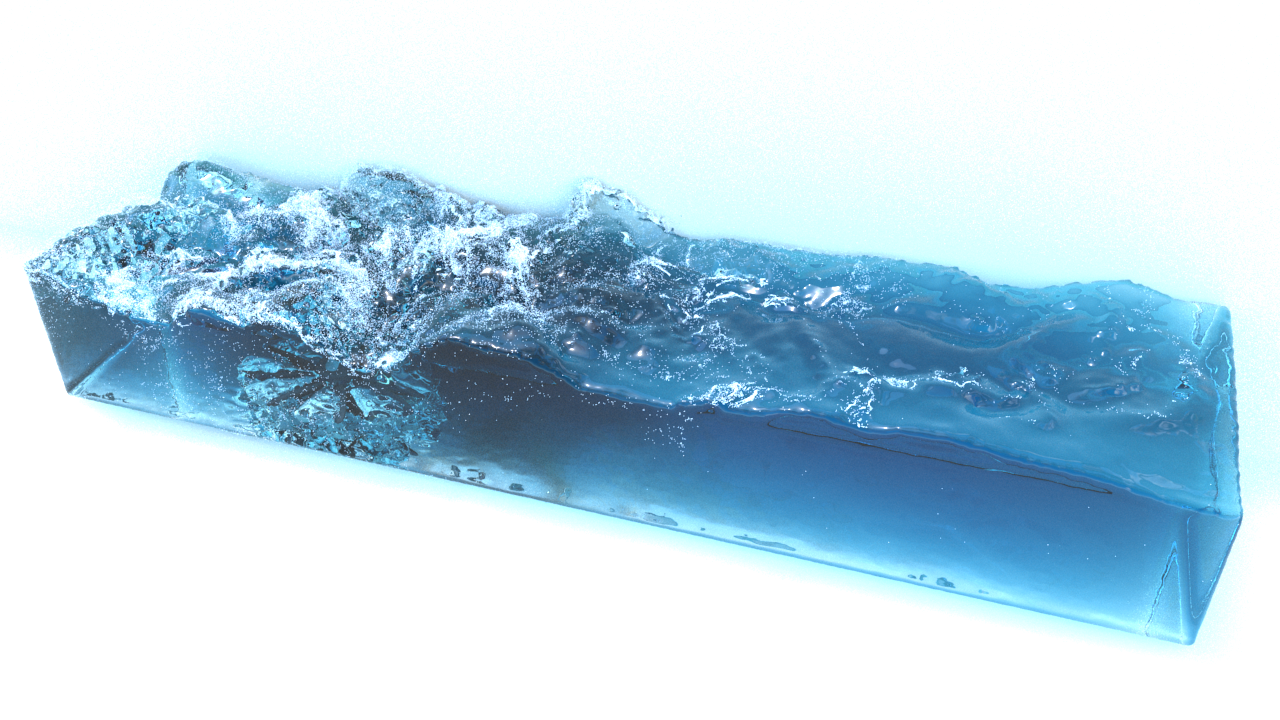}
	\caption{$1$M fluid particles interact with a fast rotating propeller resulting in highly turbulent flow.}
	\label{fig:micropolar_propeller}
\end{figure}

Bender et al.~\cite{BKKW18} demonstrated the effect of the transfer coefficient using an intuitive example (see Fig.~\ref{fig:increasing_vorticity}).
In the experiment a fluid flowing in a narrow channel was simulated.
Moreover, three obstacles were placed in the channel to provoke turbulences while the transfer coefficient was continuously increased.
In the top image we can see that that the flow is only moderately turbulent for a transfer coefficient $\KinematicViscosity_t = 0.2 \si{\square\meter\per\second}$.
For larger values the vorticity significantly increases (middle) and even tends to get unrealistic for values greater than $0.4 \si{\square\meter\per\second}$ (bottom).
Furthermore, they showcase the visual realism that can be achieved in turbulent scenarios (see Fig.~\ref{fig:micropolar_propeller}).

\subsubsection*{Discussion}

Two methods to improve the behavior of the simulation in the presence of turbulences have been explained.
In this paragraph, we would like to discuss the similarities and differences between vorticity confinement and the micropolar model.

\begin{figure*}
	\centering
	\includegraphics[width=\linewidth]{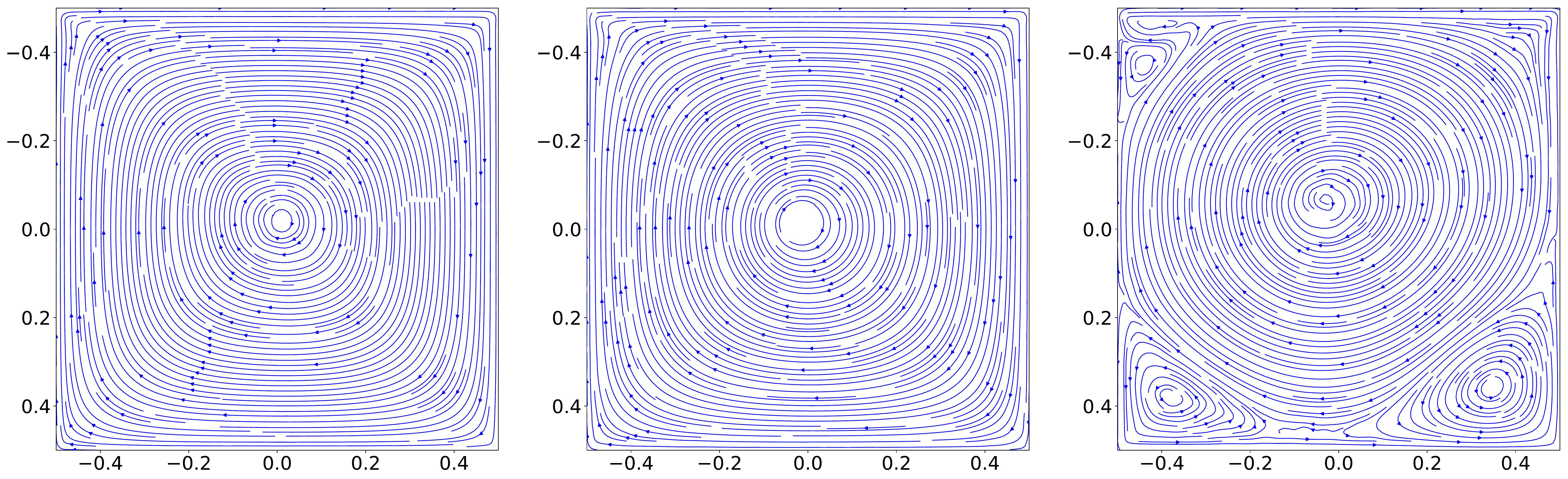}
	\caption{Velocity fields of the lid-driven cavity benchmark. Standard SPH (left) and vorticity confinement (middle) are only able to produce one large central vortex.
	In contrast, the micropolar approach yields (right) the expected result, \ie one central vortex and three smaller vortices in the corners which are rotating in the opposite direction.}
	\label{fig:LDC}
\end{figure*}

Both methods build on the concept of obtaining/maintaining a vorticity field (angular velocity field) $\AngularVelocity$ following Eq.~\eqref{eq:vorticity}.
However, the main idea of vorticity confinement is to merely identify and amplify existing vortices.
Moreover, the vorticity will always be derived from the linear field.
In contrast, the micropolar approach builds on the concept of angular momentum conservation and on modeling a constitutive model for turbulences.
In this more sophisticated setting, the velocity field and the vorticity are discretized independently and strongly coupled via the transfer terms in Eqs.~\eqref{eq:linear_micropolar_momentum} and \eqref{eq:angular_micropolar_momentum}.
In this way there is a complex interaction between both physical quantities that not only conserves existing vortices better but also facilitates the formation of new vortices.
This effect can be exemplified using the lid-driven cavity experiment carried out by Bender et al.~\cite{BKKW18} (see Fig.~\ref{fig:LDC}).
In this experiment the "lid" (top-side) of a two-dimensional domain filled with water is accelerated with constant velocity.
Given suitable model parameters, the velocity field is expected to stabilize in a big central vortex and three minor vortices rotating in the opposite direction.
This result shows that vorticity confinement successfully amplifies the vortical motion but is not able to form the additional corner vortices.
In contrast, the micropolar approach yields the expected result.


\section{Multiphase Fluids}
\label{sec:multiphase_fluids}

The simulation of multiple immiscible and miscible fluids greatly enhance visual effects in graphics. In contrast to Eulerian approaches, the particle representation of SPH offers the advantage that fluid interfaces are sharply defined. In this section, we first present how the standard SPH equations can be adapted to model density discontinuity across fluid interfaces, and introduce the resulting adapted force equations. We then discuss models for capturing complex mixing phenomena.

\subsection{Fluid Interfaces}

A simple approach to simulate multiple fluids with SPH is to assign different labels to particles of different phases, and assigning them with corresponding physical attributes such as masses and rest densities~\cite{MSKG05}.
Typically, each particle's rest volume remains constant to ensure a uniform particle sampling, thus 
$\frac{\Mass_a}{\RestDensity_a}=\frac{\Mass_b}{\RestDensity_b}$
for two fluid types $a$ and $b$. 
The momentum equation can be solved with the single flow SPH formulation presented in the previous sections, while simply using the physical attributes stored on the particles.
However, for high density ratios between phases, this can lead to instability problems that are not time step related. The desired density discontinuity across the interface is smoothed due to the nature of SPH of summing up contributions from particle neighbors. As a consequence, pressure and force fields are affected, which manifests as spurious interface tension~\cite{Hoover98,AMS+06} between the phases. 
Larger density ratios between the fluids (>10x) intensify the problems and severely degrade simulation stability regardless of the time step size. 

To capture the density discontinuity across the interface with SPH, the number density $\delta_i = \sum_j \Kernel[i][j]$ was introduced and the standard SPH equations were adapted accordingly~\cite{TM05,HA06,SP08}. The density of a particle $i$ is then computed as 
\begin{equation}
\tilde \Density_i=\Mass_i \delta_i.
\end{equation}
Like this, the density of particle $i$ is not influenced by the mass of its neighbors $j$, while still receiving the geometric contribution $\Kernel[i][j]$ from $j$. The state equation of Sections~\ref{sub:state_equation} can then be changed such that the pressure is computed with the adapted density as
\begin{equation}
\tilde \Pressure_i= k_1 \left (\left (\frac{\tilde \Density_i}{\RestDensity}\right )^{k_2} -1 \right).
\end{equation}
Solenthaler et al.~\shortcite{SP08} derived adapted forces by substituting $\tilde \Density_i$ and $\tilde \Pressure_i$ into the Navier-Stokes equations and applying the SPH formalism. 
The resulting pressure force term is then given as 
$\PressureForce= -\frac{{\nabla \tilde{\Pressure}}}{{\delta}}$.
By employing the quotient rule we then get 
$\frac{\nabla \tilde{\Pressure}}{{\delta}} =  \nabla (\frac{{\tilde{\Pressure}}}{{\delta}} ) + \frac{{\tilde{\Pressure}}}{{{\delta}^2}}\nabla \delta$.
After applying the SPH rule and replacing $V$ by $\frac{1}{\tilde \Density}$, the pressure force equation can be written as
\begin{equation}
\PressureForce_i= -\sum_j  \left( \frac{\tilde{\Pressure}_j}{{\delta_j}^2} +  \frac{\tilde{\Pressure}_i}{{\delta_i}^2}\right) \nabla \Kernel[i][j].
\end{equation}
Similar derivations can be found in~\cite{TM05,HA06}. 
The viscosity force (and other force terms) can be derived analogously and is given as
\begin{equation}
\mathbf{F}^{v}_i= \frac{1}{\delta_i} \sum_j  \frac{\mu_i + \mu_j}{2}\frac{1}{\delta_j}(\Velocity_j-\Velocity_i) \nabla^2 \Kernel[i][j].
\end{equation}

\begin{figure}[tb]
	\centering
	\includegraphics[width=\columnwidth]{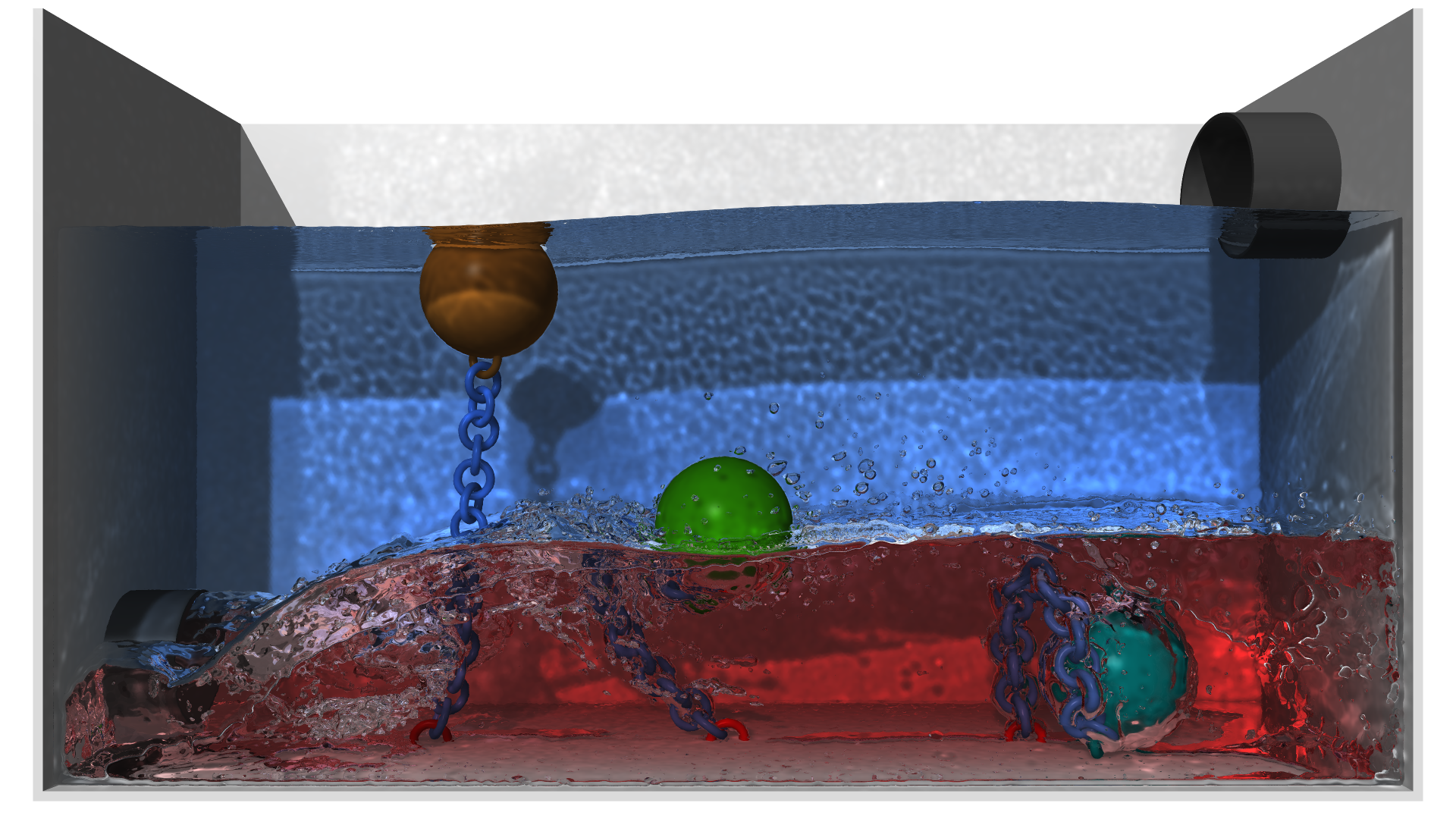} 
	\caption{Three solid buoys interacting with two IISPH fluids with different densitites~\cite{GPB+19}.}
	\label{fig:multiphase}
\end{figure}

Note that the above equations are identical to the standard SPH equations when applied to a single phase flow. For multiple fluids, however, the adapted method eliminates any spurious tension effects and notably increases stability. The method has been extended with an incompressibility condition and solid-fluid coupling~\cite{AIA+12,GPB+19}, an example is shown in Fig.~\ref{fig:multiphase}.
Moreover, the resolution at the interface has been increased using the two-scale (or multi-scale) particle simulation method in~\cite{SG11,HS13}.

The above described problems can be circumvented by replacing the summation density by the 
continuity equation that evolves the density over time (Section \ref{sec:mass_density_estimation}) and hence does not suffer from smoothing artifacts across fluid interfaces. However, this typically requires higher-order time stepping schemes and careful considerations of time step sizes to avoid accumulation of integration errors over time and thus drift from true mass conservation~\cite{SP08,SB12}. The density summation equation was also used in combination with the Shepard kernel to accurately preserve the discontinuity at the interface~\cite{GAC+09}. The method considers the volume distribution and the rate of change of the volume estimated by the continuity equation.

\subsection{Complex Mixing Phenomena}

Fluid mixing can be simulated by solving the diffusion equation $\frac{\partial C}{\partial t}=\alpha \nabla^2 C$, which evolves the concentration $C$ over time. With SPH, this equation can be written as~\cite{MSKG05}
\begin{equation}
\frac{\partial C_i}{\partial t}= \alpha\sum_j  \Mass_j \frac{C_j-C_i}{\Density_j} \nabla^2 \Kernel[i][j],
\end{equation}
where $\alpha$ defines the diffusion strength. Another SPH formulation for computing the diffusion has been presented in~\cite{LLP11}.

More complex mixing effects can be simulated by taking the flow motion and force distributions into account as demonstrated in the SPH-based mixture model of Ren et al.~\shortcite{RLY+14}. The continuity equation of the mixture model is defined as
\begin{equation}
\frac{\D\Density_m}{\D t}= \frac{\partial \Density_m}{\partial t} + \Divergence (\Density_m \Velocity_m) = 0,
\end{equation}
where $\Density_m$ is the rest density of the mixture and $\Velocity_m$ is the mixture velocity, averaged over all phases. $\Density_m$ and $\Velocity_m$ are computed using the volume fraction $\alpha_k$ of a phase $k$ with rest density $\Density_k$, \ie $\Density_m=\sum_k \alpha_k \Density_k$ and $\Velocity_m=\frac{1}{\Density_m \sum_k \alpha_k \Density_k \Velocity_k}$. 
The momentum equation for the mixture is given as
\begin{equation}
\frac{\D(\Density_m,\Velocity_m)}{\D t}= -\nabla p + \nabla \cdot (\mathbf{\tau}_m+\mathbf{\tau}_{Dm}) + \Density_m \Gravity,
\end{equation}
where $\mathbf{\tau}_m$ and $\mathbf{\tau}_{Dm}$ are the mixture's viscous stress and diffusion tensors, respectively. 

In each simulation step, the drift velocity $\Velocity_{mk} = \Velocity_{k}-\Velocity_{m}$ is computed, which represents the relative velocity of phase $k$ to the mixture. The equation can be rewritten using individual terms for slip velocity due to body forces, pressure effects that cause fluid phases to move from high to low pressure regions, and a Brownian diffusion term that represents phase drifting from high to low concentration regions. The drift velocity is then used to calculate the diffusion tensor $\tau_{Dm}$ and change in volume fraction $\D \alpha_k/\D t$.
The SPH equations for the mixture model described above can be found in the work of Ren et al.~\shortcite{RLY+14}. They demonstrate complex mixing effects including chemical reactions. The model uses WCSPH, since a divergence-free velocity field cannot be directly integrated since neither the mixture nor phase velocities are zero, even if the material is incompressible. 

Yan et al.~\shortcite{YJL+16} extended the mixture model to handle the interaction between fluid and solid phases, and demonstrated various effects including dissolution of solids, flows in porous media, and interaction with elastic materials. Another extension has been presented by Yang et al.\shortcite{YCR+15} where an energy-based model was used. The approach integrates the Cahn-Hilliard equation that describes phase separation, expanding the capability of a multi-fluid solver and enabling incompressible flows.


\section{Deformable Solids}
\label{sec:deformable_solids}

\begin{figure*}[t]
	\centering
	\includegraphics[height=5.2cm]{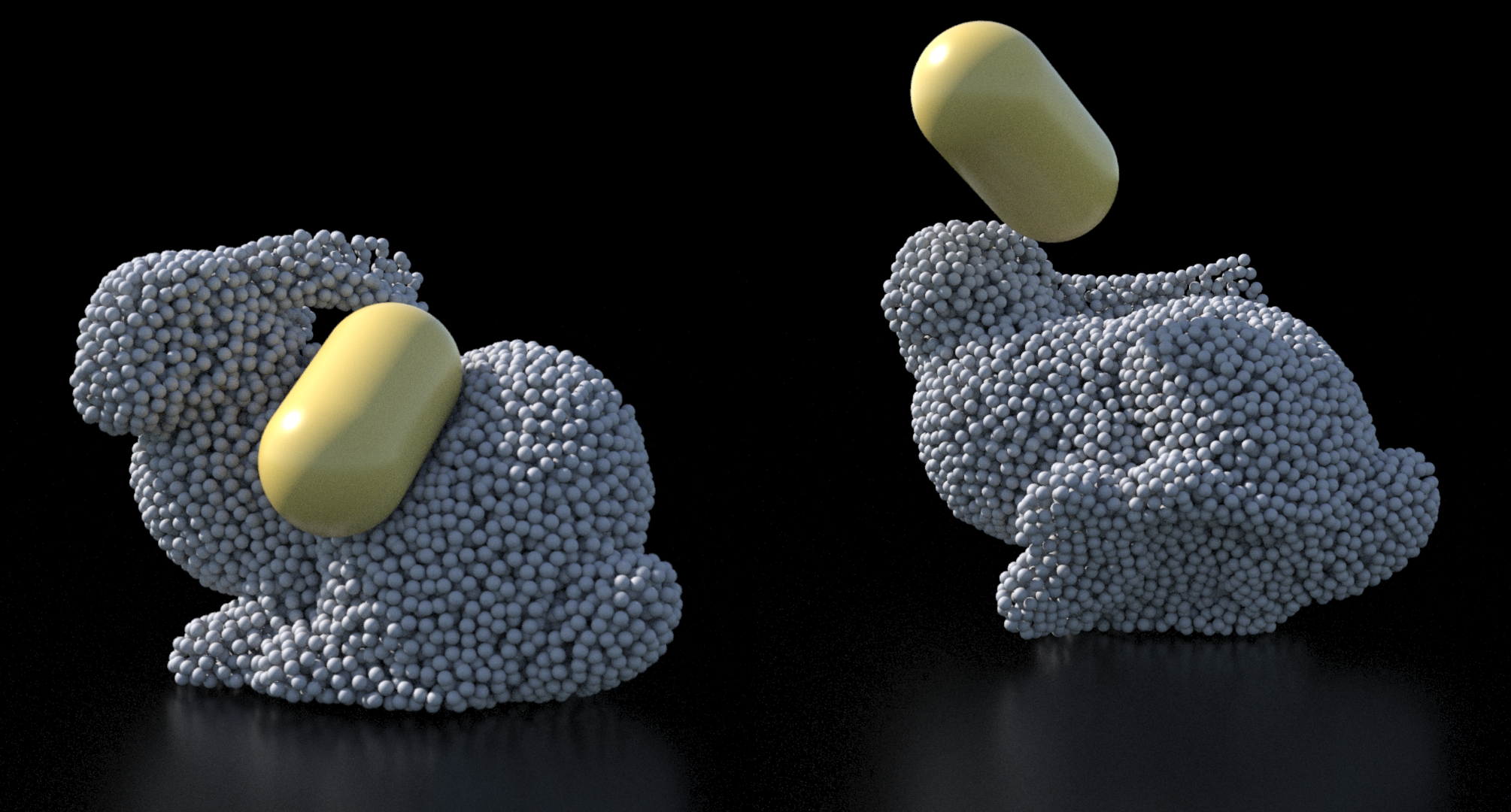} \hfill
	\includegraphics[height=5.2cm]{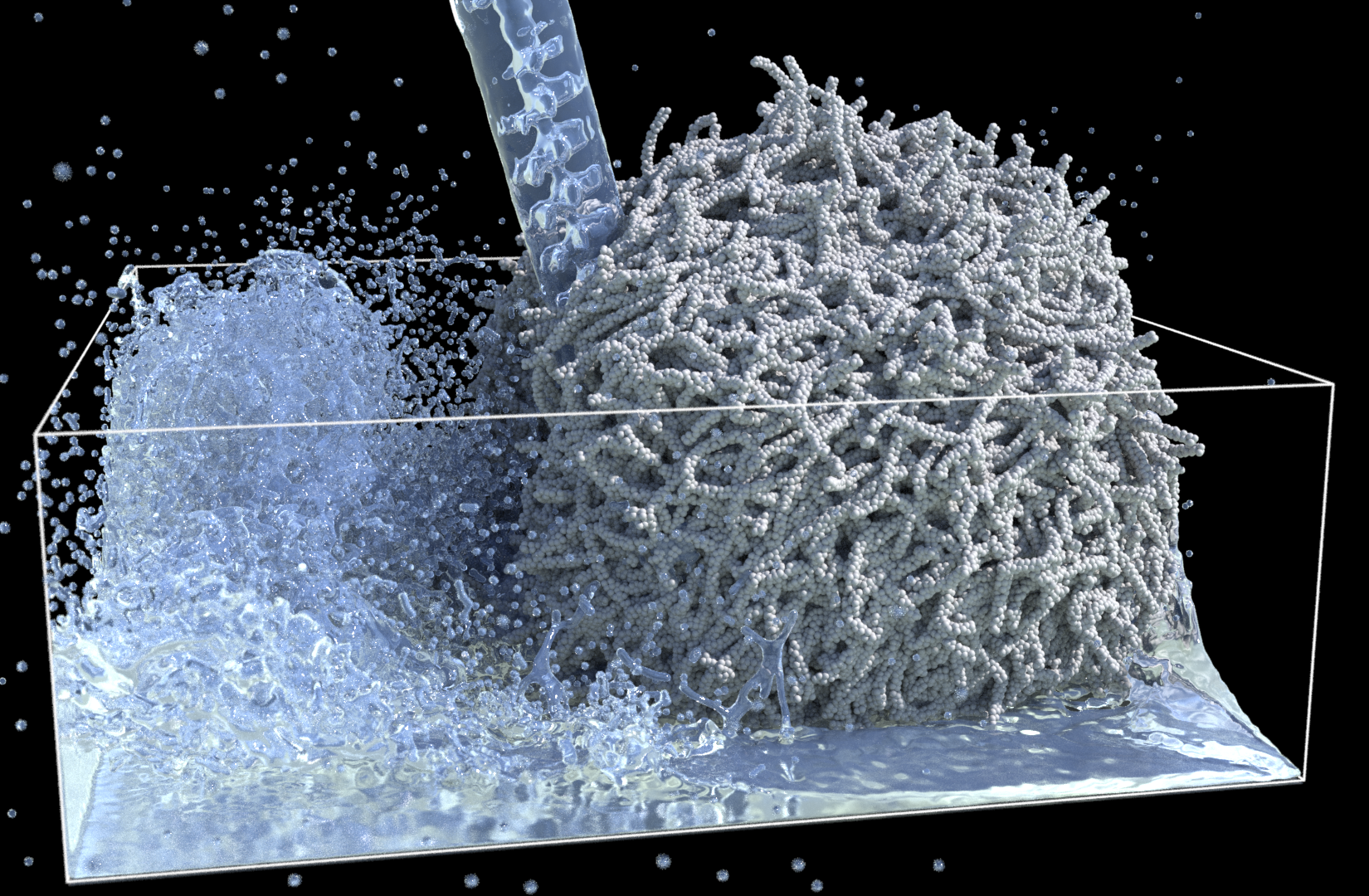} \hfill
	\caption{Deformable solids simulated using an SPH formulation~\cite{PGBT17}.	Left: An elastic bunny model is hit by a rigid capsule. Right: A deformable hairball interacts with water.
		}		
	\label{fig:deformable_solids_teaser}
\end{figure*}

The simulation of deformable solids is an active research topic in computer graphics.
The most popular simulation approaches in this area, like the finite element method (FEM)~\cite{KBT17,KKB2018} and Position-Based Dynamics (PBD)~\cite{BKCW14,BMM14,BMM17}, are mesh-based.
However, also meshless methods were investigated like the moving least squares (MLS) method~\cite{AW09}.
In this section we show that SPH is also an interesting meshless method to simulate deformable solids.
An advantage of an SPH-based simulation of deformables is that this enables a simple coupling between fluids and solids in a unified framework.

\subsection{Linear Elasticity}

In this subsection we first introduce a continuum mechanical formulation for linear elasticity. 
In the next subsection we then show how to discretize the resulting equations using SPH.

The deformation of a solid is defined by the function
\begin{equation}
	\Phi(\RefPosition) = \RefPosition + \Displacement = \Position
\label{eq:deformation_function}
\end{equation}
which maps a point $\RefPosition$ in the reference configuration to its current position $\Position$ in the deformed configuration, where $\Displacement = \Position - \RefPosition$ is the  displacement vector.
Differentiating this function with respect to the reference position $\RefPosition$ gives us the deformation gradient
\begin{equation}
	\DeformationGradient = \frac{\partial \Phi(\RefPosition)}{\partial \RefPosition} =  \frac{\partial \Position}{\partial \RefPosition} = \Identity + \frac{\partial \Displacement}{\partial \RefPosition}.
\end{equation}
This quantity can be used to measure the strain of a deformed body.
In computer graphics often a linear strain measure is used to avoid the solution of a non-linear system of equations.
For the same reason we introduce the linear infinitesimal strain tensor
\begin{equation}
 \Strain[][\DeformationGradient] = \frac12 (\DeformationGradient + \DeformationGradient^T) - \Identity.
\label{eq:linear_strain_tensor}
\end{equation}

The next step is to define a constitutive model for linear elasticity. 
We follow the work of Sifakis~\cite{Sif12} and define it in terms of the strain energy density:
\begin{equation}
	\Psi(\DeformationGradient) = \mu \Strain : \Strain + \frac{\lambda}{2}\text{tr}^2(\Strain),
\end{equation}
where $\mu$ and $\lambda$ are the Lam\'e coefficients~\cite{Sif12}.
The first Piola–Kirchhoff stress tensor is determined by differentiating the strain energy density with respect to the deformation gradient
\begin{equation}
	\FirstPiolaKirchhoffStress[][\DeformationGradient] = \frac{\partial \Psi}{\partial \DeformationGradient}.
\end{equation}
For our linear elasticity model this yields 
\begin{equation}
	\FirstPiolaKirchhoffStress[][\DeformationGradient] = 2 \mu \Strain + \lambda \text{tr}(\Strain) \Identity.
	\label{eq:linear_stress_tensor}
\end{equation}
Note that the Lam\'e coefficients can be computed from Young's modulus $k$ and Poisson's ratio $\nu$ by
\begin{equation}
	\mu = \frac{k}{2 (1 + \nu)},\quad\quad \lambda = \frac{k \nu}{(1+\nu)(1-2\nu)}.
\end{equation}
This is often more intuitive since Young's modulus is a measure of stretch resistance while Poisson ratio is a measure of incompressibility.
Finally, the elastic body forces are determined as the divergence of the stress tensor
\begin{equation}
	\BodyForce = \Divergence \FirstPiolaKirchhoffStress.
\end{equation}

In the following subsection we will discuss how this continuous elastic material model can be discretized using the SPH formulation.

\subsection{SPH Discretization}

The deformation of an elastic body is determined with respect to its reference configuration (see Eq.~\eqref{eq:deformation_function}). 
Typically the initial shape of a body is used as reference configuration in an SPH simulation. 
Since the topology of an elastic body does not change during the simulation, we store the neighborhood of each particle $i$ in the reference configuration.
In the following this reference neighborhood is denoted by $\Neighborhood_i^0$.

\paragraph*{Deformation Gradient}
A straightforward SPH discretization of the deformation gradient is given by
\begin{equation}
	\DeformationGradient[i][\Position, \RefPosition] = \sum_{j \in \Neighborhood_i^0} \Volume[j][][0] \Position[ji] \KernelGradient[][][\RefPosition[ij]][]^T,
\end{equation}
where $\Position[ji] = \Position[j] - \Position[i]$ and $\RefPosition[ij] = \RefPosition[i] - \RefPosition[j]$.
Since this SPH approximation is determined in the reference configuration, the rest volume $\Volume[j][][0]$ has to be used in the sum.
However, this formulation fails to capture rotational motion since it is not first-order consistent (see Section~\ref{sub:basics_discretization}).

\paragraph*{Kernel Gradient Correction}

Bonet and Lok~\cite{BL99} have shown that the gradient of the kernel has to fulfill the following condition to ensure that the computation is first-order consistent and therefore correctly captures rotational motion:
\begin{equation}
  \sum_{j \in \Neighborhood_i^0} \Volume[j][][0] \RefPosition[ji] \KernelGradient[][][\RefPosition[ij]][] ^T = \Identity.
\end{equation}
Now we can formulate a corrected kernel gradient which satisfies the condition by construction:
\begin{equation}
	\CorrectedKernelGradient[i][][\RefPosition[ij]][] = \KernelCorrection[i] \KernelGradient[][][\RefPosition[ij]][],
\end{equation}
where $\KernelCorrection[i]$ is a correction matrix that is defined as
\begin{equation}
	\KernelCorrection[i] = \left ( \sum_{j \in \Neighborhood_i^0} \Volume[j][][0] \KernelGradient[][][\RefPosition[ij]][] \RefPosition[ij]^T \right )^{-1}.
	\label{eq:correction_matrix}
\end{equation}
Note that this correction matrix only depends on the rest volume and the particle positions in the reference configuration. 
Therefore, this matrix is precomputed at the beginning of the simulation. 
If the matrix in Eq.~\eqref{eq:correction_matrix} is singular and cannot be inverted, \eg due to a collinear or coplanar particle configuration, the Moore–Penrose inverse is used instead.

\paragraph*{Corotated Approach}
Using the corrected kernel gradient we get a first-order consistent SPH formulation for the deformation gradient:
\begin{equation}
	\DeformationGradient[i][\Position, \RefPosition] = \sum_{j \in \Neighborhood_i^0} \Volume[j][][0] \Position[ji] \CorrectedKernelGradient[][][\RefPosition[ij]][]^T.
	\label{eq:corrected_deformation_gradient}
\end{equation}
The deformation gradient is used to compute the linear infinitesimal strain tensor by Eq.~\eqref{eq:linear_strain_tensor}. 
Note that we use a linear strain measure since in an implicit formulation it is more efficient to solve a linear system than a non-linear one. 
However, the linear strain tensor is not invariant under rotations.
In computer graphics a common solution for this problem is to use a corotational approach~\cite{BIT09,KKB2018}.
The core idea of this approach is to extract the rotation and to compute the strain measure in an unrotated frame.
In the following we will show how the rotation can be extracted and introduce the computation of a corotated deformation gradient.

The deformation gradient computed with the corrected kernel gradient (see Eq.~\eqref{eq:corrected_deformation_gradient}) is able to capture the rotation correctly. 
Hence, the per-particle rotation $\Rotation[i]$ can be directly extracted from $\DeformationGradient[i]$, \eg by using the efficient and stable method of M\"uller et al.~\cite{MBCM16}.

The extracted rotation matrix $\Rotation$ is used to rotate the reference configuration so that the resulting displacement vector $\Displacement = \Position - \Rotation \RefPosition$ contains no rotation.
Since we rotate the reference configuration, we also have to rotate the corrected kernel gradient as it depends on the reference positions.
This yields the rotated corrected kernel gradient
\begin{equation}
	\CorotatedKernelGradient[i][][\RefPosition[ij]] = \Rotation[i] \KernelCorrection[i] \KernelGradient[][][\RefPosition[ij]][].
\end{equation}
Putting all together gives us the corotated deformation gradient
\begin{equation}
	\DeformationGradient[i][\Position, \RefPosition][\ast] = \Identity + \sum_{j \in \Neighborhood_i^0} \Volume[j][][0] \left ( \Position[ji] - \Rotation[i] \RefPosition[ji] \right ) \CorotatedKernelGradient[i][][\RefPosition[ij]][]^T.
\end{equation}

Now we compute the strain tensor $\Strain[][\DeformationGradient[][][\ast]]$ using Eq.~\eqref{eq:linear_strain_tensor} and the stress tensor $\FirstPiolaKirchhoffStress[][\DeformationGradient[][][\ast]]$ using Eq.~\eqref{eq:linear_stress_tensor}.
Finally, the force is determined as the divergence of the stress tensor. 
In our SPH formulation this yields~\cite{Gan15}:
\begin{equation}
	\Force[i][\DeformationGradient[][][\ast]] = \sum_{j \in \Neighborhood_i^0} \Volume[i][][0] \Volume[j][][0] \left ( \FirstPiolaKirchhoffStress[i][\DeformationGradient[i][][\ast]] \CorotatedKernelGradient[i][][\RefPosition[ij]][] - \FirstPiolaKirchhoffStress[j][\DeformationGradient[j][][\ast]]  \CorotatedKernelGradient[j][][\RefPosition[ij]][] \right ).
\end{equation}

If we simply add these particle forces to our system, we get an explicit approach for the simulation of deformable bodies. 
However, this approach is only conditionally stable and requires small time steps when simulating stiff solids.
To improve the stability we will introduce an implicit approach in the next subsection.

\subsection{Implicit Approach}

The implicit method described in the following is based on the work of Peer et al.~\cite{PGBT17}.
Since the elastic forces depend linearly on the particle positions, an implicit formulation of the time step is straightforward
\begin{equation}
	\Velocity[][][\Time + \Delta \Time] = \Velocity[][][\Time] + \frac{\Delta \Time}{m} \Force[][\DeformationGradient[][][\Time+ \Delta \Time]],
\end{equation}
where $\DeformationGradient[][][\Time+ \Delta \Time] = \DeformationGradient[][\Position[][][\Time+ \Delta \Time], \RefPosition][\ast]$ is the deformation gradient at the end of the time step.
This means that we use the new particle positions $\Position[][][\Time+ \Delta \Time]$ to determine the elastic forces at the end of the time step.

In this formulation we have unknown positions $\Position[][][\Time+ \Delta \Time]$ and unknown velocities $\Velocity[][][\Time+ \Delta \Time]$.
In the next step we substitute the positions by $\Position[][][\Time+ \Delta \Time] = \Position[][][\Time] + \Delta \Time \Velocity[][][\Time+ \Delta \Time]$ to get a linear system for the new velocities.
Moreover, we split the computation of the force into
\begin{equation}
	\Force[][\DeformationGradient[][][\Time+ \Delta \Time]] = \Force[][\DeformationGradient[][][\Time]] + \Force[][\DeformationGradient[][][\Delta \Time]],
\end{equation}
where $\DeformationGradient[][][\Time] = \DeformationGradient[][\Position[][][\Time], \RefPosition][\ast]$ is the deformation gradient at the beginning of the time step and 
$\DeformationGradient[][][\Delta \Time] = \DeformationGradient[][\Delta t \Velocity[][][\Time + \Delta \Time], \vec 0][\ast]$ is the deformation gradient that corresponds to the position change in one time step due to the velocities $\Velocity[][][\Time + \Delta \Time]$.
Now we can bring all terms that depend on the unknown new velocities to the left hand side and the rest to the right hand side:
\begin{equation}
	\Velocity[][][\Time + \Delta \Time] - \frac{\Delta \Time}{m} \Force[][\DeformationGradient[][\Delta t \Velocity[][][\Time + \Delta \Time], \vec 0][\ast]] = \Velocity[][][\Time] + \frac{\Delta \Time}{m} \Force[][\DeformationGradient[][\Position[][][\Time], \RefPosition][\ast]].
\end{equation}
This yields a linear system for the velocities $\Velocity[][][\Time + \Delta \Time]$ which can efficiently be solved using a matrix-free conjugate gradient method.
More details about the matrix-free solver can be found in the work of Peer et al.~\cite{PGBT17}.

\subsection{Zero-Energy Mode Suppression}

\begin{figure}[t]
	\centering
	\subfloat[reference configuration\label{fig:zero_energy_modes1}]{\includegraphics[scale=0.9]{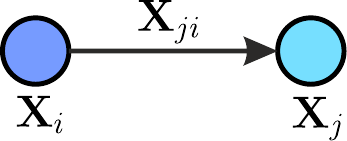}} \hspace{1cm}
	\subfloat[deformed configuration\label{fig:zero_energy_modes2}]{\includegraphics[scale=0.9]{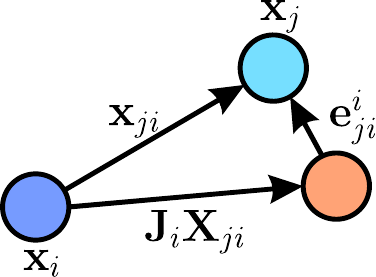}} 
		\caption{Due to numerical errors the transformed reference vector $\DeformationGradient[i] \RefPosition[ji]$ and the current deformed vector $\Position[ji]$ are typically not equal in an SPH simulation.
	The error is defined by the difference $\vec e^i_{ji} = \DeformationGradient[i] \RefPosition[ji] - \Position[ji]$.
	}
	\label{fig:zero_energy_modes}
\end{figure}

In SPH simulations zero-energy modes can occur which are similar to hour glass modes in finite element methods~\cite{Gan15}.
The displacement field in the neighborhood of a particle $i$ has to be defined exactly by its corresponding deformation gradient $\DeformationGradient[i]$.
This means that if we transform the vector $\RefPosition[ji]$ from particle $i$ to one of its neighbors $j$ in reference space using the deformation gradient $\DeformationGradient[i] \RefPosition[ji]$, this should give us the actual vector $\Position[ji]$ in the deformed configuration.
Hence, the vector 
\begin{equation}
	\vec e^i_{ji} = \DeformationGradient[i] \RefPosition[ji] - \Position[ji]
\end{equation}
should be zero.
However, this is typically not the case due to numerical errors (see Fig.~\ref{fig:zero_energy_modes}).

Ganzenm\"uller~\cite{Gan15} proposes to compute a penalty force to minimize the error vector $\vec e_{ji}$:
\begin{equation}
 \Force[i][][\text{HG}]= -\frac12 \alpha k \sum_{j \in \Neighborhood_i^0} \Volume[i][][0] \Volume[j][][0]  \frac{\Kernel[][][\RefPosition[ij]][]}{\|\RefPosition[ij]\|^2} \left ( \frac{\vec e^i_{ji} \cdot \Position[ji]}{\|\Position[ji]\|} + \frac{\vec e^j_{ij} \cdot \Position[ij]}{\|\Position[ij]\|} \right ) \frac{\Position[ji]}{\|\Position[ji]\|}, 
\end{equation}
where the coefficient $\alpha$ controls the amplitude of the zero-energy mode suppression and $k$ is the Young's modulus.
In this way the system gets more stable and hourglass modes are suppressed.


\section{Rigid Solids}
\label{sec:rigid_solids}

\begin{figure*}[t]
	\centering
	\includegraphics[width=\linewidth]{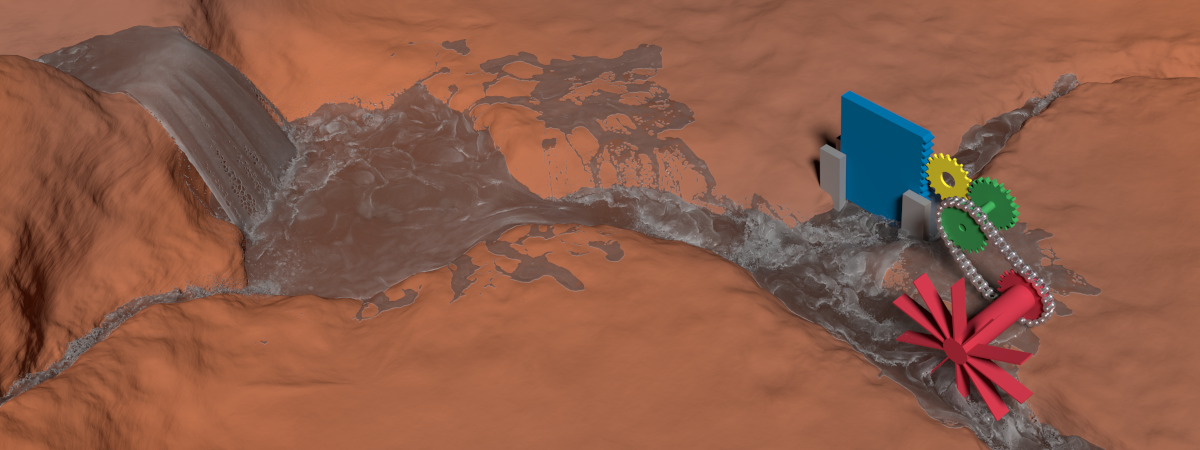} 
	\caption{The SPH-based rigid body solver enables a strong two-way coupling between fluids and rigid bodies. In this scenario 43.8M fluid particles interact with 50M static and 2.3M dynamic rigid body particles. Up to 90k simultaneous rigid-rigid contacts were resolved.
	\label{fig:rb_teaser}
		}			
\end{figure*}

Recently, it has been demonstrated that SPH can also be used to realize a rigid body simulation with contact handling~\cite{GPB+19}.
In the following we will introduce the SPH formulation for rigid bodies and show that this enables a strong two-way coupling of fluids and rigid bodies (see Fig.~\ref{fig:rb_teaser}).

The core idea of the SPH-based contact handling for rigid bodies is similar to the concept of particle-based boundary handling (see Section~\ref{sec:boundary_handling}).
Therefore, in the beginning of the simulation the surface of each rigid body is sampled by particles. 
But instead of only computing pressure forces between fluid and boundary particles, we also determine artificial pressure forces between the particles on the surfaces of the rigid bodies in order to resolve contacts.
Due to the unified particle representation of all bodies in the simulation, the neighborhood search can be used to detect the collision of rigid body particles. 
Hence, no additional collision detection method for rigid bodies is required. 

\begin{figure*}[t]
	\centering
	\includegraphics[width = 0.495\linewidth, clip = true, trim = 300pt 100pt 150pt 40pt]{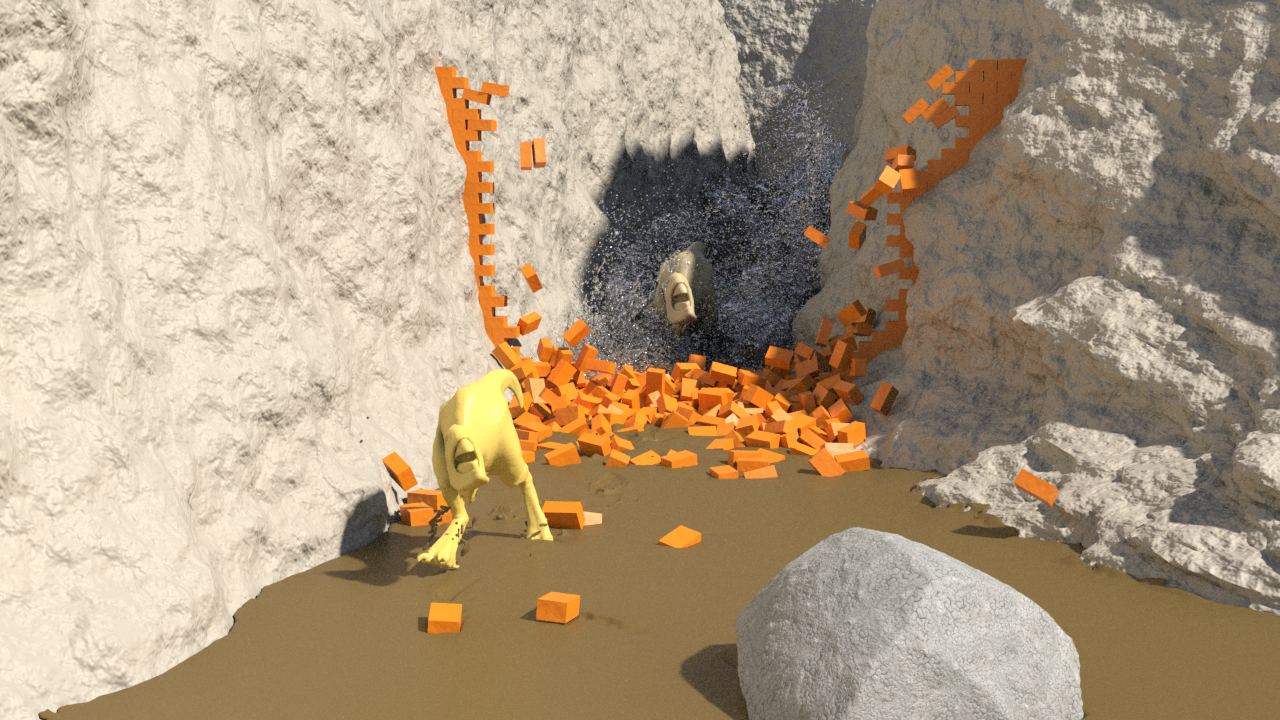} \hfill
	\includegraphics[width = 0.495\linewidth, clip = true, trim = 300pt 70pt 150pt 70pt]{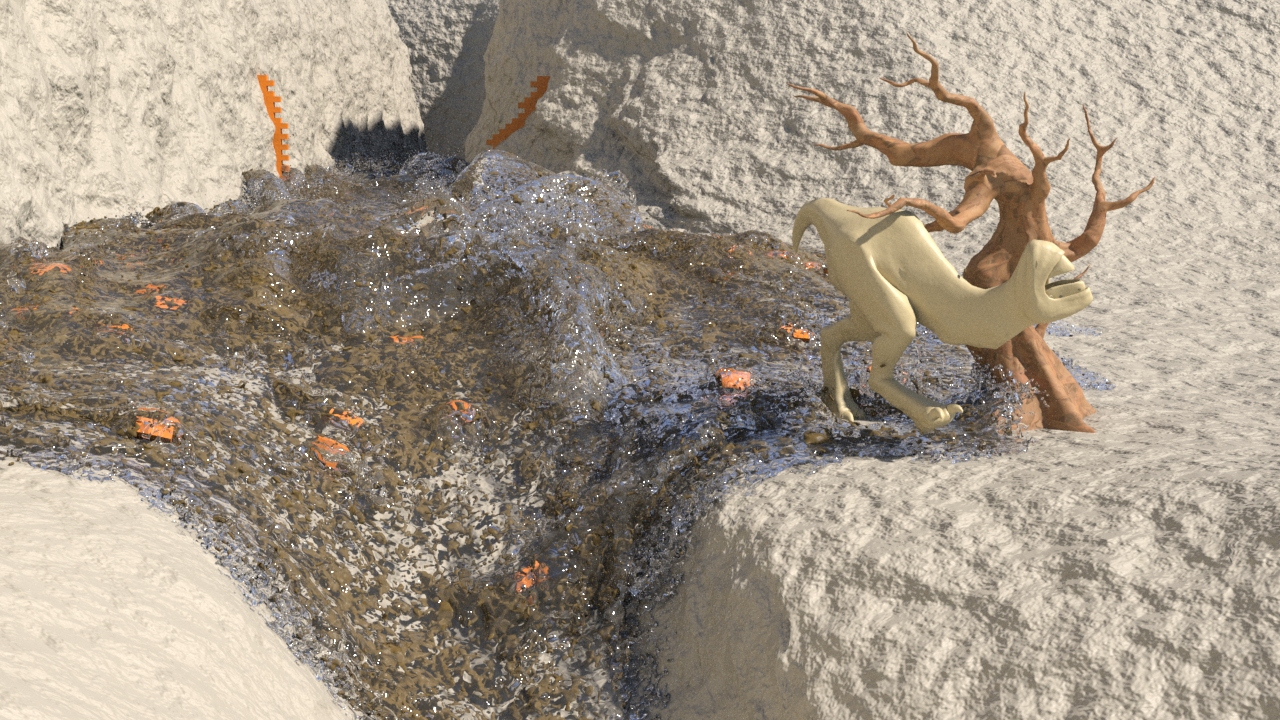}
	\caption{Two creatures run in highly viscous mud, break through a wall of rigid bodies and collide with a deformable tree.
	This complex simulation shows the power of a unified SPH solver.
	\label{fig:valley}}
\end{figure*}

Such an SPH-based rigid body solver can easily be combined with SPH discretizations of other materials.
This enables a simple two-way coupling of fluids, rigid bodies, deformable solids and highly viscous materials (see Fig.~\ref{fig:valley}).

\subsection{Rigid Body Solver}

In the following we will discuss how to compute rigid-rigid contact forces $\Force[][][\text{rr}]$.
We want to compute these forces similar to the fluid-rigid interface forces which were discussed in Section~\ref{sec:boundary_handling}.
Therefore, we first introduce an artificial rest density $\Density[r][][0] = 1$ for each rigid particle $r$. 
Note that the magnitude of the rest density can be chosen arbitrarily since we are only interested in a density deviation.
If there is a contact, we get a density deviation of $\Density[r] - \Density[r][][0] > 0$ for the corresponding particle $r$. 
In this case our goal is to determine contact forces $\Force[][][\text{rr}]$ such that $\Density[r] = \Density[r][][0]$.

Now we will derive an implicit method to compute the unknown contact forces.
We start with the continuity equation
\begin{equation}
	\frac{D \Density[r]}{D\Time} = -\Density[r] \Divergence \Velocity[r],
\end{equation}
where $\Density[r]$ and $\Velocity[r]$ are the density and the velocity of a rigid body particle $r$, respectively.
Then we use a backward difference time discretization and introduce a constant density constraint $\Density[r][][\Time+\Dt] = \Density[r][][0]$ to get
\begin{equation}
	\frac{\Density[r][][0] - \Density[r]}{\Dt} = -\Density[r] \Divergence \Velocity[r][][\Time + \Dt],
	\label{eq:rb_discretized_continuity_eq}
\end{equation}
where $\Velocity[r][][\Time + \Dt]$ is the velocity of the rigid body particle $r$ at time $\Time + \Dt$.
This velocity vector can be written as
\begin{equation}
	\Velocity[r][][\Time + \Dt] = \Velocity[R][][\Time + \Dt] + \AngularVelocity[R][][\Time + \Dt] \times \vec r_r^{\Time + \Dt}, 
	\label{eq:rb_point_vel}
\end{equation}
where $\Velocity[R][][\Time + \Dt]$ and $\AngularVelocity[R][][\Time + \Dt]$ are the linear and angular velocity of the rigid body $R$ at time $\Time + \Dt$, respectively, and $\vec r_r^{\Time + \Dt}$ is the vector from the center of mass of the rigid body to the position of the particle $r$. 
The new velocities are determined by an Euler integration step
\begin{align}
	\Velocity[R][][\Time + \Dt] &= \Velocity[R][][] + \Dt \frac{1}{\Mass[R]} \left ( \Force[R] + \sum_{k \in \RigidParticles} \Force[k][][\text{rr}] \right )  \\
	\AngularVelocity[R][][\Time + \Dt]  &= \AngularVelocity[R][][] + \Dt \InertiaTensor[R][][-1] \left ( \Torque[R] + (\InertiaTensor[R] \AngularVelocity[R][][]) \times \AngularVelocity[R][][] + \sum_{k \in \RigidParticles} \vec r_k \times \Force[k][][\text{rr}]\right ),  \label{eq:rb_angular_vel}
\end{align}
where $\InertiaTensor[R]$ is the inertia tensor of the rigid body. 
The vectors $\Force[R]$ and $\Torque[R]$ contain all forces and torques acting on the body except the unknown rigid-rigid contact forces $\Force[k][][\text{rr}]$.
$\RigidParticles$ denotes the set of all particles of rigid body $R$.
Note that all quantities on the right hand side are at time $t$. 
For improved readability we omitted the time parameter for all quantities at the current time $t$.

In the next step we substitute Eqs.~\eqref{eq:rb_point_vel}-\eqref{eq:rb_angular_vel} in Eq.~\eqref{eq:rb_discretized_continuity_eq} to get a linear system for the unknown rigid-rigid contact forces 
\begin{align}
	\frac{\Density[r][][0] - \Density[r]}{\Dt} = &-\Density[r] \Divergence \left ( \Velocity[R] +  \frac{\Dt}{\Mass[R]} \Force[R] \right ) - \Density[r] \Divergence \left ( \frac{\Dt}{\Mass[R]} \sum_{k \in \RigidParticles}  \Force[k][][\text{rr}] \right )  \nonumber \\ \nonumber
	&-\Density[r] \Divergence \left ( \left ( \AngularVelocity[R] + \Dt \InertiaTensor[R][][-1] \left (  \Torque[R] +  (\InertiaTensor[R] \AngularVelocity[R][][]) \times \AngularVelocity[R][][] \right ) \right ) \times \vec r_r^{\Time+\Dt} \right )  \\ 
	&-\Density[r] \Divergence \left ( \Dt \left ( \InertiaTensor[R][][-1] \sum_{k \in \RigidParticles} \vec r_k \times \Force[k][][\text{rr}] \right ) \times \vec r_r^{\Time+\Dt} \right ).
	\label{eq:rb_system}
\end{align}
To simplify this system we use the approximation $\vec r_r^{\Time+\Dt} = \vec r_r$.
Moreover, we introduce the velocity vector
\begin{equation}
	\Velocity[r][][s] = 	\Velocity[R] +  \frac{\Dt}{\Mass[R]} \Force[R] + \left ( \AngularVelocity[R] + \Dt \InertiaTensor[R][][-1] \left (  \Torque[R] +  (\InertiaTensor[R] \AngularVelocity[R][][]) \times \AngularVelocity[R][][] \right ) \right ) \times \vec r_r.
\end{equation}
This vector determines the new velocity of a particle $r$ after a time step which considers all forces and torques except the unknown contact forces. 
In this way we can write the right-hand side of our linear system in a compact form:
\begin{equation}
	s_r = \frac{\Density[r][][0] - \Density[r]}{\Dt} + \Density[r] \Divergence \Velocity[r][][s].
\end{equation}
The left-hand side contains all terms of Eq.~\eqref{eq:rb_system} that depend on the rigid-rigid contact forces $\Force[k][][\text{rr}]$.
The resulting linear system has the form
\begin{equation}
	-\Density[r] \Divergence \left ( \frac{\Dt}{\Mass[R]} \sum_{k \in \RigidParticles} \Force[k][][\text{rr}] + \left ( \Dt \InertiaTensor[R][][-1] \sum_{k \in \RigidParticles} \vec r_k \times \Force[k][][\text{rr}] \right ) \times \vec r_k \right ) = s_r. 
\end{equation}
The left-hand side can further be simplified by introducing the matrix 
\begin{equation}
	\mat K_{rk} = \frac{1}{\Mass[R]}\Identity - \tilde{\vec r}_r \InertiaTensor[R][][-1] \tilde{\vec r}_k,
\end{equation}
where $\tilde{\vec r}_r$ is the cross product matrix of $\vec r_r$ to get
\begin{equation}
	-\Density[r] \Divergence \left ( \Dt \sum_{k \in \RigidParticles} \mat K_{rk}  \Force[k][][\text{rr}] \right ) = s_r .
\end{equation}
Note that the matrix $\mat K_{rk}$ is well-known in the area of rigid body solvers~\cite{Mir96a,BET14}.

Our goal is to resolve the contacts by a pressure force. 
Therefore, we define 
\begin{equation}
	\Force[k][][\text{rr}] = -\Volume[k] \Gradient \Pressure[k],
\end{equation}
where $\Volume[k]$ is an artificial volume of particle $k$ and $\Pressure[k]$ is an unknown pressure which is used to resolve the collision.
This yields the final linear system
\begin{equation}
	\Density[r] \Divergence \left ( \Dt \sum_{k \in \RigidParticles} \Volume[k] \mat K_{rk}  \Gradient \Pressure[k] \right ) = \frac{\Density[r][][0] - \Density[r]}{\Dt} + \Density[r] \Divergence \Velocity[r][][s].
\end{equation}

Solving the linear system gives us the unknown pressure values for all rigid particles.
Note that the linear system contains one equation for each rigid particle. 
However, if a particle $r$ has no contact to a particle of another rigid body, we can remove the corresponding equation from the system and set $\Pressure[r] = 0$ as no contact must be resolved in this case.

\subsection{Implementation}

In the following we describe how the quantities in the derived linear system are computed.

The artificial rest volume of a rigid particle $r$ is determined as
\begin{equation}
	\Volume[r][][0] = \frac{0.7}{\sum_{k \in \RigidParticles} \Kernel[r][k]}.
\end{equation}
A detailed discussion about this computation can be found in the work of Gissler et al.~\cite{GPB+19}.
Together with the artificial rest density $\Density[r][][0] = 1$ the actual density of a rigid particle is computed as $\Density[r] = \sum_k \Volume[r][][0] \Density[r][][0] \Kernel[r][k]$, where the sum considers the particles $k$ of all rigid bodies within the support radius of the kernel.
Due to the sum over the particles of neighboring rigid bodies, we get a density deviation of $\Density[r] - \Density[r][][0] > 0$ in case of a contact.
In this case we compute the actual volume of a rigid particle $r$ as $\Volume[r] = \frac{\Density[r][][0] \Volume[r][][0]}{\Density[r]}$.

The divergence on the right-hand side of the linear system is determined as
\begin{equation}
	\Divergence \Velocity[r][][s] = \frac{1}{\Density[r]} \sum_{k \in \RigidParticles} \Volume[k] \Density[k] \left ( \Velocity[k][][s] - \Velocity[r][][s]\right ) \cdot \KernelGradient[r][k].
\end{equation}
Finally, we solve the linear system using a relaxed Jacobi solver and update the pressure in iteration $l+1$ as
\begin{equation}
	\Pressure[r][][l+1] = \Pressure[r][][l] + \frac{\beta_r^\text{RJ}}{b_r} \left( s_r - \Density[r] \Divergence \left ( \Dt \sum_{k \in \RigidParticles} \Volume[k] \mat K_{rk} \Gradient \Pressure[k][][l] \right ) \right ),	
\end{equation}
where $b_r$ is the diagonal element of the linear system and $\beta_r^\text{RJ}$ is the relaxation coefficient which is set to $\beta_r^\text{RJ} = \frac{0.5}{\text{num\_contacts}}$.

\subsection{Conclusion}

\begin{figure}[t]
		\centering
		\begin{tikzpicture}
			\node[anchor = south west, inner sep = 0] (image) at (0,0) {\includegraphics[width = \columnwidth]{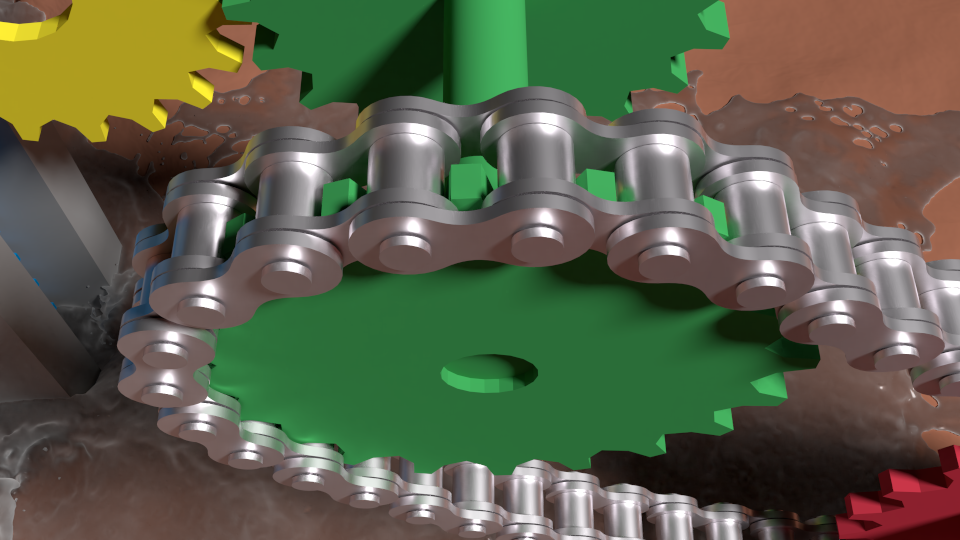}};
			\begin{scope}[x = {(image.south east)}, y = {(image.north west)}]
				\draw[red, ultra thick] (0.275, 0.45) rectangle (0.725, 0.875);
			\end{scope}
		\end{tikzpicture}\\
		\includegraphics[width = \columnwidth, clip = true, trim = 200pt 175pt 200pt 50pt]{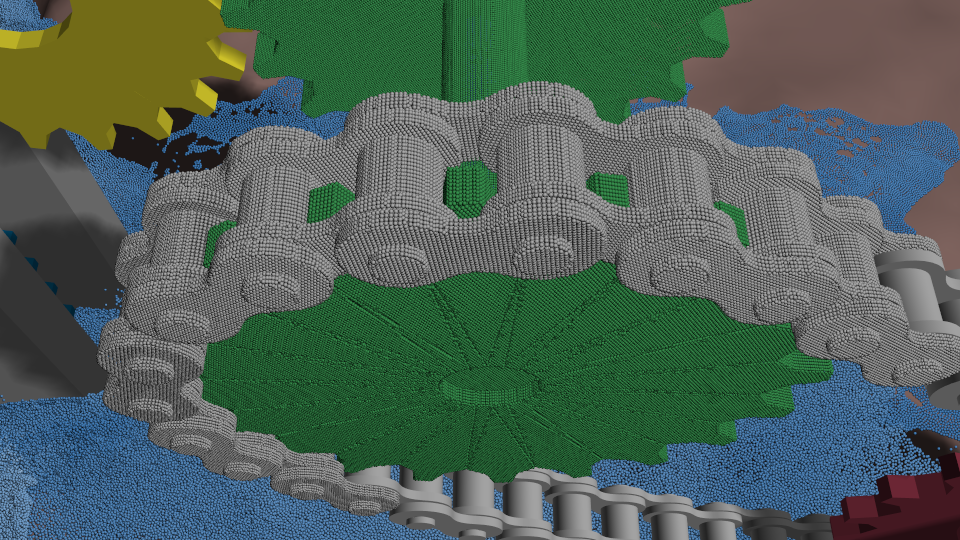}
		\caption{
		Closeup of the water gate scene in Fig.~\ref{fig:rb_teaser}.
		The single chain elements are connected only by rigid-rigid contacts. 
		}
		\label{fig:rigid_bodies}
\end{figure}

The introduced SPH-based rigid body solver enables a strong coupling between fluids and rigid bodies. 
As shown in Fig.~\ref{fig:rigid_bodies} the solver is able to accurately handle complex scenarios with thousands of simultaneous contacts.
It can be easily extended to simulate friction effects~\cite{GPB+19}.
Finally, together with the SPH-based simulation of deformable solids (see Section~\ref{sec:deformable_solids}), it can be combined to a unified SPH solver which supports the coupling of fluids, rigid bodies, deformable solids and highly viscous materials (see Fig.~\ref{fig:valley}).


\section{Data Driven Fluid Simulation}
\label{sec:data_drive_fluids}
\begin{figure}[tb]
	\centering
	\includegraphics[width=\linewidth]{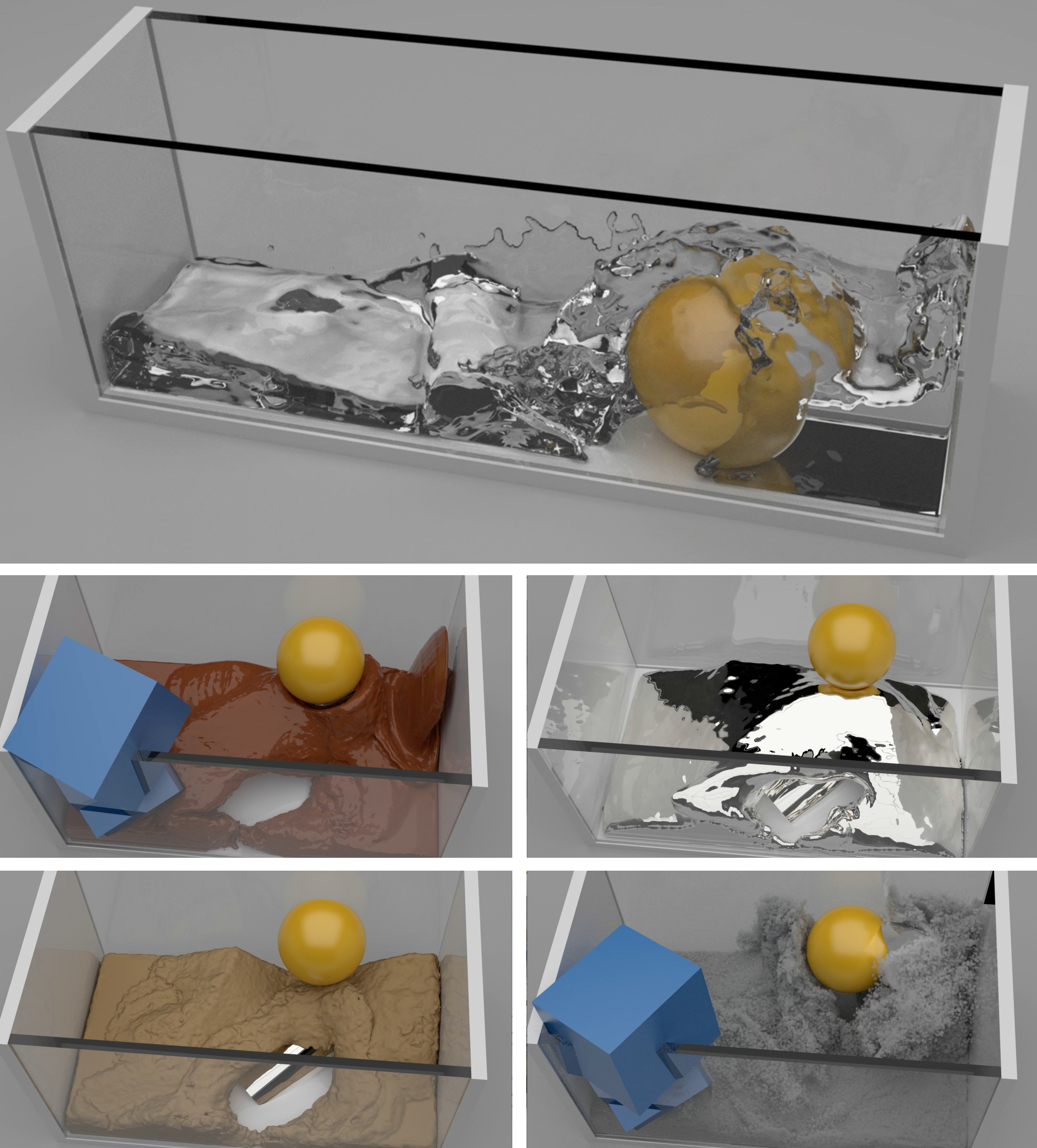} 
	\caption{The regression forest approach for SPH~\cite{LJS+15} enables real-time simulations of over a million particles. At runtime, additional external forces can be added to mimic different material properties without retraining the model.}
	\label{fig:regression_forest}
\end{figure}
Using machine learning for fluid simulations is a largely unexplored research area, but first results are promising and indicate the potential of such data-driven approaches.
In the Lagrangian context, the seminal work of Ladick\'{y} et al.~\shortcite{LJS+15} employed Regression forests to infer particle accelerations (and velocities) using handcrafted, SPH-based features. We discuss this work in more detail below. Um et al.~\shortcite{UHT18} presented a method to augment simulations with learned splashes from a high-resolution FLIP simulation, but included also an example where SPH training data was used. Somewhat related to SPH, Schenk et al.~\shortcite{SF18} proposed a differentiable PBF solver~\cite{MM13} for deep neural networks. They have presented convolution layers for summing up contributions from neighbors and for fluid-object interaction, which potentially can be adapted to SPH fluids as well.
Other work mainly focused on Eulerian simulations, for example to substitute the pressure projection step with a CNN~\cite{TSS+16}, to synthesize flow simulations from a set of reduced parameters~\cite{KAT19}, to compute smoke super-resolution with GAN networks~\cite{XFC+18}, or to predict pressure field changes for multiple subsequent time-steps with LSTM~\cite{WBT18}.

\subsection{Regression Fluid}
In the following, we give an overview of the regression forest approach for SPH presented by Ladick\'{y} et al.~\shortcite{LJS+15}. The work aimed at enabling real-time applications of millions of particles for games and virtual reality applications. The main idea is to formulate an SPH solver as a regression problem, where the acceleration (or velocity) of each particle at time $\Time+\Dt$ is efficiently estimated given the state at time $\Time$. 

As input to the regressor, a feature vector $\Phi_{\Position[i]}$ is evaluated for each particle. The features are designed such that they represent the individual forces and constraints of the Navier-Stokes equations: the used features model pressure, incompressibility, viscosity and surface tension. In order to evaluate features without using an explicit neighbor search step, context-based integral features are computed that are defined as flat-kernel sums of rectangular regions surrounding a particle. The different box sizes allow to capture the behavior of both close and distant particles, and the features can be evaluated in constant time and are robust to small deviations. More details on the computation of context-based integral features can be found in~\cite{LJS+15}.
Three different ways were considered for the learning strategy:

\noindent \textbf{1. Learning na\"{\i}ve prediction:} The first approach directly learns particle accelerations $\Acceleration$, given the evaluated features at state $t$. The regression problem is formulated as 
\begin{equation}
\Acceleration[i](t):=Reg(\Phi_{\Position[i]}),
\end{equation}
where $Reg(.)$ is the learned regression function. Velocities and positions are then integrated with
\begin{eqnarray}
\Velocity[i](t+\Dt) &=& \Velocity[i](t) + \Acceleration[i](t) \Dt 
\label{eq:advection_learning1}\\
\Position[i](t+\Dt) &=& \Position[i](t) + \frac{\Velocity[i](t) + \Velocity[i](t+\Dt)}{2}\Dt.
\label{eq:advection_learning2}
\end{eqnarray}
This formulation mimics standard SPH and hence does not consider incompressibility. 

\noindent \textbf{2. Learning prediction with hindsight:} The second strategy first computes external forces, advects particles, and applies collision handling. Then, this intermediate state with particle positions $\Position[i]^*$ is used to compute integral features. 
The regression learns a corrective acceleration and is defined as
\begin{equation}
\Acceleration[i](t):=Reg(\Phi_{\Position[i]^*}),
\end{equation}
followed by advection as in Eqs.~\eqref{eq:advection_learning1} and~\eqref{eq:advection_learning2}. Unlike the na\"{\i}ve approach, the regressor is able to predict compressions and hence to counteract those with a corrective acceleration. Conceptually, this approach mimics PCISPH~\cite{SP09}. 

\noindent \textbf{3. Learning correction:} The third approach starts similarly as the second one, but instead of learning accelerations, corrective velocities are computed. The regression problem is defined as
\begin{equation}
\Delta \Velocity[i]^{corr}:=Reg(\Phi_{\Position[i]^*}),
\end{equation}
and positions and velocities are updated with 
\begin{eqnarray}
\Velocity[i](t+\Dt) &=& \Velocity[i]^*(t) + \Delta \Velocity[i]^{corr} \\
\Position[i](t+\Dt) &=& \Position[i]^*(t) + \frac{\Delta \Velocity[i]^{corr}}{2}\Dt.
\end{eqnarray}
This approach counteracts compressions as well, and conceptually mimics PBF~\cite{MM13}. Unlike PBF, the regressor takes into account information from a larger neighborhood, and hence does not require several iterations to converge.

For training the regression forest, 165 scenes consisting of 1-6 million particles and moving obstacles (sphere, box, cylinder) were randomly generated and computed for 6 seconds. The training time was 4 days on 12 CPUs, and the size of the resulting model was about 40 MB. With the regression fluid approach it is possible to simulate 1 to 1.5 million particles in real-time, and hence this approach represents an attractive alternative to traditional solvers for games and virtual reality applications (Fig.~\ref{fig:regression_forest}). 

With the na\"{\i}ve prediction, strong compression artifacts are visible. The system cannot self-correct in the next frames since the model has never seen such distorted states during the training. Both prediction with hindsight and learning corrections can handle incompressibility well, however the third approach seems to lead to smaller errors compared to the ground truth data. Additionally, with the second and third approaches, external forces can be added without retraining the model. 
This allows adding surface tension, friction, or drag effects at runtime to mimic different material properties as illustrated in Fig.~\ref{fig:regression_forest}. 
The disadvantage of the regression fluids approach - and in fact of all machine learning based strategies - is that learning methods are not capable to extrapolate the model far outside the training data (\eg when domain size or fluid resolution change).


\section{SPlisHSPlasH}
\label{sec:splishsplash}

\begin{figure*}[t]
	\centering
	\includegraphics[height=3.9cm]{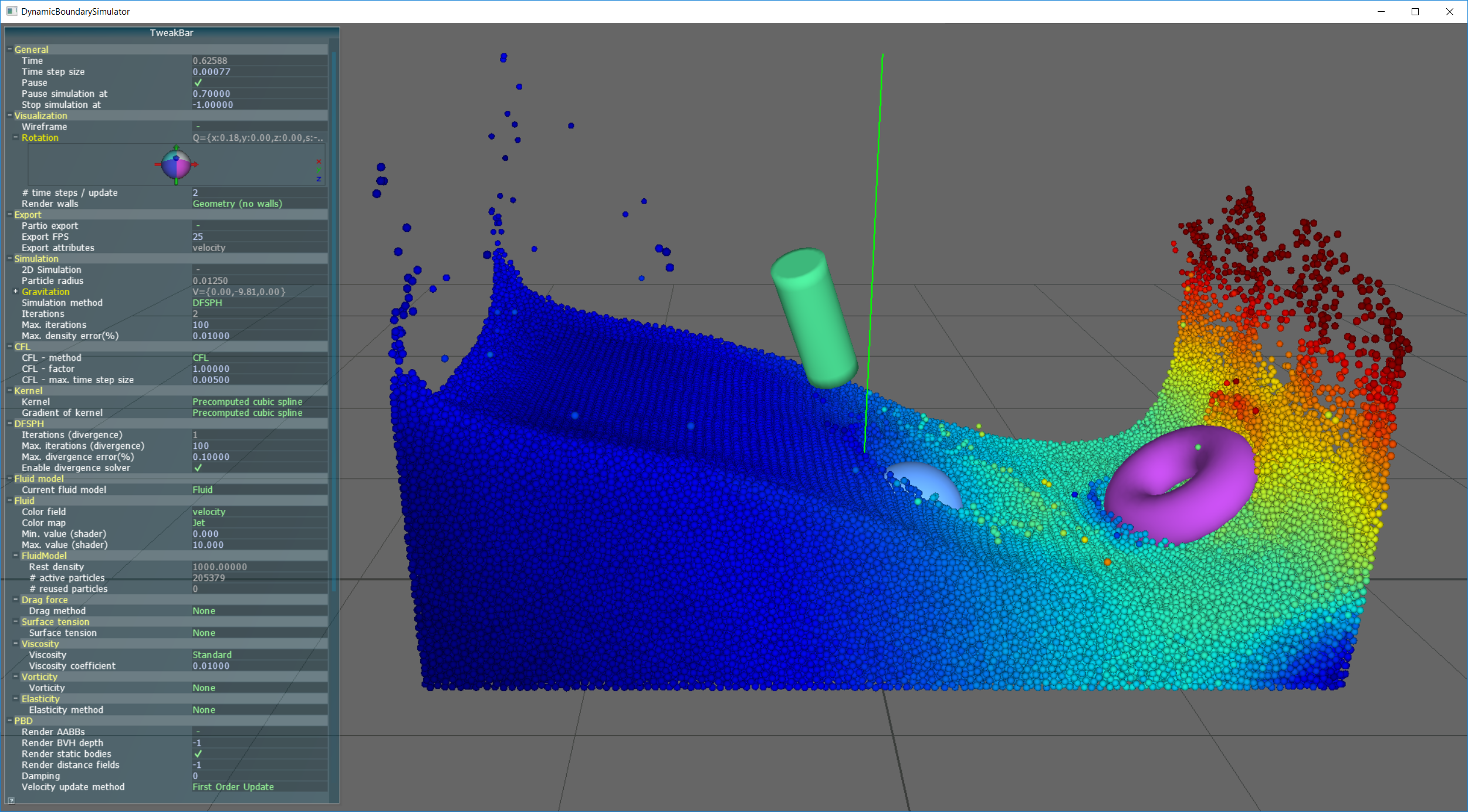} \hfill
	\includegraphics[height=3.9cm]{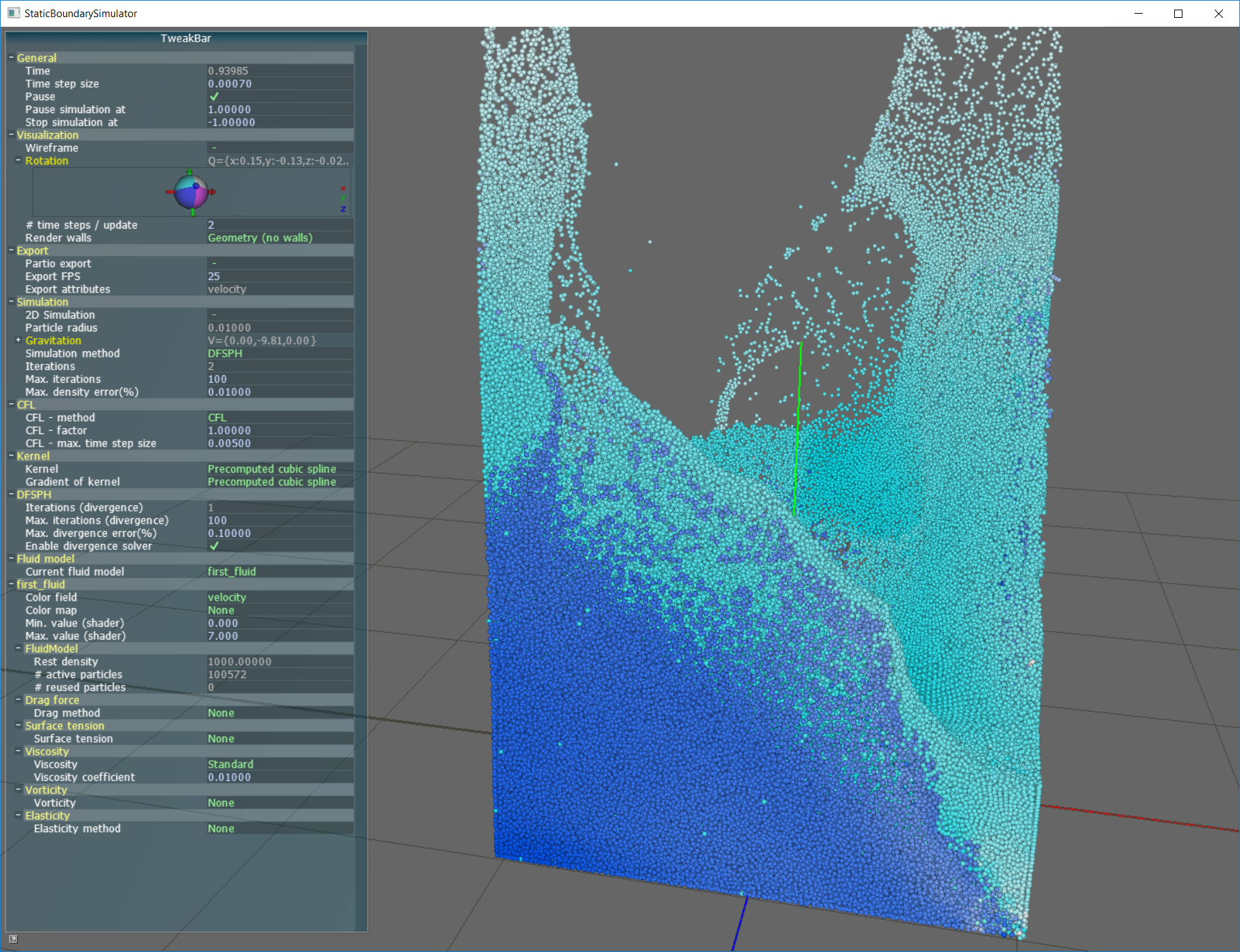} \hfill
	\includegraphics[height=3.9cm]{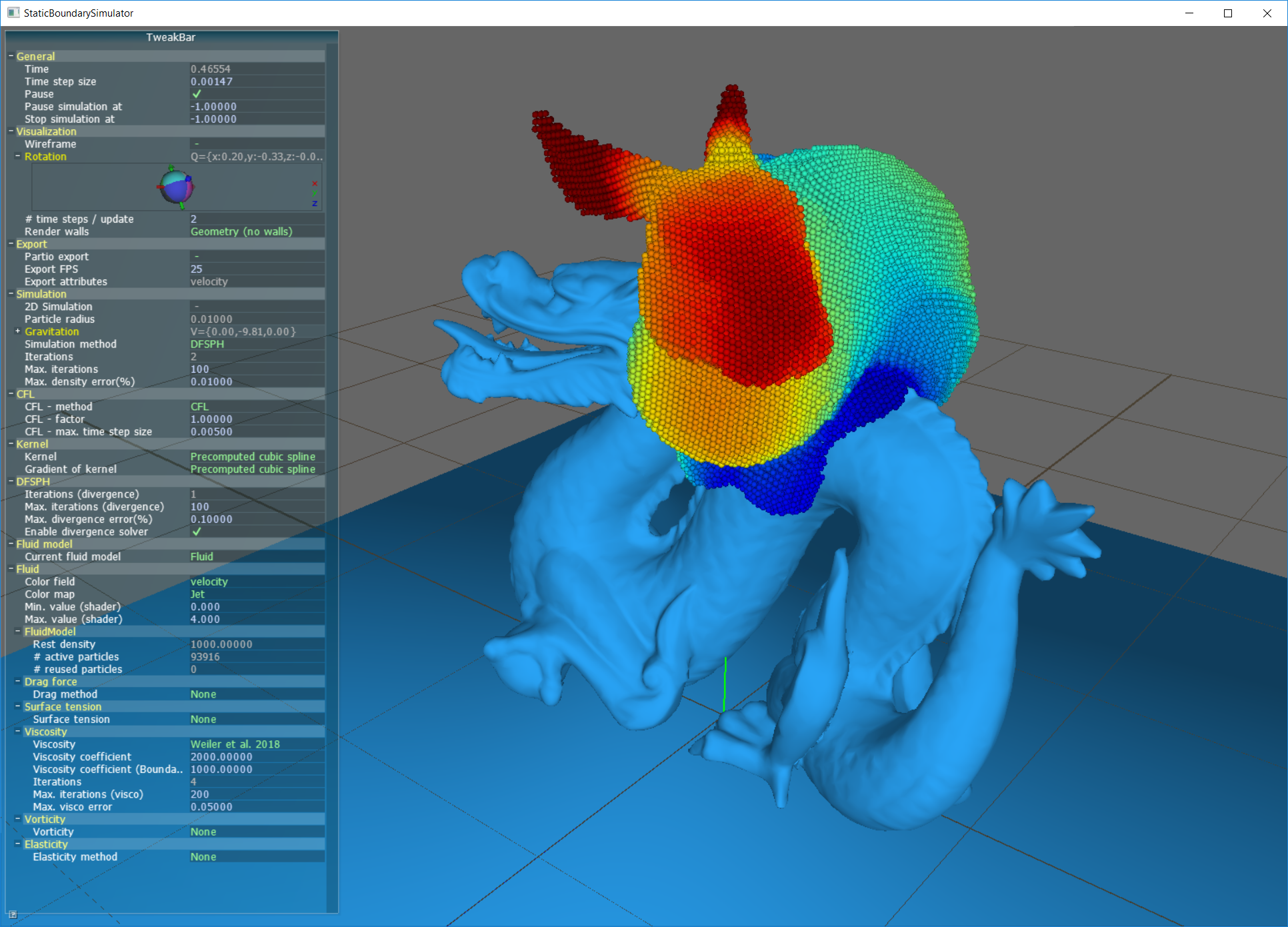} 
	\caption{Screenshots of the SPlisHSPlasH fluid simulation framework. Left: Two-way coupling of dynamic rigid bodies and a color-coded fluid. Center: Two-phase double dam break simulation. Right: Highly viscous bunny collides with a static dragon model.
		}		
	\label{fig:SPlisHSPlasH}
\end{figure*}

In this section we want to introduce SPlisHSPlasH~\cite{SPlisHSPlasH} which is an SPH-based open-source library for the physically-based simulation of fluids (see Fig.~\ref{fig:SPlisHSPlasH}).
The SPlisHSPlasH framework contains a reference implementation of many of the methods introduced in this tutorial and several simulations shown in the figures were performed using this library.
Therefore, we think that our open-source framework perfectly supplements these course notes.

In the current version SPlisHSPlasH implements six of the most popular explicit and implicit pressure solvers~\cite{BT07,SP09,MM13,ICS+14,BK15,WKB16} which enable the simulation of incompressible fluids with several million particles. 
Moreover, the explicit~\cite{SB12,Mon92} and implicit viscosity methods~\cite{TDF+15,PICT15,PT16,BK17,WKBB18} introduced in Section~\ref{sec:viscosity} are implemented. 
Hence, the library supports the simulation of low viscous flow and highly viscous materials.
Surface tension effects (see Section~\ref{sec:surface_tension}) can also be simulated using SPlisHSPlasH.
The framework currently implements microscopic and macroscopic approaches~\cite{BT07,AAT13,HWZ+14}.
To simulate turbulent fluids, SPlisHSPlasH implements vorticity confinement~\cite{MM13} and the micropolar model~\cite{BKKW17} which were discussed in Section~\ref{sec:vorticity}.

Aside from forces which act within the same phase, SPlisHSPlasH also supports multiphase simulations~\cite{SP08} and provides functionality to couple different materials.
The interaction between air phase and fluid phase is realized using drag forces~\cite{MMCK14,GBP+17a}.
The framework implements the approach of Akinci et al.~\cite{AIA+12} to simulate the coupling between rigid bodies and fluids. 
The required surface sampling of the bodies is performed automatically using a Poisson disk sampling.
For the simulation of dynamic rigid bodies SPlisHSPlasH uses the open-source PositionBasedDynamics library~\cite{PBD}. 
This library simulates the rigid bodies using a position-based approach~\cite{DCB14}.
Collisions between the bodies are efficiently detected using signed distance fields~\cite{KDBB17} while the contacts are resolved using a projected Gauss-Seidel method~\cite{BET14}.
Finally, SPlisHSPlasH implements different methods for the simulation of deformable solids~\cite{BIT09,PGBT17} using an SPH formulation (see Section~\ref{sec:deformable_solids}). 
Since an SPH formulation is used, the two-way coupling between solids and fluids is simply handled by the implemented multiphase method.

SPlisHSPlasH uses a neighborhood search based on the compact hashing approach of Ihmsen et al.~\cite{IABT11}.
This approach is discussed in more detail in Section~\ref{sec:neighborhood_search}.
The neighborhood search is implemented in our open-source library CompactNSearch~\cite{CompactNSearch}.

The SPlisHSPlasH framework has many more features like emitters, adaptive time-stepping or the support of different kernel functions.
Moreover, the library has some useful tools like volume sampling of closed geometries or the export of particle data for Maya or Houdini.
New simulation scenarios can be created easily using a JSON-based scene file format.
Finally, due to a modular concept it is simple to extend the library and to integrate own SPH methods. 
Therefore, we think that SPlisHSPlasH is a good starting point for all beginners in the area of SPH-based simulations.


\section{Conclusion}
\label{sec:conclusion}

This tutorial introduced state-of-the-art SPH techniques for the physics based simulation of fluids and solids in graphics and presented practical guidelines for implementations.
Various concepts that are particularly relevant for graphics applications were discussed.
We showed that with the recent improvements SPH models have matured and ultimately emerged as a competitive alternative to Eulerian fluid simulations or hybrid approaches.
Particular challenges of SPH concerning neighborhood search algorithms, pressure solvers, or versatile fluid-solid interaction techniques have been overcome.
With the improved robustness and efficiency, millions of particles can today be simulated on a single desktop computer.
Accordingly, the Lagrangian SPH method reaches an unprecedented level of visual quality, where fine-scale surface effects and flow details are reliably captured. 

The SPH community -- in graphics as well as in other research disciplines -- is very active and the field advances quickly.
Each community contributes to different aspects of SPH simulations, and the research often finds applications across disciplines. 
For graphics applications, it was especially important to efficiently enforce incompressibility on unstructured particles and hence to eliminate the severe time step restrictions of standard SPH techniques. We have presented a practical introduction to various SPH concepts that enforce volume conservation and/or divergence-free velocity fields.
A current difficulty is that these approaches render time stepping more challenging, since the largest possible time step does not necessarily result in the best overall performance.
Future work is certainly necessary to establish a CFL condition for these methods, as well as to overcome the current main performance bottleneck which is still the time step restriction especially when using millions of particles.

Using large particle numbers is one of the key components for high visual quality and production level results.
Such high-resolution simulations pose new challenges and existing concepts might need to be revisited. Especially speed, flexibility and controllability are core aspects, for which solutions are still largely missing. This problem, however, affects not only the SPH field but the entire fluid community in graphics likewise, and has triggered research on pre- and post-processing methods or data-driven approaches.

Our tutorial introduced the SPH-based open-source library SPlisHSPlasH that contains reference implementations of many concepts that we discussed.
This implementation is an excellent starting point for students, researchers and practitioners, and may serve as a valuable tool for future research.

	\bibliographystyle{eg-alpha}
	\bibliography{main}

\newcommand{\etalchar}[1]{$^{#1}$}
\begin{thebibliography}{\uppercase{BGFAO17}}

\bibitem[AAT13]{AAT13}
\textsc{Akinci N., Akinci G., Teschner M.}:
\newblock {Versatile surface tension and adhesion for SPH fluids}.
\newblock \emph{ACM Transactions on Graphics 32}, 6 (2013), 1--8.

\bibitem[Abe12]{Abe12}
\textsc{Abeyaratne R.}:
\newblock \emph{Continuum Mechanics: Volume II of Lecture Notes on The
  Mechanics of Elastic Solids}.
\newblock techreport, MIT Department of Mechanical Engineering, 2012.

\bibitem[AIA{\etalchar{*}}12]{AIA+12}
\textsc{Akinci N., Ihmsen M., Akinci G., Solenthaler B., Teschner M.}:
\newblock {Versatile rigid-fluid coupling for incompressible SPH}.
\newblock \emph{ACM Transactions on Graphics 31}, 4 (July 2012), 1--8.

\bibitem[AMS{\etalchar{*}}07]{AMS+06}
\textsc{Agertz O., Moore B., Stadel J., Potter D., Miniati F., Read J., Mayer
  L., Gawryszczak A., Kravtsov A., Monaghan J., Nordlund A., Pearce F., Quilis
  V., Rudd D., Springel V., Stone J., Tasker E., Teyssier R., Wadsley J.,
  Walder R.}:
\newblock Fundamental differences between {SPH} and grid methods.
\newblock \emph{Mon. Not. R. Astron. Soc. 380}, 3 (2007), 963--978.

\bibitem[AW09]{AW09}
\textsc{Adams B., Wicke M.}:
\newblock {Meshless Approximation Methods and Applications in Physics Based
  Modeling and Animation}.
\newblock In \emph{Proceedings of the Eurographics conference} (2009), EG '09,
  Eurographics Association, pp.~213--239.

\bibitem[Ben19a]{PBD}
\textsc{Bender J.}:
\newblock {PositionBasedDynamics Library}.
\newblock \\
  \url{https://github.com/InteractiveComputerGraphics/PositionBasedDynamics},
  2019.

\bibitem[Ben19b]{SPlisHSPlasH}
\textsc{Bender J.}:
\newblock {SPlisHSPlasH Library}.
\newblock \\ \url{https://github.com/InteractiveComputerGraphics/SPlisHSPlasH},
  2019.

\bibitem[BET14]{BET14}
\textsc{Bender J., Erleben K., Trinkle J.}:
\newblock {Interactive Simulation of Rigid Body Dynamics in Computer Graphics}.
\newblock \emph{Computer Graphics Forum 33}, 1 (2014), 246--270.

\bibitem[BGFAO17]{BGAO17}
\textsc{Barreiro H., Garc\'{\i}a-Fern\'andez I., Aldu\'an I., Otaduy M.~A.}:
\newblock Conformation constraints for efficient viscoelastic fluid simulation.
\newblock \emph{ACM Transactions on Graphics 36}, 6 (2017), 221.1--221.11.

\bibitem[BGI{\etalchar{*}}18]{BGI+18}
\textsc{Band S., Gissler C., Ihmsen M., Cornelis J., Peer A., Teschner M.}:
\newblock Pressure boundaries for implicit incompressible sph.
\newblock \emph{ACM Transactions on Graphics 37}, 2 (Feb. 2018), 14:1--14:11.

\bibitem[BGPT18]{BGPT18}
\textsc{Band S., Gissler C., Peer A., Teschner M.}:
\newblock {MLS} pressure boundaries for divergence-free and viscous {SPH}
  fluids.
\newblock \emph{Computers \& Graphics 76} (nov 2018), 37--46.

\bibitem[BIT09]{BIT09}
\textsc{Becker M., Ihmsen M., Teschner M.}:
\newblock {Corotated SPH for deformable solids}.
\newblock In \emph{Proceedings of Eurographics Conference on Natural Phenomena}
  (2009), pp.~27--34.

\bibitem[BK15]{BK15}
\textsc{Bender J., Koschier D.}:
\newblock {Divergence-Free Smoothed Particle Hydrodynamics}.
\newblock In \emph{ACM SIGGRAPH/Eurographics Symposium on Computer Animation}
  (2015), pp.~1--9.

\bibitem[BK17]{BK17}
\textsc{Bender J., Koschier D.}:
\newblock {Divergence-Free SPH for Incompressible and Viscous Fluids}.
\newblock \emph{IEEE Transactions on Visualization and Computer Graphics 23}, 3
  (2017), 1193--1206.

\bibitem[BKCW14]{BKCW14}
\textsc{Bender J., Koschier D., Charrier P., Weber D.}:
\newblock {Position-Based Simulation of Continuous Materials}.
\newblock \emph{Computers \& Graphics 44}, 1 (2014), 1--10.

\bibitem[BKKW17]{BKKW17}
\textsc{Bender J., Koschier D., Kugelstadt T., Weiler M.}:
\newblock A micropolar material model for turbulent sph fluids.
\newblock In \emph{ACM SIGGRAPH/Eurographics Symposium on Computer Animation}
  (July 2017), pp.~1--8.

\bibitem[BKKW18]{BKKW18}
\textsc{Bender J., Koschier D., Kugelstadt T., Weiler M.}:
\newblock Turbulent micropolar sph fluids with foam.
\newblock \emph{IEEE Transactions on Visualization and Computer Graphics}
  (2018).

\bibitem[BL99]{BL99}
\textsc{Bonet J., Lok T.-S.}:
\newblock Variational and momentum preservation aspects of smooth particle
  hydrodynamic formulations.
\newblock \emph{Computer Methods in Applied Mechanics and Engineering 180}, 1
  (1999), 97 -- 115.

\bibitem[BLS12]{BLS12}
\textsc{Bodin K., Lacoursi{\`{e}}re C., Servin M.}:
\newblock {Constraint fluids}.
\newblock \emph{IEEE Transactions on Visualization and Computer Graphics 18}
  (2012), 516--526.

\bibitem[BMM14]{BMM14}
\textsc{Bender J., M{\"{u}}ller M., Macklin M.}:
\newblock {A Survey on Position-Based Simulation Methods in Computer Graphics}.
\newblock \emph{Computer Graphics Forum 33}, 6 (2014), 228--251.

\bibitem[BMM17]{BMM17}
\textsc{Bender J., M{\"u}ller M., Macklin M.}:
\newblock A survey on position based dynamics, 2017.
\newblock In \emph{EUROGRAPHICS 2017 Tutorials} (2017), Eurographics
  Association.

\bibitem[Bri15]{Bri15}
\textsc{Bridson R.}:
\newblock \emph{Fluid Simulation for Computer Graphics, Second Edition}.
\newblock Taylor \& Francis, 2015.

\bibitem[Bro85]{Bro85}
\textsc{Brookshaw L.}:
\newblock A method of calculating radiative heat diffusion in particle
  simulations.
\newblock \emph{Publications of the Astronomical Society of Australia 6}, 2
  (1985), 207–210.

\bibitem[BT07]{BT07}
\textsc{Becker M., Teschner M.}:
\newblock {Weakly compressible SPH for free surface flows}.
\newblock In \emph{ACM SIGGRAPH/Eurographics Symposium on Computer Animation}
  (2007), pp.~1--8.

\bibitem[CBG{\etalchar{*}}18]{CBG+18}
\textsc{Cornelis J., Bender J., Gissler C., Ihmsen M., Teschner M.}:
\newblock An optimized source term formulation for incompressible {SPH}.
\newblock \emph{The Visual Computer} (Feb. 2018).

\bibitem[CIPT14]{CIPT14}
\textsc{Cornelis J., Ihmsen M., Peer A., Teschner M.}:
\newblock {IISPH}-{FLIP} for incompressible fluids.
\newblock \emph{Computer Graphics Forum 33}, 2 (may 2014), 255--262.

\bibitem[CM99]{CM99}
\textsc{Cleary P.~W., Monaghan J.~J.}:
\newblock Conduction modelling using smoothed particle hydrodynamics.
\newblock \emph{Journal of Computational Physics 148}, 1 (1999), 227 -- 264.

\bibitem[{Com}16a]{Video:DFSPH}
\textsc{{Computer Animation, RWTH Aachen University}}:
\newblock Divergence-free sph for incompressible and viscous fluids.
\newblock \url{www.youtube.com/watch?v=tl4mx0TtaAc}, 2016.

\bibitem[{Com}16b]{Video:500million}
\textsc{{Computer Graphics, University of Freiburg}}:
\newblock Terrain 2 - up to 500 million particles with {PreonLab} ({FIFTY2}).
\newblock \url{www.youtube.com/watch?v=4y-VBLzA9Mw}, 2016.

\bibitem[{Com}17]{Video:Viscous}
\textsc{{Computer Graphics, University of Freiburg}}:
\newblock Ship under attack.
\newblock \url{www.youtube.com/watch?v=_O6fqLOCTew}, 2017.

\bibitem[{Com}18]{Video:Elastic}
\textsc{{Computer Graphics, University of Freiburg}}:
\newblock An implicit {SPH} formulation for incompressible linearly elastic
  solids.
\newblock \url{www.youtube.com/watch?v=qd3gKVX89qo}, 2018.

\bibitem[DCB14]{DCB14}
\textsc{Deul C., Charrier P., Bender J.}:
\newblock Position-based rigid-body dynamics.
\newblock \emph{Computer Animation and Virtual Worlds 27}, 2 (Sept. 2014),
  103--112.

\bibitem[dGWH{\etalchar{*}}15]{dGWH+15}
\textsc{de~Goes F., Wallez C., Huang J., Pavlov D., Desbrun M.}:
\newblock {Power Particles: An incompressible fluid solver based on power
  diagrams}.
\newblock \emph{ACM Transactions on Graphics 34}, 4 (2015), 50:1--50:11.

\bibitem[ER03]{ER03}
\textsc{Espanol P., Revenga M.}:
\newblock Smoothed dissipative particle dynamics.
\newblock \emph{Physical Review E 67}, 2 (2003), 026705.

\bibitem[FAW17]{FAW17}
\textsc{Fürstenau J.-P., Avci B., Wriggers P.}:
\newblock {A comparative numerical study of pressure-Poisson-equation
  discretization strategies for SPH}.
\newblock In \emph{12th International SPHERIC Workshop} (2017).

\bibitem[{FIF}16]{Video:PreonLab}
\textsc{{FIFTY2 Technology}}:
\newblock {PreonLab Promotion}.
\newblock \url{www.youtube.com/watch?v=giS2r5JPgy0}, 2016.

\bibitem[FLR{\etalchar{*}}13]{FLR+13}
\textsc{Ferrand M., Laurence D.~R., Rogers B.~D., Violeau D., Kassiotis C.}:
\newblock {Unified semi-analytical wall boundary conditions for inviscid,
  laminar or turbulent flows in the meshless SPH method}.
\newblock \emph{International Journal for Numerical Methods in Fluids 71}, 4
  (Feb. 2013), 446--472.

\bibitem[FM15]{FM15}
\textsc{Fujisawa M., Miura K.~T.}:
\newblock {An Efficient Boundary Handling with a Modified Density Calculation
  for SPH}.
\newblock \emph{Computer Graphics Forum 34}, 7 (2015), 155--162.

\bibitem[FMH{\etalchar{*}}94]{FMH+94}
\textsc{{Flebbe} O., {Muenzel} S., {Herold} H., {Riffert} H., {Ruder} H.}:
\newblock {Smoothed Particle Hydrodynamics: Physical viscosity and the
  simulation of accretion disks}.
\newblock \emph{The Astrophysical Journal 431} (Aug. 1994), 754--760.

\bibitem[GAC{\etalchar{*}}09]{GAC+09}
\textsc{Grenier N., Antuono M., Colagrossi A., Touz\'{e} D.~L., Alessandrini
  B.}:
\newblock An {H}amiltonian interface {SPH} formulation for multi-fluid and free
  surface flows.
\newblock \emph{Journal of Computational Physics 228}, 22 (2009), 8380 -- 8393.

\bibitem[Gan15]{Gan15}
\textsc{Ganzenmüller G.~C.}:
\newblock An hourglass control algorithm for lagrangian smooth particle
  hydrodynamics.
\newblock \emph{Computer Methods in Applied Mechanics and Engineering 286} (apr
  2015), 87--106.

\bibitem[GBP{\etalchar{*}}17]{GBP+17a}
\textsc{Gissler C., Band S., Peer A., Ihmsen M., Teschner M.}:
\newblock Generalized drag force for particle-based simulations.
\newblock \emph{Computers \& Graphics 69} (dec 2017), 1--11.

\bibitem[GM77]{GM77}
\textsc{Gingold R.~a., Monaghan J.}:
\newblock {Smoothed Particle Hydrodynamics: Theory and Application to
  Non-Spherical Stars}.
\newblock \emph{Monthly Notices of the Royal Astronomical Society}, 181 (1977),
  375--389.

\bibitem[GPB{\etalchar{*}}19]{GPB+19}
\textsc{Gissler C., Peer A., Band S., Bender J., Teschner M.}:
\newblock Interlinked sph pressure solvers for strong fluid-rigid coupling.
\newblock \emph{ACM Transactions on Graphics 38}, 1 (Jan. 2019), 5:1--5:13.

\bibitem[HA06]{HA06}
\textsc{Hu X., Adams N.}:
\newblock A multi-phase {SPH} method for macroscopic and mesoscopic flows.
\newblock \emph{Journal of Computational Physics 213}, 2 (2006), 844--861.

\bibitem[HKK07a]{HKK07}
\textsc{Harada T., Koshizuka S., Kawaguchi Y.}:
\newblock {Smoothed particle hydrodynamics in complex shapes}.
\newblock In \emph{Spring Conference on Computer Graphics} (2007),
  pp.~191--197.

\bibitem[HKK07b]{HKK07a}
\textsc{Harada T., Koshizuka S., Kawaguchi Y.}:
\newblock {Smoothed Particle Hydrodynamics on GPUs}.
\newblock In \emph{Computer Graphics International} (2007), pp.~63--70.

\bibitem[HLL{\etalchar{*}}12]{HLL+12}
\textsc{He X., Liu N., Li S., Wang H., Wang G.}:
\newblock {Local Poisson SPH for Viscous Incompressible Fluids}.
\newblock \emph{Computer Graphics Forum 31} (2012), 1948--1958.

\bibitem[Hoo98]{Hoover98}
\textsc{Hoover W.}:
\newblock Isomorphism linking smooth particles and embedded atoms.
\newblock \emph{Physica A: Statistical Mechanics and its Applications 260}, 3
  (1998), 244--254.

\bibitem[HS13]{HS13}
\textsc{Horvath C.~J., Solenthaler B.}:
\newblock Mass preserving multi-scale {SPH}.
\newblock Pixar Technical Memo 13-04, Pixar Animation Studios, 2013.

\bibitem[HWZ{\etalchar{*}}14]{HWZ+14}
\textsc{He X., Wang H., Zhang F., Wang H., Wang G., Zhou K.}:
\newblock {Robust Simulation of Sparsely Sampled Thin Features in SPH-Based
  Free Surface Flows}.
\newblock \emph{ACM Transactions on Graphics 34}, 1 (2014), 7:1--7:9.

\bibitem[IAAT12]{IAAT12}
\textsc{Ihmsen M., Akinci N., Akinci G., Teschner M.}:
\newblock Unified spray, foam and air bubbles for particle-based fluids.
\newblock \emph{The Visual Computer 28}, 6-8 (2012), 669--677.

\bibitem[IABT11]{IABT11}
\textsc{Ihmsen M., Akinci N., Becker M., Teschner M.}:
\newblock {A Parallel SPH Implementation on Multi-Core CPUs}.
\newblock \emph{Computer Graphics Forum 30}, 1 (Mar. 2011), 99--112.

\bibitem[IAGT10]{IAGT10}
\textsc{Ihmsen M., Akinci N., Gissler M., Teschner M.}:
\newblock {Boundary handling and adaptive time-stepping for PCISPH}.
\newblock In \emph{Virtual Reality Interactions and Physical Simulations}
  (2010), pp.~79--88.

\bibitem[ICS{\etalchar{*}}14]{ICS+14}
\textsc{Ihmsen M., Cornelis J., Solenthaler B., Horvath C., Teschner M.}:
\newblock {Implicit incompressible SPH}.
\newblock \emph{IEEE Transactions on Visualization and Computer Graphics 20}, 3
  (2014), 426--435.

\bibitem[IOS{\etalchar{*}}14]{IOS+14}
\textsc{Ihmsen M., Orthmann J., Solenthaler B., Kolb A., Teschner M.}:
\newblock {SPH Fluids in Computer Graphics}.
\newblock \emph{Eurographics (State of the Art Reports)} (2014), 21--42.

\bibitem[JSD04]{JSD04}
\textsc{Jubelgas M., Springel V., Dolag K.}:
\newblock {Thermal conduction in cosmological SPH simulations}.
\newblock \emph{{Monthly Notices of the Royal Astronomical Society} 351}, 2
  (2004), 423--435.

\bibitem[KAT{\etalchar{*}}19]{KAT19}
\textsc{Kim B., Azevedo V., Thuerey N., Gross M., Solenthaler B.}:
\newblock Deep fluids: A generative network for parameterized fluid
  simulations.
\newblock \emph{Computer Graphics Forum} (2019).

\bibitem[KB17]{KB17}
\textsc{Koschier D., Bender J.}:
\newblock Density maps for improved sph boundary handling.
\newblock In \emph{ACM SIGGRAPH/Eurographics Symposium on Computer Animation}
  (July 2017), pp.~1--10.

\bibitem[KBT17]{KBT17}
\textsc{Koschier D., Bender J., Thuerey N.}:
\newblock {Robust eXtended Finite Elements for Complex Cutting of Deformables}.
\newblock \emph{ACM Transactions on Graphics 36}, 4 (2017), 55:1--55:13.

\bibitem[KDBB17]{KDBB17}
\textsc{Koschier D., Deul C., Brand M., Bender J.}:
\newblock An hp-adaptive discretization algorithm for signed distance field
  generation.
\newblock \emph{IEEE Transactions on Visualization and Computer Graphics 23},
  10 (2017), 2208--2221.

\bibitem[KKB18]{KKB2018}
\textsc{Kugelstadt T., Koschier D., Bender J.}:
\newblock Fast corotated {FEM} using operator splitting.
\newblock \emph{Computer Graphics Forum 37}, 8 (2018).

\bibitem[Kos19]{CompactNSearch}
\textsc{Koschier D.}:
\newblock {CompactNSearch Library}.
\newblock \\
  \url{https://github.com/InteractiveComputerGraphics/CompactNSearch}, 2019.

\bibitem[Lau11]{lautrup2011physics}
\textsc{Lautrup B.}:
\newblock \emph{Physics of Continuous Matter}.
\newblock Taylor \& Francis, 2011.

\bibitem[LJS{\etalchar{*}}15]{LJS+15}
\textsc{Ladick\'{y} L., Jeong S., Solenthaler B., Pollefeys M., Gross M.}:
\newblock Data-driven fluid simulations using regression forests.
\newblock \emph{ACM Transactions on Graphics 34}, 6 (Oct. 2015), 199:1--199:9.

\bibitem[LKR09]{LKR10}
\textsc{Lai M., Krempl E., Ruben D.}:
\newblock \emph{{Introduction to Continuum Mechanics}}.
\newblock Butterworth-Heinemann, 2009.

\bibitem[LL10]{LL10}
\textsc{Liu M., Liu G.}:
\newblock {Smoothed Particle Hydrodynamics (SPH): an Overview and Recent
  Developments}.
\newblock \emph{Archives of Computational Methods in Engineering 17}, 1 (2010),
  25--76.

\bibitem[LLP11]{LLP11}
\textsc{Liu S., Liu Q., Peng Q.}:
\newblock Realistic simulation of mixing fluids.
\newblock \emph{The Visual Computer 27}, 3 (2011), 241--248.

\bibitem[{\L}uk99]{L99}
\textsc{{\L}ukaszewicz G.}:
\newblock \emph{{Micropolar Fluids}}.
\newblock Modeling and Simulation in Science, Engineering and Technology.
  Birkh{\"{a}}user Boston, 1999.

\bibitem[MBCM16]{MBCM16}
\textsc{M\"{u}ller M., Bender J., Chentanez N., Macklin M.}:
\newblock A robust method to extract the rotational part of deformations.
\newblock In \emph{Proceedings of ACM SIGGRAPH Conference on Motion in Games}
  (2016), MIG '16, ACM.

\bibitem[MFK{\etalchar{*}}15]{MFK+15}
\textsc{Mayrhofer A., Ferrand M., Kassiotis C., Violeau D., Morel F.-X.}:
\newblock {Unified semi-analytical wall boundary conditions in SPH: analytical
  extension to 3-D}.
\newblock \emph{Numerical Algorithms 68}, 1 (Jan. 2015), 15--34.

\bibitem[Mir96]{Mir96a}
\textsc{Mirtich B.~V.}:
\newblock \emph{Impulse-based dynamic simulation of rigid body systems}.
\newblock PhD thesis, University of California at Berkeley, 1996.

\bibitem[MM13]{MM13}
\textsc{Macklin M., M{\"{u}}ller M.}:
\newblock {Position Based Fluids}.
\newblock \emph{ACM Transactions on Graphics 32}, 4 (2013), 1--5.

\bibitem[MMCK14]{MMCK14}
\textsc{Macklin M., M{\"{u}}ller M., Chentanez N., Kim T.-Y.}:
\newblock {Unified Particle Physics for Real-Time Applications}.
\newblock \emph{ACM Transactions on Graphics 33}, 4 (2014), 1--12.

\bibitem[Mon92]{Mon92}
\textsc{Monaghan J.}:
\newblock {Smoothed Particle Hydrodynamics}.
\newblock \emph{Annual Review of Astronomy and Astrophysics 30}, 1 (1992),
  543--574.

\bibitem[Mon05]{Mon05}
\textsc{Monaghan J.~J.}:
\newblock {Smoothed Particle Hydrodynamics}.
\newblock \emph{Reports on Progress in Physics 68}, 8 (2005), 1703--1759.

\bibitem[MSKG05]{MSKG05}
\textsc{M{\"{u}}ller M., Solenthaler B., Keiser R., Gross M.}:
\newblock {Particle-based fluid-fluid interaction}.
\newblock In \emph{ACM SIGGRAPH/Eurographics Symposium on Computer Animation}
  (2005), p.~237.

\bibitem[{Nex}17]{Video:RealFlow}
\textsc{{NextLimit}}:
\newblock {RealFlow Showreel 2017}.
\newblock \url{www.youtube.com/watch?v=nnv-95w1d5A}, 2017.

\bibitem[PGBT17]{PGBT17}
\textsc{Peer A., Gissler C., Band S., Teschner M.}:
\newblock An implicit sph formulation for incompressible linearly elastic
  solids.
\newblock \emph{Computer Graphics Forum} (2017), n/a--n/a.

\bibitem[PICT15]{PICT15}
\textsc{Peer A., Ihmsen M., Cornelis J., Teschner M.}:
\newblock {An Implicit Viscosity Formulation for SPH Fluids}.
\newblock \emph{ACM Transactions on Graphics 34}, 4 (2015), 1--10.

\bibitem[Pri12]{Pri12a}
\textsc{Price D.~J.}:
\newblock Smoothed particle hydrodynamics and magnetohydrodynamics.
\newblock \emph{Journal of Computational Physics 231}, 3 (Feb. 2012), 759--794.

\bibitem[PT16]{PT16}
\textsc{Peer A., Teschner M.}:
\newblock Prescribed velocity gradients for highly viscous {SPH} fluids with
  vorticity diffusion.
\newblock \emph{IEEE Transactions on Visualization and Computer Graphics}
  (2016), 1--9.

\bibitem[RL96]{RL96}
\textsc{Randles P., Libersky L.}:
\newblock Smoothed particle hydrodynamics: Some recent improvements and
  applications.
\newblock \emph{Computer Methods in Applied Mechanics and Engineering 139}, 1
  (1996), 375 -- 408.

\bibitem[RLY{\etalchar{*}}14]{RLY+14}
\textsc{Ren B., Li C., Yan X., Lin M.~C., Bonet J., Hu S.-M.}:
\newblock {Multiple-Fluid SPH Simulation Using a Mixture Model}.
\newblock \emph{ACM Transactions on Graphics 33}, 5 (2014), 1--11.

\bibitem[SB12]{SB12}
\textsc{Schechter H., Bridson R.}:
\newblock {Ghost SPH for animating water}.
\newblock \emph{ACM Transactions on Graphics 31}, 4 (2012), 61:1--61:8.

\bibitem[SF18]{SF18}
\textsc{Schenck C., Fox D.}:
\newblock Spnets: Differentiable fluid dynamics for deep neural networks.
\newblock In \emph{CoRL} (2018), vol.~87 of \emph{Proceedings of Machine
  Learning Research}, {PMLR}, pp.~317--335.

\bibitem[SG11]{SG11}
\textsc{Solenthaler B., Gross M.}:
\newblock Two-scale particle simulation.
\newblock \emph{TOG 30}, 4 (2011), 72:1--72:8.

\bibitem[Sif12]{Sif12}
\textsc{Sifakis E.}:
\newblock \emph{{SIGGRAPH 2012 Course Notes FEM Simulation of 3D Deformable
  Solids Part 1}}.
\newblock Tech. rep., University of Wisconsin-Madison, 2012.

\bibitem[SP08]{SP08}
\textsc{Solenthaler B., Pajarola R.}:
\newblock {Density Contrast SPH Interfaces}.
\newblock In \emph{ACM SIGGRAPH/Eurographics Symposium on Computer Animation}
  (2008), pp.~211--218.

\bibitem[SP09]{SP09}
\textsc{Solenthaler B., Pajarola R.}:
\newblock {Predictive-corrective incompressible SPH}.
\newblock \emph{ACM Transactions on Graphics 28}, 3 (2009), 40:1--40:6.

\bibitem[TDF{\etalchar{*}}15]{TDF+15}
\textsc{Takahashi T., Dobashi Y., Fujishiro I., Nishita T., Lin M.}:
\newblock {Implicit Formulation for SPH-based Viscous Fluids}.
\newblock \emph{Computer Graphics Forum 34}, 2 (2015), 493--502.

\bibitem[THM{\etalchar{*}}03]{THM+03}
\textsc{Teschner M., Heidelberger B., M{\"u}ller M., Pomerantes D., Gross
  M.~H.}:
\newblock Optimized spatial hashing for collision detection of deformable
  objects.
\newblock In \emph{Vmv} (2003), vol.~3, pp.~47--54.

\bibitem[TM05]{TM05}
\textsc{Tartakovsky A., Meakin P.}:
\newblock Modeling of surface tension and contact angles with smoothed particle
  hydrodynamics.
\newblock \emph{Physical Review E 72}, 2 (2005), 026301.

\bibitem[TSSP16]{TSS+16}
\textsc{Tompson J., Schlachter K., Sprechmann P., Perlin K.}:
\newblock {Accelerating Eulerian Fluid Simulation With Convolutional Networks},
  jul 2016.

\bibitem[UHT18]{UHT18}
\textsc{Um K., Hu X., Thuerey N.}:
\newblock {Liquid splash modeling with neural networks}.
\newblock \emph{CGF 37}, 8 (2018), 171--182.

\bibitem[WBF{\etalchar{*}}96]{WBF+96}
\textsc{Watkins S.~J., Bhattal A.~S., Francis N., Turner J.~A., Whitworth
  A.~P.}:
\newblock A new prescription for viscosity in smoothed particle hydrodynamics.
\newblock \emph{Astron. Astrophys. Suppl. Ser. 119}, 1 (1996), 177--187.

\bibitem[WBT18]{WBT18}
\textsc{Wiewel S., Becher M., Thuerey N.}:
\newblock {Latent-space Physics: Towards Learning the Temporal Evolution of
  Fluid Flow}, feb 2018.

\bibitem[WKB16]{WKB16}
\textsc{Weiler M., Koschier D., Bender J.}:
\newblock Projective fluids.
\newblock In \emph{ACM Motion in Games} (2016), pp.~1--6.

\bibitem[WKBB18]{WKBB18}
\textsc{Weiler M., Koschier D., Brand M., Bender J.}:
\newblock A physically consistent implicit viscosity solver for sph fluids.
\newblock \emph{Computer Graphics Forum 37}, 2 (2018).

\bibitem[XFCT18]{XFC+18}
\textsc{Xie Y., Franz E., Chu M., Thuerey N.}:
\newblock tempo{GAN}: A temporally coherent, volumetric {GAN} for
  super-resolution fluid flow.
\newblock \emph{TOG 37}, 4 (2018), 95.

\bibitem[YCR{\etalchar{*}}15]{YCR+15}
\textsc{Yang T., Chang J., Ren B., Lin M.~C., Zhang J.~J., Hu S.-M.}:
\newblock Fast multiple-fluid simulation using helmholtz free energy.
\newblock \emph{ACM Transactions on Graphics 34}, 6 (Oct. 2015), 201:1--201:11.

\bibitem[YJL{\etalchar{*}}16]{YJL+16}
\textsc{Yan X., Jiang Y.-T., Li C.-F., Martin R.~R., Hu S.-M.}:
\newblock Multiphase sph simulation for interactive fluids and solids.
\newblock \emph{ACM Transactions on Graphics 35}, 4 (July 2016), 79:1--79:11.

\end{thebibliography}

\end{document}